\documentclass[preprint,prc,aps,tightenlines,float,showpacs,showkeys,superscriptaddress,nofootinbib]{revtex4-1}
\usepackage{amsmath}
\usepackage{amssymb}
\usepackage[dvips]{graphicx,color}
\usepackage{dcolumn}
\usepackage{epsfig}
\usepackage{bm}

\usepackage{color}

\newcommand{\ket}[1]{|{#1}\rangle}
\newcommand{\bra}[1]{\langle{#1}|}
\newcommand{\inp}[2]{\langle{#1}|{#2}\rangle}

\newcommand{\slas}[1]{\not\!{#1}}
\newcommand{\sla}[1]{\not\!\!{#1}}

\def\ohalf{{\textstyle{1\over 2}}}
\def\thalf{{\textstyle{3\over 2}}}
\def\fhalf{{\textstyle{5\over 2}}}
\def\shalf{{\textstyle{7\over 2}}}
\def\nhalf{{\textstyle{9\over 2}}}

\begin{document}

\title{Nucleon resonances within a dynamical coupled-channels model of $\pi N$ and $\gamma N$ reactions}
\author{H. Kamano}
\affiliation{Research Center for Nuclear Physics, Osaka University, Ibaraki, Osaka 567-0047, Japan}
\author{S.X. Nakamura}
\affiliation{Yukawa Institute for Theoretical Physics, Kyoto University, Kyoto, 606-8502, Japan}
\author{T.-S. H. Lee}
\affiliation{Physics Division, Argonne National Laboratory, Argonne, Illinois 60439, USA}
\author{T. Sato}
\affiliation{Department of Physics, Osaka University, Toyonaka, Osaka 560-0043, Japan}

\begin{abstract}
The  nucleon resonances are investigated within a dynamical coupled-channels
model of $\pi N$ and $\gamma N$ reactions up to the invariant mass $W = 2$ GeV.
The meson-baryon ($MB$) channels included in the calculations are 
$MB =$ $\pi N$, $\eta N$, $K\Lambda$,  $K\Sigma$, and 
$\pi\pi N$ that has $\pi\Delta$, $\rho N$, and $\sigma N$ resonant components.
The meson-baryon amplitudes $T_{M'B',MB}(W)$ are calculated from solving a set of
coupled-channels integral equations defined by an interaction Hamiltonian consisting of 
(a) meson-exchange interactions  $v_{M'B',MB}$ derived from phenomenological Lagrangian, and
(b) vertex interactions $N^* \to MB$ for describing the
transition of a bare excited nucleon state $N^*$ to a meson-baryon channel $MB$.
The parameters of $v_{M'B',MB}$ are mainly constrained by the fit to the data of 
$\pi N \to \pi N$ in the low-energy region up to  $W=1.4$ GeV. 
The bare masses of $N^*$ and the $N^*\to MB$ parameters are 
then determined in simultaneous fits to the data of $\pi N \to \pi N$ up to $W=2.3$ GeV
and those of $\pi N \to \eta N, K\Lambda, K\Sigma$ 
and $\gamma N \to \pi N, \eta N, K\Lambda, K\Sigma$ up to $W = 2.1$ GeV.
The pole positions and residues of nucleon resonances are extracted
by analytically continuing the meson-baryon amplitudes $T_{M'B',MB}(W)$ 
to the complex Riemann energy surface.
From the extracted residues, we have determined 
the $N^* \to \pi N, \gamma N, \eta N, K\Lambda, K\Sigma$
transition amplitudes at resonance poles.
We compare the resonance pole positions from our analysis with 
those given by the Particle Data Group
and the recent coupled-channels analyses by the J\"ulich and Bonn-Gatchina groups.
Four results agree well only for the first $N^*$ in each spin-parity-isospin 
$(J^P,I)$ channel. 
For higher mass states, the number of states and their resonance
positions from four results do not agree well. 
We discuss the possible sources of the discrepancies and the need of
additional data from new hadron facilities such as J-PARC.
\end{abstract}


\maketitle

\section{Introduction}

One of the important problems in hadron physics is to understand the structure
of the nucleon within Quantum Chromodynamics (QCD).
Because the nucleon is a composite particle, its properties are closely
related to the spectrum and the structure of its excited states.
From the available data, we know that all of
the excited nucleon states (denoted collectively as $N^*$)
are unstable and couple strongly with
the meson-baryon continuum states to form resonances in
$\pi N$ and $\gamma N$ reactions. 
Therefore, the study of the $N^*$ resonances 
in $\pi N$ and $\gamma N$ reactions has been a well-recognized important task
in advancing our understanding of the structure of baryons.
It can also provide important information for understanding how 
the confinement and chiral symmetry breaking emerge from QCD.

In this work, we report on the results from an investigation of $N^*$ resonances with
an extension of the dynamical coupled-channels (DCC) model developed in Ref.~\cite{msl07}.
Schematically, the following coupled-channels integral equations in each partial
wave are solved within the DCC model of Ref.~\cite{msl07},
\begin{eqnarray}
T_{\beta,\alpha}(p_\beta,p_\alpha;W) = 
V_{\beta,\alpha}(p_\beta,p_\alpha; W)
+ \sum_{\gamma} \int p^{2}d p  V_{\beta,\gamma}(p_\beta, p; W)
G_{\gamma}(p; W) T_{\gamma,\alpha}(p , p_\alpha; W)  \,,
\label{eq:teq}
\end{eqnarray}
with
\begin{eqnarray}
V_{\beta,\alpha}(p_\beta,p_\alpha; W) = 
v_{\beta,\alpha}(p_\beta,p_\alpha)
+ \sum_{N^*}\frac{\Gamma^{\dagger}_{N^*,\beta}(p_\beta)
 \Gamma_{N^*,\alpha}(p_\alpha)} {W-M^0_{N^*}} \,,
\label{eq:veq}
\end{eqnarray}
where $\alpha,\beta,\delta =$ $\gamma N$, $\pi N$, $\eta N$, and
$\pi\pi N$ that has the unstable $\pi \Delta$, $\rho N$, and $\sigma N$ components,
$G_\delta (p;W)$ is the Green's function of the channel $\delta$, 
$M^0_{N^*}$ is the mass of a bare excited nucleon state $N^*$, 
$v_{\beta,\alpha}$ is defined by the meson-exchange mechanisms, and 
the vertex interaction $\Gamma_{N^*,\alpha}$ defines the $\alpha \to N^*$ transition.
We describe in Sec.~\ref{sec:dcc}
how Eqs.~(\ref{eq:teq}) and~(\ref{eq:veq}) can be cast into a form that is most
convenient for extracting the nucleon resonances from the amplitude 
$T_{\beta,\alpha}(p_\beta,p_\alpha;W)$.
In the past few years, we have applied this DCC model to analyze $\pi N$ and $\gamma N$ 
reactions with $\pi N$~\cite{jlms07,jlmss08,jklmss09}, $\eta N$~\cite{djlss08}, 
and $\pi\pi N$~\cite{kjlms09-1,kjlms09-2} final states. 
The method for extracting the nucleon resonances within the considered DCC model
was developed in Refs.~\cite{ssl09,ssl10} with the results presented 
in Refs.~\cite{sjklms10,ssl10,knls10}.
During this developing stage, the DCC model parameters were not determined by 
\textit{simultaneous} fits to all of the considered data.
In addition, the very extensive data of $K\Lambda$ and $K\Sigma$ photoproduction 
reactions were not included in the analysis.
As a step in improving our analysis, we have extended these earlier
efforts to perform a \textit{combined} analysis of the available data 
for $\pi N, \gamma N\rightarrow \pi N, \eta N, K\Lambda, K\Sigma$.
The purpose of this paper is to report on the results from this effort.

The starting point of our analysis is to extend the model, defined by 
Eqs.~(\ref{eq:teq}) and~(\ref{eq:veq}), to include the $K\Lambda$ and $K\Sigma$ channels. 
We then apply the same numerical procedures detailed in 
our previous publications~\cite{msl07,jlms07,jlmss08,ssl09,ssl10}
to perform the calculations. 
Our main effort is to determine the parameters of the interactions
$V_{\beta,\alpha}$ of Eq.~(\ref{eq:veq}) by fitting simultaneously all
of the rather extensive data, as explained later.
The nucleon resonances are then extracted from the resulting model 
by using the analytic continuation method developed in Refs.~\cite{ssl09,ssl10}.

The extraction of nucleon resonances has a long history and several 
different approaches have been developed.
To see the main features of our approach, as well as the other 
dynamical models~\cite{juelich,juelich13-1,pj-91,gross,sl,pasc,pitt-ky,fuda,ntuanl},
we briefly discuss how the nucleon resonances are extracted by the other analysis groups.
It is common to parametrize~\cite{said,bonn,bg2012,ksu,maid,jlab-yeve,mokeev,giessen,kvi}
the partial-wave amplitudes in terms of polynomial functions, the Breit-Wigner forms, 
the tree diagrams of phenomenological Lagrangian, or various combinations of them. 
The $K$-matrix method is used in these analyses to unitarize the constructed amplitudes.
In most cases, the resulting forms of partial-wave amplitudes depend algebraically on
the energy variable $W$ and hence it is rather efficient numerically to fit the data 
and extract the resonances.
The analyses~\cite{cmb,zegrab,pitt-anl} based on the unitary Carnegie-Mellon-Berkeley model 
also only involve solving algebraic equations in extracting the resonances from 
the partial-wave amplitudes.
In an approach based on a DCC model, such as the one defined by
Eqs.~(\ref{eq:teq}) and~(\ref{eq:veq}), the partial-wave amplitudes 
are calculated by solving a set of coupled-channels integral equations. 
Thus the computation effort needed in fitting the meson production data 
is considerably more complex than that of the $K$-matrix analyses. 
Furthermore, the resulting partial-wave amplitudes, 
defined by the coupled-channels integral equations such as Eqs.~(\ref{eq:teq}) and~(\ref{eq:veq}) 
used in our analysis, have complicated analytic structure and must be analyzed carefully 
to develop a correct procedure for extracting the nucleon resonances.

Compared with the models~\cite{said,bonn,bg2012,ksu,cmb, zegrab,pitt-anl} with polynomial
parameterizations of the partial-wave amplitudes, 
our approach as well as all dynamical models have
an important constraint in fitting the data.
In the polynomial fits, the parameters in each partial-wave of each channel
are adjusted independently. 
However, the dynamical models have much less
freedom in adjusting the parameters to fit the data
because the partial-wave amplitudes in all partial waves and in all reaction channels are
related to the same parameters of the meson-exchange mechanisms.

Obviously, a DCC approach is much more complex and difficult than the other 
approaches~\cite{cmb,said,ksu,zegrab,pitt-anl,maid,jlab-yeve,mokeev,giessen,kvi,bonn,bg2012}. 
This however is needed to investigate the dynamical origin and
the internal structure of the nucleon resonances. 
As can be seen from the ingredients of the interaction 
$V_{\beta,\alpha}(p_\beta,p_\alpha;W)$ in Eq.~(\ref{eq:veq}),
the dynamical model considered in our approach is aimed at exploring 
a question whether a nucleon resonance can be interpreted as a system 
of a core state surrounded by a meson cloud, 
a molecule-like meson-baryon state, or a mixture of them. 
Such an interpretation 
has been obtained for the $\Delta$ (1232) resonance
in various meson-exchange models of $\pi N$ and $\gamma N$ reactions
up to $W = 1.3$ GeV.
An example can be found in Ref.~\cite{sl}, 
where $\Delta(1232)$ was interpreted as
a baryon made of a core state and a pion cloud.
The resulting core state can be identified, qualitatively, with the
$\Delta$ of a hadron structure model with only constituent-quark degrees of freedom.
Our earlier DCC analysis~\cite{sjklms10,knls10} has also provided useful 
information on the dynamical origin of the Roper resonance,
and has provided an interpretation of the mass of the first excited nucleon state with
isospin-spin-parity $I(J^P)=1/2(1/2^+)$ predicted 
by most of the hadron structure models such as the model based on 
the Dyson-Schwinger equation of QCD~\cite{rccr11}.
The DCC analysis performed in this work is a necessary step toward improving 
our understanding of the dynamical origins and the structure of all nucleon 
resonances with mass below 2 GeV.

In Sec.~\ref{sec:dcc}, we briefly describe the formulation of the DCC model used in our analysis.
The formula for extracting nucleon resonances developed in Refs.~\cite{ssl09,ssl10} are reviewed 
in Sec.~\ref{sec:res-ex}.
The procedures for determining the model parameters are explained in Sec.~\ref{sec:fit-proc}. 
The fits to the data are presented and discussed in Sec.~\ref{sec:results}. 
The extracted nucleon resonances are given in Sec.~\ref{sec:res-para}. 
In Sec.~\ref{sec:summary}, we discuss the possible future developments.
For completeness, we also explain the essential details of our calculations
in Appendices~\ref{app:self}-\ref{app:pot-em}.
The determined model parameters are given in Appendix~\ref{app:model-para}.

\section{Dynamical Coupled-Channels Model}
\label{sec:dcc}

Because the formulation of the DCC model employed in this work has been given in detail
in Refs.~\cite{msl07,jlms07,jlmss08,jklmss09,kjlms09-1,kjlms09-2}, here we
only briefly describe the relevant equations that are needed to define
the notations for presenting our results. 
We also indicate several improvements 
we have made for performing the combined analysis
of pion- and photon-induced $\pi N$, $\eta N$, $K\Lambda$, and $K\Sigma$
production reactions.

\subsection{Hadronic amplitudes}
\label{sec:dcc-had}

Within the formulation of Ref.~\cite{msl07}, 
we apply the projection operator method~\cite{feshbach}
to cast the partial-wave components of the $T$ matrix elements of the meson-baryon reactions, 
$M(\vec k) + B(-\vec k) \to M'(\vec k') + B'(-\vec k')$, into the form
\begin{equation}
T_{M'B',MB}(k',k;W) = t_{M'B',MB}(k',k;W) + t^R_{M'B',MB}(k',k;W),
\label{eq:tmbmb}
\end{equation}
where $W$ is the total energy, $k$ and $k'$ are the 
meson-baryon relative momenta in the center of mass frame, and 
$MB, M'B' = \pi N, \eta N, \pi\Delta, \rho N, \sigma N, K\Lambda, K\Sigma$
are the reaction channels included in this analysis.
[The label ``$MB$'' also specifies quantum numbers (spin, parity, isospin etc)
associated with the channel $MB$.]

The ``non-resonant'' amplitude $t_{M'B',MB}(k',k;W)$ in Eq.~(\ref{eq:tmbmb})
is defined by a set of coupled-channels integral equations\footnote{
Because of the perturbative nature of the electromagnetic interactions, it is
only necessary to solve the coupled-channels equations in the channel space
excluding the $\gamma N$ channel. The $\gamma N \rightarrow \pi N$ amplitude
up to the order of $e=\sqrt{4\pi/137}$ can
then be calculated  perturbatively, as can be seen in Sec.~\ref{sec:dcc-ele}.
}
\begin{eqnarray}
t_{M'B',MB}(k',k;W) &=& V_{M'B',MB}(k',k;W) 
\nonumber \\
&& 
+ \sum_{M^{''} B^{''}} \int_{C_{M''B''}} k''^{2}dk'' V_{M'B',M''B''}(k',k'';W)
\nonumber\\
&&
\qquad\qquad\qquad
\times
G_{M''B''}(k'';W) t_{M''B'',MB}(k'',k;W).
\label{eq:cc-eq}
\end{eqnarray}
Here $C_{M''B''}$ is the integration path,
which is taken from $0$ to $\infty$ for the physical $W$;
the summation $\sum_{MB}$ runs over the orbital angular momentum and total spin indices
for all $MB$ channels allowed in a given partial wave;
$G_{M''B''}(k;W)$ are the meson-baryon Green's functions.
Defining  $E_\alpha(k)=[m^2_\alpha + k^2]^{1/2}$ with $m_\alpha$ being
the mass of a particle $\alpha$,
the meson-baryon Green's functions in the above equations are:
\begin{eqnarray}
G_{MB}(k;W)=\frac{1}{W-E_M(k)-E_B(k) + i\epsilon},
\label{eq:prop-stab}
\end{eqnarray}
for the stable $\pi N$, $\eta N$, $K\Lambda$, and $K\Sigma$  channels, and
\begin{equation}
G_{MB}(k;W)=\frac{1}{W-E_M(k)-E_B(k) -\Sigma_{MB}(k;W)},
\label{eq:prop-unstab}
\end{equation}
for the unstable $\pi\Delta$, $\rho N$, and $\sigma N$ channels.
The details of the self energy $\Sigma_{MB}(k;W)$ in Eq.~(\ref{eq:prop-unstab})
are given in Appendix~\ref{app:self}.
The branch points of the meson-baryon Green's functions $G_{MB}(k;W)$,
as defined by zeros of the denominator of Eqs.~(\ref{eq:prop-stab}) and~(\ref{eq:prop-unstab}),
are related closely to the search of resonance positions on the complex energy surface,
as explained in Sec.~\ref{sec:res-ex}.
In Table~\ref{tab:mbgreen-branch}, we list the branch points of 
the $\pi \Delta$, $\rho N$, and $\sigma N$ channels within the model considered here.

\begin{table}[t]
\caption{\label{tab:mbgreen-branch}
Branch points of the meson-baryon Green's functions $G_{MB}(k;W)$ 
for unstable channels $MB=\pi\Delta,\sigma N,\rho N$.
For the calculation of the Green's functions, we take $m_N=938.5$ MeV and $m_\pi=138.5$ MeV.
}
\begin{ruledtabular}
\begin{tabular}{lr}
            & Branch point (MeV)  \\\hline
$\pi\Delta$ & $(1211.5 -i55.0) + m_\pi$  \\
$\sigma N$  & $(483.7-i185.8)  + m_N$   \\
            & $(1032.3-i247.7) + m_N$ \\
$\rho N$    & $(765.5 -i75.0)  + m_N$  
\end{tabular}
\end{ruledtabular}
\end{table}
\begin{figure}[t]
\includegraphics[clip,width=0.8\textwidth]{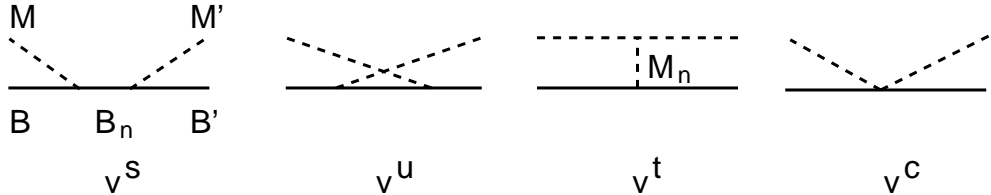}
\caption{\label{fig:mex} Meson-exchange mechanisms.}
\end{figure}
\begin{figure}[t]
\includegraphics[clip,width=0.8\textwidth]{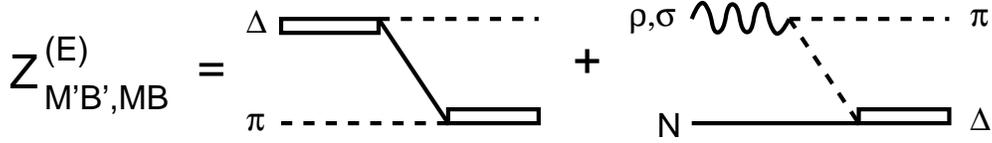}
\caption{\label{fig:z-diag} $Z$-diagram  mechanisms.}
\end{figure}

The driving terms of Eq.~(\ref{eq:cc-eq}) are
\begin{equation}
V_{M'B',MB}(k',k;W) = v_{M'B',MB}(k',k) + Z^{(E)}_{M'B',MB}(k',k;W). 
\label{eq:veff-mbmb}
\end{equation}
Here the potentials $v_{M'B',MB}(k',k)$ are meson-exchange interactions
derived from the tree-diagrams, as illustrated in Fig.~\ref{fig:mex},   
of phenomenological Lagrangian. 
Within the unitary transformation method~\cite{sl,sko,jklmss09a,sljpg} 
used in the derivation, 
those potentials are energy independent and free of singularities.
In Appendix~\ref{app:lag}, we list the Lagrangian used in our derivations.
The resulting forms of $v_{M'B',MB}(k',k)$ and their partial-wave expansions
are given in Appendix~\ref{app:pot}.

The energy-dependent $Z^{(E)}_{M'B',MB}(k',k;W)$ terms in
Eq.~(\ref{eq:veff-mbmb}), as illustrated in Fig.~\ref{fig:z-diag},
contain the {\it moving} singularities owing to the three-body $\pi\pi N$ cut. 
These $Z$-diagram terms were neglected in the fit of Ref.~\cite{jlms07}
and are now included in this combined analysis.
With this inclusion of the $Z$-diagram terms,
we have confirmed that our model satisfies the three-body $\pi\pi N$ unitarity perfectly 
within the numerical accuracy.
The procedures for evaluating the partial-wave
matrix elements of $Z^{(E)}_{M'B',MB}(k',k;W)$
are explained in detail in Appendix~E of Ref.~\cite{msl07}.

The second term on the right-hand-side of Eq.~(\ref{eq:tmbmb}) 
is the $N^*$-excitation term defined by
\begin{eqnarray} 
t^R_{M'B',MB}(k',k;W)= \sum_{N^*_n, N^*_m}
\bar{\Gamma}_{M'B',N^*_n}(k';W) [D(W)]_{n,m}
\bar{\Gamma}_{N^*_m, MB}(k;W) .
\label{eq:tmbmb-r} 
\end{eqnarray}
Here the dressed $N^\ast \to MB$ and $MB \to N^\ast$ vertices are, respectively, defined by
\begin{equation}
\bar\Gamma_{MB,N^\ast}(k;W) =
\Gamma_{MB,N^\ast}(k) +
\sum_{M'B'}\int q^2 dq t_{MB,M'B'}(k,q;W) G_{M'B'}(q,W)\Gamma_{M'B',N^\ast}(q),
\label{eq:dress-1}
\end{equation}
\begin{equation}
\bar\Gamma_{N^\ast,MB}(k;W) =
\Gamma_{N^\ast,MB}(k) +
\sum_{M'B'}\int q^2 dq \Gamma_{N^\ast,M'B'}(q) G_{M'B'}(q,W) t_{M'B',MB}(q,k;W),
\label{eq:dress-2}
\end{equation}
with $\Gamma_{MB,N^\ast}(k)$ being the bare $N^\ast\to MB$ decay vertex
[note that $\Gamma_{N^\ast,MB}(k)= \Gamma_{MB,N^\ast}^\dag(k)$];
the inverse of the dressed $N^\ast$ propagator is defined by
\begin{equation}
[D^{-1}(W)]_{n,m} = (W - M^0_{N^*_n})\delta_{n,m} - [\Sigma_{N^\ast}(W)]_{n,m},
\label{eq:nstar-g}
\end{equation}
where $M_{N^*}^0$ is the mass of the bare $N^*$ state 
and the $N^\ast$ self-energies $\Sigma_{N^\ast}(W)$ are given by 
\begin{eqnarray}
[\Sigma_{N^\ast}(W)]_{n,m}
&=&\sum_{MB}\int_{C_{MB}}k^2 dk\Gamma_{N^*_n,MB}(k) G_{MB}(k;W)
\bar{\Gamma}_{MB,N^*_m}(k;W).
\label{eq:nstar-sigma}
\end{eqnarray}
We  emphasize here  that the $N^\ast$ propagator $D(W)$ can have off-diagonal
terms owing to the meson-baryon interactions.

Equations~(\ref{eq:tmbmb})-(\ref{eq:nstar-sigma}) define the DCC 
model used in our analysis.
In the absence of theoretical input, the DCC model, as well as all hadron reaction models,
has parameters that can only be determined phenomenologically
from fitting the data. 
The meson-exchange interactions $v_{M'B',MB}$ depend on the coupling constants 
and the cutoffs of form factors that regularize their matrix elements. 
While the values of some of the coupling constants
can be estimated from the flavor SU(3) relations,
we allow most of them to vary in the fits.
The $s$-channel and $u$-channel mechanisms of $v_{M'B',MB}$ 
($v^s$ and $v^u$ in Fig.~\ref{fig:mex}) include at each meson-baryon-baryon
vertex a form factor of the form
\begin{eqnarray}
F(\vec{k},\Lambda)=\left(\frac{\Lambda^2}{\vec{k}^2+\Lambda^2}\right)^2,
\label{eq:ff}
\end{eqnarray}
with $\vec{k}$ being the meson momentum. For the meson-meson-meson
vertex of $t$-channel mechanism ($v^t$), the form Eq.~(\ref{eq:ff}) is
also used with $\vec{k}$ being the momentum of the exchanged meson.
For the contact term ($v^c$) we regularize it by
$F(\vec{k'},\Lambda')F(\vec{k},\Lambda)$.
The bare vertex functions in Eqs.~(\ref{eq:dress-1}) 
and~(\ref{eq:dress-2}) are parametrized as
\begin{eqnarray}
{\Gamma}_{MB(LS),N^\ast}(k)
&=& \frac{1}{(2\pi)^{3/2}}\frac{1}{\sqrt{m_N}}C_{MB(LS),N^\ast}
\left(\frac{\Lambda_{N^\ast}^2}{\Lambda_{N^\ast}^2 + k^2}\right)^{(2+L/2)}
\left(\frac{k}{m_\pi}\right)^{L} ,
\label{eq:gmb}
\end{eqnarray}
where $L$ and $S$ denote the orbital angular momentum and spin of the $MB$ state, respectively.
All of the possible $(L,S)$ states in each partial wave 
included in our coupled-channels calculations are listed in Table.~\ref{tab:pw}.
The vertex function~(\ref{eq:gmb}) behaves as $k^L$ at $k\sim 0$ and $k^{-4}$ for $k\to\infty$.
The coupling constant $C_{MB(LS),N^*}$ and the cutoff $\Lambda_{N^*}$
are adjusted along with the bare masses $M^0_{N^*}$  in the fits.
It is noted that in our early analysis~\cite{jlms07},
the different cutoffs were introduced for each $MB(LS)$ state 
of a given bare $N^\ast$, and those were allowed to vary independently in the fit.
In this analysis, however, we
use a single cutoff $\Lambda_{N^\ast}$ for all $MB(LS)$ states.
This drastically reduces the number of parameters associated with the hadronic interaction
of the bare $N^\ast$ states.
\begin{table}
\caption{
The orbital angular momentum $(L)$ and total spin ($S$) of each $MB$ channel allowed in a given partial wave.
In the first column, partial waves are denoted with the conventional notation $l_{2I2J}$ as well as ($I$,$J^P$).
\label{tab:pw}}
\begin{ruledtabular}
\begin{tabular}{ccccccccccc}
$l_{2I2J}$ $(I,J^P)$ &\multicolumn{10}{c}{$(L,S)$ of the considered partial waves}\\
\cline{2-11}
& $\pi N$     & $\eta N$   &\multicolumn{2}{c}{$\pi \Delta $}& $\sigma N$   &\multicolumn{3}{c}{$\rho N$} & $K\Lambda$   &$K\Sigma$ \\
&             &            &$(\pi \Delta)_1$&$(\pi \Delta)_2$&              &$(\rho N)_1$&$(\rho N)_2$& $(\rho N)_3$&              &          \\
\hline
$S_{11}$ $(1,\ohalf^-)$&($0,\ohalf$)&($0,\ohalf$)&($2,\thalf$)& --         &($1,\ohalf$)&($0,\ohalf$)&($2,\thalf$)& --         &($0,\ohalf$)&($0,\ohalf$)\\
$S_{31}$ $(3,\ohalf^-)$&($0,\ohalf$)&   --       &($2,\thalf$)&--          &    --      &($0,\ohalf$)&($2,\thalf$)& --         &--          &($0,\ohalf$)\\
$P_{11}$ $(1,\ohalf^+)$&($1,\ohalf$)&($1,\ohalf$)&($1,\thalf$)&--          &($0,\ohalf$)&($1,\ohalf$)&($1,\thalf$)& --         &($1,\ohalf$)&($1,\ohalf$)\\
$P_{13}$ $(1,\thalf^+)$&($1,\ohalf$)&($1,\ohalf$)&($1,\thalf$)&($3,\thalf$)&($2,\ohalf$)&($1,\ohalf$)&($1,\thalf$)&($3,\thalf$)&($1,\ohalf$)&($1,\ohalf$)\\
$P_{31}$ $(3,\ohalf^+)$&($1,\ohalf$)&   --       &($1,\thalf$)&--          &    --      &($1,\ohalf$)&($1,\thalf$)& --         &--          &($1,\ohalf$)\\
$P_{33}$ $(3,\thalf^+)$&($1,\ohalf$)&   --       &($1,\thalf$)&($3,\thalf$)&     --     &($1,\ohalf$)&($1,\thalf$)&($3,\thalf$)&--          &($1,\ohalf$)\\
$D_{13}$ $(1,\thalf^-)$&($2,\ohalf$)&($2,\ohalf$)&($0,\thalf$)&($2,\thalf$)&($1,\ohalf$)&($2,\ohalf$)&($0,\thalf$)&($4,\thalf$)&($2,\ohalf$)&($2,\ohalf$)\\
$D_{15}$ $(1,\fhalf^-)$&($2,\ohalf$)&($2,\ohalf$)&($2,\thalf$)&($4,\thalf$)&($3,\ohalf$)&($2,\ohalf$)&($2,\thalf$)&($4,\thalf$)&($2,\ohalf$)&($2,\ohalf$)\\
$D_{33}$ $(3,\thalf^-)$&($2,\ohalf$)&    --      &($0,\thalf$)&($2,\thalf$)& --         &($2,\ohalf$)&($0,\thalf$)&($2,\thalf$)&--          &($2,\ohalf$)\\
$D_{35}$ $(3,\fhalf^-)$&($2,\ohalf$)&    --      &($2,\thalf$)&($4,\thalf$)&    --      &($2,\ohalf$)&($2,\thalf$)&($4,\thalf$)&--          &($2,\ohalf$)\\
$F_{15}$ $(1,\fhalf^+)$&($3,\ohalf$)&($3,\ohalf$)&($1,\thalf$)&($3,\thalf$)&($2,\ohalf$)&($3,\ohalf$)&($1,\thalf$)&($3,\thalf$)&($3,\ohalf$)&($3,\ohalf$)\\
$F_{17}$ $(1,\shalf^+)$&($3,\ohalf$)&($3,\ohalf$)&($3,\thalf$)&($5,\thalf$)&($4,\ohalf$)&($3,\ohalf$)&($3,\thalf$)&($5,\thalf$)&($3,\ohalf$)&($3,\ohalf$)\\
$F_{35}$ $(3,\fhalf^+)$&($3,\ohalf$)&    --      &($1,\thalf$)&($3,\thalf$)& --         &($3,\ohalf$)&($1,\thalf$)&($3,\thalf$)&--          &($3,\ohalf$)\\
$F_{37}$ $(3,\shalf^+)$&($3,\ohalf$)&  --        &($3,\thalf$)&($5,\thalf$)&  --        &($3,\ohalf$)&($3,\thalf$)&($5,\thalf$)&--          &($3,\ohalf$)\\     
$G_{17}$ $(1,\shalf^-)$&($4,\ohalf$)&($4,\ohalf$)&($2,\thalf$)&($4,\thalf$)&($3,\ohalf$)&($4,\ohalf$)&($2,\thalf$)&($4,\thalf$)&($4,\ohalf$)&($4,\ohalf$)\\
$G_{19}$ $(1,\nhalf^-)$&($4,\ohalf$)&($4,\ohalf$)&($4,\thalf$)&($6,\thalf$)&($5,\ohalf$)&($4,\ohalf$)&($4,\thalf$)&($6,\thalf$)&($4,\ohalf$)&($4,\ohalf$)\\
$G_{37}$ $(3,\shalf^-)$&($4,\ohalf$)&   --       &($2,\thalf$)&($4,\thalf$)&--          &($4,\ohalf$)&($2,\thalf$)&($4,\thalf$)&--          &($4,\ohalf$)\\
$G_{39}$ $(3,\nhalf^-)$&($4,\ohalf$)&  --        &($4,\thalf$)&($6,\thalf$)&  --        &($4,\ohalf$)&($4,\thalf$)&($6,\thalf$)&--          &($4,\ohalf$)\\   
$H_{19}$ $(1,\nhalf^+)$&($5,\ohalf$)&($5,\ohalf$)&($3,\thalf$)&($5,\thalf$)&($4,\ohalf$)&($5,\ohalf$)&($3,\thalf$)&($5,\thalf$)&($5,\ohalf$)&($5,\ohalf$)\\
$H_{39}$ $(3,\nhalf^+)$&($5,\ohalf$)&($5,\ohalf$)&($3,\thalf$)&($5,\thalf$)&--          &($5,\ohalf$)&($3,\thalf$)&($5,\thalf$)&--          &($5,\ohalf$)      
\end{tabular}
\end{ruledtabular}
\end{table}

\subsection{Electromagnetic amplitudes}
\label{sec:dcc-ele}

\begin{figure}[t]
\includegraphics[clip,width=0.67\textwidth]{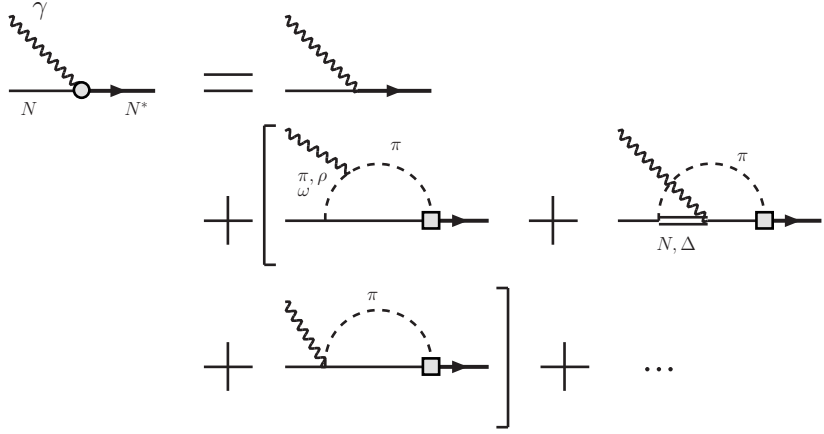}
\caption{\label{fig:gn-nstar}
Dressed $\gamma N \rightarrow N^*$ vertex defined by
Eq.~(\ref{eq:nstargn}).
}
\end{figure}

With the hadronic amplitudes $t_{M'B',MB}(k',k;W)$ defined in Eq.~(\ref{eq:cc-eq}),
the partial wave amplitudes for the $\gamma(\vec q) + N(-\vec q) \to M' (\vec k') + B'(-\vec k')$ 
reactions are expressed as~\cite{msl07},
\begin{eqnarray}
T_{M'B',\gamma N} (k',q;W) &=& t_{M'B',\gamma N} (k',q;W) + t^R_{M'B',\gamma N} (k',q;W),
\label{eq:tmbgn}
\end{eqnarray}
with
\begin{eqnarray}
t_{M'B',\gamma N} (k',q;W) &=& v_{M'B', \gamma N} (k',q)
\nonumber\\
&&
+\sum_{M''B''} \int p^2 dp \, t_{M'B',M''B''}(k',p;W)G_{M''B''}(p;W)v_{M''B'', \gamma N} (p,q),
\\
t^R_{M' B',\gamma N}(k',q; W) &=&
\sum_{n,m} \bar\Gamma_{M'B',N^\ast_n}(k';W) [D(W)]_{n,m} \bar\Gamma_{N^\ast_m,\gamma N}(q;W),
\\
\bar\Gamma_{N^\ast,\gamma N}(q;W) &=& 
\Gamma_{N^\ast,\gamma N}(q) 
+ \sum_{M'B'}\int p^2 dp \Gamma_{N^*,M'B'}(p) G_{M'B'}(p,W) t_{M'B',\gamma N}(p,q;W).
\nonumber\\
&&
\label{eq:nstargn}
\end{eqnarray}
Here $v_{MB,\gamma N}$ is the meson-exchange potential for 
the $\gamma N \to MB$ processes, and $\bar \Gamma_{N^*,\gamma N}(q;W)$
is the dressed $\gamma N \to N^*$ vertex (Fig.~\ref{fig:gn-nstar}).
The procedures for calculating $v_{MB,\gamma N}$ are detailed in Ref.~\cite{msl07}
and also given here in Appendix~\ref{app:pot-em}. 
In the latter, the ingredients associated with $KY$ channels are newly added. 

For the bare $\gamma N \to N^*$ vertex, we depart from the simple parametrization
given in  Ref.~\cite{msl07} and write it in the helicity representation as
\begin{eqnarray}
\Gamma_{N^\ast,\gamma N}(q) = \frac{1}{(2\pi)^{3/2}}
\sqrt{\frac{m_N}{E_N(q)}}\sqrt{\frac{q_R}{|q_0|}} A^{N^\ast}_\lambda\delta_{\lambda,(\lambda_\gamma-\lambda_N)},
\label{eq:nstar-gn}
\end{eqnarray}
where $\lambda_\gamma$ ($\lambda_N$) is the helicity quantum number of the photon (nucleon)
and $q_R$ and $q_0$ are defined by $M_{N^\ast}=q_R+E_N(q_R)$ and $W=q_0+E_N(q)$, respectively. 
The helicity amplitudes $A^{N^*}_\lambda$ in the above equation
are related to the multipole amplitudes $E^{N^\ast}_{l\pm}$ and $M^{N^\ast}_{l\pm}$
of $\gamma N \to N^*$ processes as
\begin{eqnarray}
A^{N^*}_{3/2}&=& \frac{\sqrt{l(l+2)}}{2}[-M^{N^\ast}_{l+} + E^{N^\ast}_{l+}],
\label{eq:a32-me+}
\\
A^{N^*}_{1/2}&=&-\frac{1}{2}[lM^{N^\ast}_{l+} + (l+2) E^{N^\ast}_{l+}],
\label{eq:a12-me+}
\end{eqnarray}
for $j = l + 1/2$, and
\begin{eqnarray}
A^{N^*}_{3/2}&=& -\frac{\sqrt{(l-1)(l+1)}}{2}[M^{N^\ast}_{l-} + E^{N^\ast}_{l-}],
\label{eq:a32-me-}
\\
A^{N^*}_{1/2}&=&+\frac{1}{2}[(l+1)M^{N^\ast}_{l-} - (l-1) E^{N^\ast}_{l-}],
\label{eq:a12-me-}
\end{eqnarray}
for $j = l - 1/2$. 
The multipole amplitudes are parametrized as
\begin{eqnarray}
M^{N^\ast}_{l\pm}(q)&=& \left(\frac{q}{m_\pi}\right)^l
\left(\frac{(\Lambda^{\text{e.m.}}_{N^\ast})^2 + m_\pi^2}{(\Lambda_{N^\ast}^{\text{e.m.}})^2 + q^2}\right)^{(2+l/2)} \tilde M^{N^\ast}_{l\pm},
\label{eq:M-para}
\\
E^{N^\ast}_{l\pm}(q)&=& \left(\frac{q}{m_\pi}\right)^{(l\pm 1)}
\left(\frac{(\Lambda^{\text{e.m.}}_{N^\ast})^2 + m_\pi^2}{(\Lambda_{N^\ast}^{\text{e.m.}})^2 + q^2}\right)^{[2+(l\pm 1)/2]} \tilde E^{N^\ast}_{l\pm},
\label{eq:E-para}
\end{eqnarray}
where the cutoff $\Lambda^{\text{e.m.}}_{N^\ast}$ and the coupling constants
$\tilde M_{l\pm}^{N^*}$ and $\tilde E_{l\pm}^{N^*}$ are determined in fitting the data.
One significant difference between this work and our previous analysis~\cite{jlmss08}
is that the multipole amplitudes, 
or equivalently the helicity amplitudes, for
the $\gamma N\to N^\ast$ processes now have momentum dependence.
To compare with our previous works~\cite{sl} on $\Delta(1232)$, however,
we depart from the above parametrization and use the following forms
for the first bare state in $P_{33}$:
\begin{eqnarray}
A^{\text{1st}P_{33}}_{3/2} &=& -x_{A_{3/2}}\frac{\sqrt{3}}{2} A\left[G^{SL}_{M}(0) - (1-N)G_E^{SL}(0)\right] ,
\label{eq:1p33-a32}
\\
A^{\text{1st}P_{33}}_{1/2} &=&-x_{A_{1/2}}\frac{1       }{2} A\left[G^{SL}_{M}(0) - (1+N)G_E^{SL}(0)\right] ,
\label{eq:1p33-a12}
\end{eqnarray}
with
\begin{eqnarray}
A   &=& 
\frac{e}{2m_N} \sqrt{\frac{Wq}{m_N}} \frac{W+m_N}{M_{N^*}+m_N} ,
\\
N   &=&
2
\left(\frac{W-m_N}{M_{N^*}-m_N}\right)^2 ,
\end{eqnarray}
where $G^{SL}_M(0) = 1.85$ and $G^{SL}_E(0) = 0.025$~\cite{sl}.
The factors $x_{A_{3/2}}$ and $x_{A_{1/2}}$ in Eqs.~(\ref{eq:1p33-a32}) and~(\ref{eq:1p33-a12})
are treated as free parameters in our fitting processes.

\section{Extractions of Nucleon Resonances}
\label{sec:res-ex}

We follow the earlier works~\cite{bohm,dalitz} to define that
a nucleon resonance with a complex mass $M_R$ is
an ``eigenstate'' of a Hamiltonian $H\ket{\psi^R_{N^*}} = M_R \ket{\psi^R_{N^*}}$
under the so-called outgoing boundary condition.
Then from the spectral expansion of the Low equation for reaction amplitude
$T(W)= H' + H'(W-H)^{-1}H'$, where we have defined $H'=H-H_0$ with
$H_0$ being the non-interacting free Hamiltonian, we have
\begin{eqnarray}
T_{M'B', MB}(k^0_{M'B'}, k^0_{MB} ; W\to M_R)
&=&\frac{\bra{k^0_{M'B'}}H'\ket{\psi^R_{N^*}}\bra{\psi^R_{N^*}}H'\ket{k^0_{MB}}}{W-M_R} + \cdot\cdot,
\label{eq:em-exp-2}
\end{eqnarray}
where $k^0_{MB}$ and $k^0_{M'B'}$ are the on-shell momenta defined by
\begin{eqnarray}
W&=&E_{M}(k^0_{MB})+E_{B}(k^0_{MB}) \nonumber \\
&=&E_{M'}(k^0_{M'B'})+
E_{B'}(k^0_{M'B'})\,. \label{eq:on-k}
\end{eqnarray}
Therefore the resonance masses $M_R$ can be defined as the pole positions of
the meson-baryon amplitude $T_{M'B',MB}(k^0_{M'B'}, k^0_{MB}; W)$
on the complex Riemann $W$ surface.
Because of the quadratic relation between the energy and momentum
variables, each $MB$ channel for a given  $W$ can have a physical ($p$)
sheet characterized by $\mathrm{Im}(k^0_{MB}) >0$ and an unphysical ($u$) sheet by
$\mathrm{Im}(k^0_{MB}) < 0$.
Like all previous works, we only look for the poles close to the physical region 
and/or having large effects on scattering  observables. 
All of these poles are on the unphysical sheet of the $\pi N$ channel, 
but could be on either  $(u)$ or $(p)$ sheets of other channels.
To find the resonance poles, we analytically continue 
Eqs.~(\ref{eq:tmbmb})-(\ref{eq:nstar-sigma})
to the complex $W$ plane by using the method
detailed in Refs.~\cite{ssl09,ssl10}. 
The main step is to choose appropriate momentum-integration paths
in solving Eqs.~(\ref{eq:tmbmb})-(\ref{eq:nstar-sigma}).

As derived in Refs.~\cite{ssl09,ssl10}, we can write Eq.~(\ref{eq:em-exp-2})
in terms of $t_{M'B',MB}(k',k;W)$, which is the solution of Eq.~(\ref{eq:cc-eq}),
$t^R_{M'B',MB}(k',k;W)$ defined by Eq.~(\ref{eq:tmbmb-r}), and 
$\bar{\Gamma}_{MB,N^*}$ [$\bar{\Gamma}_{N^*,MB}$] defined by 
Eq.~(\ref{eq:dress-1}) [Eq.~(\ref{eq:dress-2})].
Explicitly, as the energy approaches a resonance position in the complex $W$ plane, 
the total meson-baryon amplitudes can be written as
\begin{eqnarray}
T_{M'B',MB}(k^R_{M'B'},k^R_{MB}; W \to M_R) =  \tilde B_{M'B',MB}(M_R)
+ \frac{\tilde R_{M'B',MB}(M_R)}{W-M_R}\,,
\label{eq:res-pole}
\end{eqnarray}
where the on-shell momenta $k^R_{MB}$ and $k^R_{M'B'}$ 
are defined by Eq.~(\ref{eq:on-k}) with $W=M_R$, and
\begin{eqnarray}
\tilde B_{M'B',MB}(M_R) & = & t_{M'B',MB}(k^R_{M'B'},k^R_{MB};  M_R) 
\nonumber \\
&& 
+ \frac{d}{dW}\left[(W - M_R) t^R_{M'B',MB}(k^R_{M'B'},k^R_{MB};W)\right]_{W=M_R}\,, 
\\
\tilde R_{M'B',MB}(M_R)&=& \bar{\Gamma}^R_{M'B'}(k^R_{M'B'},M_R) \bar{\Gamma}^R_{MB}(k^R_{MB},M_R)\,,
\label{eq:res-r}
\end{eqnarray}
with
\begin{equation}
\bar{\Gamma}^R_{MB}(k^R_{MB},M_R) = \sum_{i}\chi_i\bar{\Gamma}_{MB,N^*_i}(k^R_{MB},M_R).
\label{eq:res-barg}
\end{equation}
Here $\chi_i$ represents $i$th ``bare'' resonance component of the dressed $N^*$
and satisfies
\begin{eqnarray}
\sum_j \left[D^{-1}(M_R)\right]_{ij}\chi_j & = &
\sum_j [(M_R - M_{N^*_i})\delta_{ij} - \Sigma(M_R)_{ij}]\chi_j = 0.
\label{eq:nstar-wf}
\end{eqnarray}
If there is only one bare $N^*$ state,
with $D^{-1}(W)=W-M_{N^*}-\Sigma(W)$,
it is easy to see that
\begin{eqnarray}
\chi & = & \frac{1}{\sqrt{1 - \Sigma'(M_R)}} \,,
\end{eqnarray}
where $\Sigma'(M_R) = [d\Sigma/dW]_{W=M_R}$.

With the normalizations we employ, the $S$-matrix in each
partial wave is related to $T_{M'B',MB}(k^0_{M'B'},k^0_{MB}; W)$ by
\begin{eqnarray}
S_{M'B',MB} (W) = \delta_{M'B',MB} + 2 i F_{M'B',MB}(W),
\label{eq:s-mx}
\end{eqnarray}
where
\begin{eqnarray}
F_{M'B',MB}(W) =- [\rho_{M'B'}(W)]^{1/2} T_{M'B',MB}(k^0_{M'B'},k^0_{MB}; W) [\rho_{MB}(W)]^{1/2},
\label{eq:f-t}
\end{eqnarray}
with
\begin{eqnarray}
\rho_{MB}(W) = \pi \frac{k^0_{MB}E_M(k^0_{MB})E_B(k^0_{MB})}{W}.
\end{eqnarray}

For the later use, we denote the residue of the scattering amplitude $F_{M'B',MB}(W)$
at a resonance pole as $R_{M'B',MB}$,
\begin{eqnarray}
F_{M'B',MB}(W \rightarrow M_R) &=& 
\frac{S_{M'B',MB}(W \rightarrow M_R)-\delta_{M'B',MB}}{2i} \nonumber \\
&=&- \left[\frac{R_{M'B',MB}}{W-M_R}\right]_{W\rightarrow M_R}.
\label{eq:f-mbmb}
\end{eqnarray}
From Eqs.~(\ref{eq:res-pole}), (\ref{eq:res-r}),
(\ref{eq:s-mx})-(\ref{eq:f-t}), and~(\ref{eq:f-mbmb})
we have
\begin{eqnarray}
R_{M'B',MB} & = & 
[\rho_{M'B'}(M_R)]^{1/2} 
\bar{\Gamma}^R_{M'B'}(k^R_{M'B'},M_R) \bar{\Gamma}^R_{MB}(k^R_{MB},M_R)
[\rho_{MB}(M_R)]^{1/2}.
\label{eq:pin-resid}
\end{eqnarray}
Here it is noted that we have fixed a sign freedom of $\chi_i$ in 
$\bar{\Gamma}^R_{MB}(k^R_{MB},M_R)$
in such a way that $\bar \Gamma^R_{\pi N}(k^R_{\pi N}, M_R)$ 
is given by $[\rho_{\pi N}(M_R)]^{1/2} \bar \Gamma^R_{\pi N}(k^R_{\pi N}, M_R) = [R_{\pi N, \pi N}]^{1/2}$ with 
$-\pi < \mathrm{arg}(R_{\pi N,\pi N}) < \pi$.
The $\pi N$ elasticity  of a resonance is then defined as
\begin{eqnarray}
\eta_e= \frac{|R_{\pi N,\pi N}|}{-\mathrm{Im} (M_R)} \,.
\label{eq:elasticity}
\end{eqnarray}

With a similar procedure, we have the 
helicity amplitudes of $\gamma N \rightarrow N^*$ at the resonance pole $M_R$ 
as~\cite{ssl10}
\begin{eqnarray}
A_{3/2} & = & C\times \bar{\Gamma}^R_{\gamma N}(M_R,\lambda_\gamma=1,\lambda_N=-1/2),
\label{eq:a32}\\
A_{1/2} & = & C \times
\bar{\Gamma}^R_{\gamma N}(M_R,\lambda_\gamma=-1,\lambda_N=-1/2) \, \label{eq:a12},
\end{eqnarray}
where  $\lambda_N$ and $\lambda_\gamma$ are the helicities of the initial
nucleon and photon, respectively, and
\begin{eqnarray}
C=\sqrt{\frac{E_N(\vec{q})}{m_N }} \frac{1}{\sqrt{2K}} 
\sqrt{\frac{(2J^R+1)(2\pi)^3(2q_0)}{4\pi}},
\label{eq:coef-c}
\end{eqnarray}
where $J^R$ is the spin of the resonance state,
$q_0 = |\vec{q}|$ and $K  =  (M_R^2 - m_N^2)/(2M_R)$.

\section{Fitting procedures}
\label{sec:fit-proc}

The model parameters specified in Sec.~\ref{sec:dcc} are determined by fitting 
the data of pion- and photon-induced $\pi N$, $\eta N$, $K\Lambda$, 
and $K\Sigma$ production reactions off a proton.
One of the difficult tasks in the analysis is to decide the data we fit.
For the $\pi N \rightarrow \pi N$ data, we rely on the database of SAID~\cite{saiddb}. 
For simplicity, we fit their single energy solutions of $\pi N$ partial-wave 
amplitudes instead of fitting the original $\pi N$ elastic scattering data.
There is always a question on whether this is a reliable procedure. 
We justify this by observing that the $\pi N$ partial-wave amplitudes determined 
by using a very different approach based on the dispersion-relations~\cite{hohler,cmb-pin} 
are not so different from that of SAID. 
Thus the SAID amplitudes, which are extracted from much more data than 
the previous analyses, can be considered as fairly reliable representations 
of the original $\pi N$ elastic scattering data 
(about 30,000 data points as compiled in SAID).

In this work, we want to determine the spectrum of the $N^\ast$ resonances with masses only 
up to 2 GeV and widths less than  400 MeV.
Because the  $N^*$ resonances 
can affect the observables over the energy
range of their widths,
we include the data of $\pi N \to \pi N$ up to $W = 2.3$ GeV to make sure that
the $N^\ast$ resonances with masses near about 1.9 GeV are properly identified within 
our model.
In Table~\ref{tab:pwa-data}, we list the data points of 
the SAID energy-independent solutions included in our fits.
In the fitting process, we check the resulting parameters
by comparing the predicted observables with the original $\pi N$ scattering data.

For the data of $\pi N \to \eta N, K\Lambda, K\Sigma$ and 
$\gamma N \to \pi N, \eta N, K\Lambda, K\Sigma$,
we  use the database of the Bonn-Gatchina group~\cite{bg2012} with some differences: 
(i) we only include the data up to $W = 2.1$ GeV;
(ii) following the discussion in Ref.~\cite{djlss08} 
(the references of the data are also found in Ref.~\cite{djlss08}), 
the data for $\pi^- p \to \eta n$ differential cross sections
used in our analysis are different from those of Bonn-Gatchina group; and
(iii) we include the new data of the polarization $G$ of $\gamma p \to \pi^0 p$ 
from CBELSA/TAPS Collaboration~\cite{cbelsa2012}.
The data of $\pi N \to \eta N, K\Lambda, K\Sigma$ 
and $\gamma N \to \pi N, \eta N, K\Lambda, K\Sigma$
included in our fit are listed in Tables~\ref{tab:ha-data} and~\ref{tab:em-data}. 
It is known~\cite{tabakin,shkl11} that
the determinations of the multipole amplitudes of pseudoscalar meson photoproductions
need at least eight observables, i.e., $d\sigma/d\Omega$, $\Sigma$, $T$, $P$, and 
appropriately chosen four double-spin polarizations.
We see from Table~\ref{tab:em-data} that the polarization observables, 
in particular the double-spin polarizations, do not have enough data points.
Thus it is rather difficult to fit these data because they
have much less weights in the $\chi^2$-minimization processes.

We next discuss how we perform the minimization of $\chi^2$. We follow the
most commonly used definition
\begin{equation}
\chi^2 = \sum_{O}\sum_{i,j}
\frac{[O^{\text{model}}(W_i,\theta_j)-O^{\text{exp.}}(W_i,\theta_j)]^2}
     {[\delta O^{\text{exp.}}(W_i,\theta_j)]^2},
\label{eq:chi2}
\end{equation}
where $O^{\text{model}}(W_i,\theta_j)$ is the observable $O$ 
at energy $W_i$ and angle $\theta_j$ calculated from our DCC model, 
while $O^{\text{exp.}}(W_i,\theta_j)$ and 
$\delta O^{\text{exp.}}(W_i,\theta_j)$ are the central value and 
the statistical error of the experimental data, respectively.
(Note that the $\pi N$ partial-wave amplitudes depend only on the energy.)
There are more sophisticated minimization procedures accounting 
separately for the systematic and statistical errors. 
Thus some of the discrepancies between our final
results and the data, as presented in the next section, 
could be partly attributed to our use of Eq.~(\ref{eq:chi2}) for $\chi^2$. 
Such a more careful fitting procedure will be desirable when we move to our next
analysis including more complete data from future experiments, as
discussed later.
\begin{table}
\caption{\label{tab:pwa-data} 
Number of the data points of $\pi N \rightarrow \pi N$ amplitudes 
included in our fits. The data are from SAID analysis~\cite{saiddb}.
}
\begin{ruledtabular}
\begin{tabular}{lrlrrr}
Partial wave  &       & Partial wave&&\\
\hline
$S_{11}$ & 65$\times$2 & $S_{31}$ & 65$\times$2 &\\
$P_{11}$ & 65$\times$2 & $P_{31}$ & 61$\times$2 &\\
$P_{13}$ & 61$\times$2 & $P_{33}$ & 65$\times$2 &\\
$D_{13}$ & 61$\times$2 & $D_{33}$ & 59$\times$2 &\\
$D_{15}$ & 61$\times$2 & $D_{35}$ & 40$\times$2 &\\
$F_{15}$ & 48$\times$2 & $F_{35}$ & 43$\times$2 &\\
$F_{17}$ & 32$\times$2 & $F_{37}$ & 44$\times$2 &\\
$G_{17}$ & 42$\times$2 & $G_{37}$ & 32$\times$2 &\\
$G_{19}$ & 28$\times$2 & $G_{39}$ & 32$\times$2 &\\
$H_{19}$ & 34$\times$2 & $H_{39}$ & 31$\times$2 &\\
& &  & &\\
Sum     & 994 &&944&1938
\end{tabular}
\end{ruledtabular}
\end{table}
\begin{table}
\caption{\label{tab:ha-data}
Number of data points of hadronic processes included in our fits.
See Refs.~\cite{bg2012,djlss08} for the data references.
}
\begin{ruledtabular}
\begin{tabular}{lrrrr}
                          & $d\sigma/d\Omega$ & $P$  &$\beta$&Sum \\
\hline
$\pi^- p \to \eta p$      & 294               &  --  &--& 294\\
                          &                   &      &&    \\
$\pi^- p \to K^0 \Lambda$ & 544               &  262 & 43 & 849\\
$\pi^- p \to K^0 \Sigma^0$& 160               &  70  &--& 230\\
$\pi^+ p \to K^+ \Sigma^+$& 552               &  312 &7& 871\\
                             &       &   &&    \\
 Sum                         & 1550      & 644    &50& 2244 
\end{tabular}
\end{ruledtabular}
\end{table}
\begin{table}[t]
\caption{\label{tab:em-data}
The number of data points of photoproduction processes included in our fits.
See Refs.~\cite{bg2012,cbelsa2012} for the data references.
 }
\begin{ruledtabular}
\begin{tabular}{lrrrrrrrrrrrr}
                         &$d\sigma/d\Omega$&$\Sigma$&$T$ &$P$ &$\hat E$&$G$&$H$&$O_{x'}$& $O_{z'}$ &$C_{x}$& $C_{z}$ &Sum \\
\hline
$\gamma p\to \pi^0 p$    &4381             & 1128   &380 &589 &140     &125&49 &7       &7         &--      &--        &6806\\
$\gamma p\to \pi^+ n$    &2315             &  747   &678 &222 &231     &86 &128&     -- &--        &--      &--        &4407\\
                         &                 &        &    &    &        &   &   &        &          &        &          &    \\
$\gamma p\to \eta p$     &3221             &  235   &50  & -- &--      &-- &-- &--      & --       &     -- &    --    &3506 \\
                         &                 &        &    &    &        &   &   &        &          &        &          &    \\
$\gamma p\to K^+\Lambda$ & 800             & 86     &66  & 865&--      &-- & --&     66 & 66       &79      &     79   &2107\\
$\gamma p\to K^+\Sigma^0$& 758             & 62     &--  & 169&--      &-- & --&--      &--        &40      & 40       &1069\\
$\gamma p\to K^0\Sigma^+$&220              & 15     &--  &  36&--      &-- & --&--      &--        &--      & --       & 271\\
                         &                 &        &    &    &        &   &   &        &          &        &          &    \\
Sum                      & 11695           & 2273   &1174&1881&371     &211&177&    73  & 73       &119     &119       &18166
\end{tabular}
\end{ruledtabular}
\end{table}

In the absence of theoretical input, our main challenge is to determine 
the bare $N^*$ mass $M^0_{N^*}$, the strong $N^* \to MB$ bare vertex functions 
defined by Eq.~(\ref{eq:gmb}), and the electromagnetic $N^* \to \gamma N$ 
bare vertex functions defined by Eq.~(\ref{eq:nstar-gn}).
For each partial wave, the number of the $N^*$ parameters contained is 
$N_{N^\ast} \times (3 + N^{\text{str.}}_{\text{c.c.}} + N^{\text{e.m.}}_{\text{c.c.}} )$.
Here $N_{N^*}$ is the number of the bare $N^*$ states included in the partial wave,
and the numbers in the parentheses are those of the parameters each bare $N^\ast$ state has:
3 comes from the bare mass $M^0_{N^*}$ and the
cutoffs $\Lambda_{N^*}$ and $\Lambda^{\text{e.m.}}_{N^*}$ of the hadronic and electromagnetic 
decay vertex functions, and $N^{\text{str.}}_{\text{c.c.}}$ ($N^{\text{e.m.}}_{\text{c.c.}}$) 
is the number of coupling constants of the hadronic (electromagnetic) interactions, 
which can be $5\leq N^{\text{str.}}_{\text{c.c.}}\leq 10$ 
($N^{\text{e.m.}}_{\text{c.c.}}= 1~\text{or}~2$). 
We thus  face  a many-parameters problem in fitting the data, which is also present 
in all existing coupled-channels analysis of nucleon resonances. 
This common problem poses difficulties in assigning the errors for the determined model parameters.

Here we have additional difficulties owing to the long computation time 
in solving the coupled-channels integral equations~(\ref{eq:cc-eq}), 
as mentioned in the previous sections.
We thus follow all previous works on dynamical model analyses 
and do not try to resolve this difficult problem here. 
We therefore are not able to provide the
errors of the determined resonance parameters. 
This must be improved in the future.

Owing to the many-parameters problem mentioned above and the limitation
of the current computation power in the $\chi^2$ minimization,
it is practically not possible to get convergent results 
if all of the model parameters are allowed to vary \textit{simultaneously} in 
the $\chi^2$-minimization processes.
We therefore take the following strategy. 
First we determine the parameters
associated with the meson-exchange potentials $v_{M'B',MB}$ to some extent.
The parameters associated with the bare $N^\ast$ states are then determined.
In the latter step the parameters of $v_{M'B',MB}$ may be varied 
 only when it is necessary to fine tune the fits.
This two-step procedure is essential because
the most time-consuming part of the computation is to calculate the
meson-exchange amplitude $t_{M'B',MB}$ from 
the meson-exchange potentials $v_{M'B',MB}$ by solving
the coupled-channels equations~(\ref{eq:cc-eq}). 
Hence  the computation
time is drastically increased if the parameters of $v_{M'B',MB}$
are varied in the $\chi^2$ minimization processes.

Concretely, our fitting procedure is as follows. 
Guided by the success of the meson-exchange 
models~\cite{juelich,juelich13-1,pj-91,gross,sl,pasc,pitt-ky,fuda,ntuanl}
in the $\Delta(1232)$ region, we first adjust the parameters
associated with the meson-exchange potentials $v_{M'B',MB}$ to
fit the data of the $\pi N$ partial-wave amplitudes at low energies with 
$W \leq 1.4$ GeV, where one bare $N^\ast$ state in the $P_{33}$
partial wave is included to incorporate the $\Delta(1232)$ contribution.
Fortunately, we find that most of the parameters in $v_{M'B',MB}$ are heavily
constrained by the $\pi N$ partial-wave amplitudes at low energies and do not 
have to be varied too much in the later fitting processes.
Once a good $\pi N$ partial-wave amplitudes in the $W \leq 1.4$ GeV region is obtained, 
we extend the fits of the amplitudes to $W = 2.3$ GeV
by including the bare $N^\ast$ states in each partial wave.
To minimize the number of the bare $N^*$ states,
we first include only one bare $N^*$ state in each partial wave
and try to fit the data of the $\pi N$ partial-wave amplitudes 
in the entire considered energy region 
by adjusting its bare mass, $M^0_{N^*}$, and
vertex function parameters, $C_{N^*,MB(LS)}$ and $\Lambda_{N^*}$.
If this fails, we then also allow the parameters associated 
with $v_{M'B',MB}$ to vary in some limited ranges.
If this fails again, we then include one more bare 
$N^*$ state in some partial waves and repeat the process.
After completing the fit of the $\pi N$ partial-wave amplitudes, we extend our global fit 
step by step: first include the $\pi N$ reaction data of Table~\ref{tab:ha-data},
and then include the $\gamma N$ reaction data of Table~\ref{tab:em-data}.
This procedure has to be repeated many times to make sure that
we have reached the limitation of the model in the $\chi^2$ minimization.
The resulting masses, coupling constants, and cutoff parameters for the meson-exchange
potentials are given in Tables~\ref{tab:model-para-mass}-\ref{tab:nres-cu} 
of Appendix~\ref{app:model-para}.
We find that the considered data can be fitted to a very large extent
with  one or two bare $N^*$ states in each partial wave.  
All the resulting cutoff parameters for the nonresonant and $N^*$ interactions 
are in the range of $500$-$2000$ MeV, 
which are similar to those in typical meson-exchange
models~\cite{juelich,juelich13-1,pj-91,gross,sl,pasc,pitt-ky,fuda,ntuanl}.

As seen in Tables~\ref{tab:pwa-data}-\ref{tab:em-data}, 
the numbers of the data points of each reaction are very different.
The observables with much fewer data points are hard to fit because they have
little effects on $\chi^2$ minimization. 
Thus, in calculating Eq.~(\ref{eq:chi2}) for the fitting, we need to put
``artificial'' weights on $\chi^2$ for those observables such as 
polarization observables. 
This procedure, which is highly undesirable, is, however, necessary 
before those data become more extensive.
Also for this reason, we cannot give meaningful $\chi^2$ values.
To show the quality of our fits, in the next section we present 
detailed comparisons with the data for all processes considered.
Hopefully, the situation, in particular the scarce data for 
the pion-induced inelastic reactions, can be improved with 
new experiments at J-PARC.

\section{Results}
\label{sec:results}

In this section we present the results from our fits to
22,348 data points listed in Tables~\ref{tab:pwa-data}-\ref{tab:em-data}.
The resulting values of the model parameters are presented 
in Tables \ref{tab:model-para-mass}-\ref{tab:res-em} of Appendix~\ref{app:model-para}.
The formulas for calculating the considered
observables of the $\pi N, \gamma N \rightarrow \pi N, \eta N, K\Lambda, K\Sigma$
from the partial-wave amplitudes defined in Sec.~\ref{sec:dcc} can be 
straightforwardly derived following the formulas given  
in Refs.~\cite{msl07,shkl11} and will not be given here. 
Here it should be noted that in the literature various notations 
have been employed for the polarization observables
of pseudoscalar photoproduction reactions (see Ref.~\cite{rosetta}
for the summary of such notations used by different analysis groups).
In this work, we follow the notation defined in Ref.~\cite{shkl11}.
In the following sections, we discuss separately the results for each considered reaction.

\subsection{$\pi N \to \pi N$}

\begin{figure}[t]
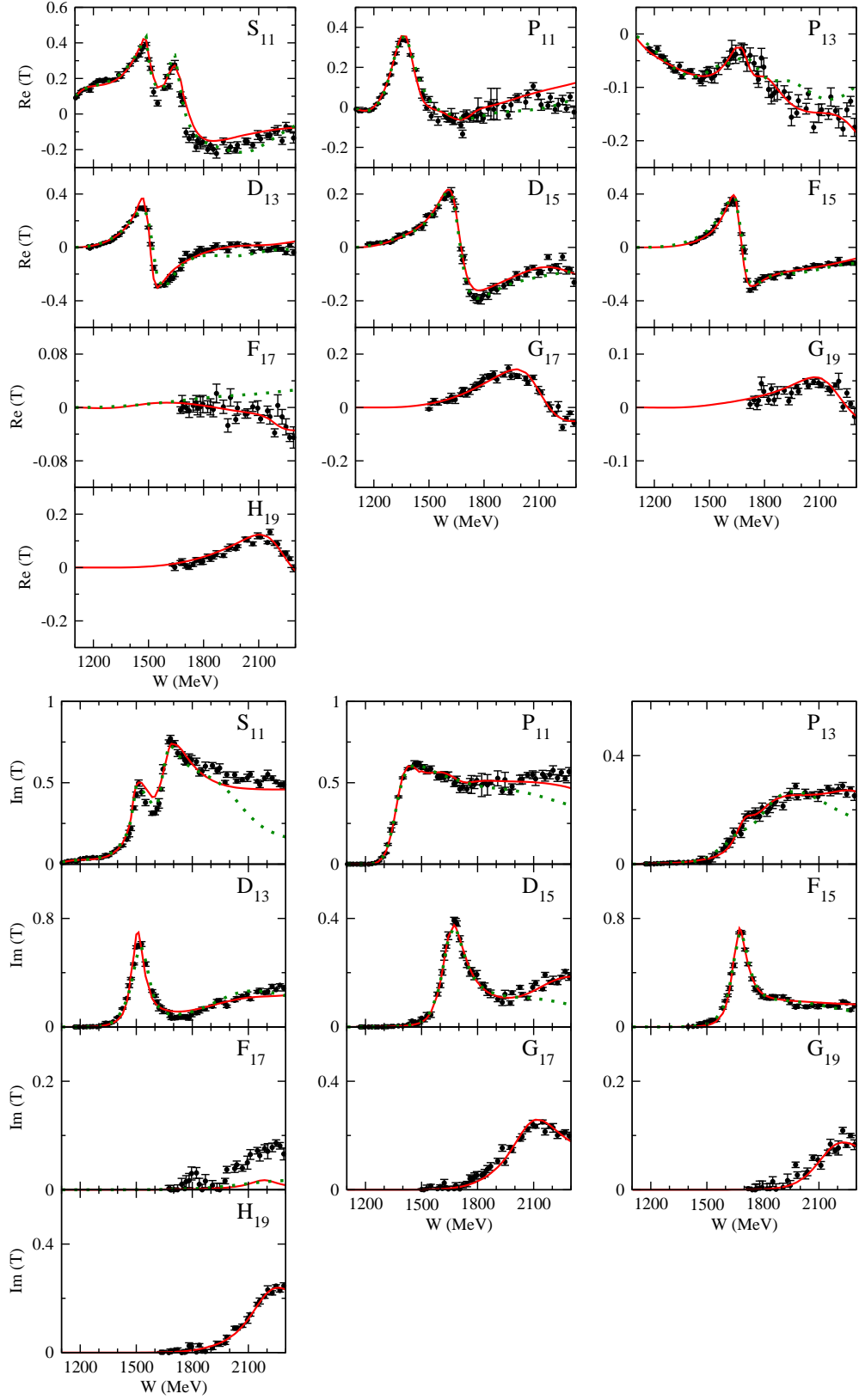

\includegraphics[clip,width=0.8\textwidth]{pwa-re1}
\includegraphics[clip,width=0.8\textwidth]{pwa-im1}
\caption{(Color online)
Partial-wave amplitudes of $\pi N$ scattering with isospin $I=1/2$.
Upper (lower) panels are for real (imaginary) parts of the amplitudes.
(Red solid curves) current results; 
(green dotted curves) results from our previous analysis~\cite{jlms07}.
}
\label{fig:pin-amp-1}
\end{figure}

In Figs.~\ref{fig:pin-amp-1} and~\ref{fig:pin-amp-3},
we present our results for the $\pi N \to \pi N$ partial-wave amplitudes up to $W=2.3$ GeV.
Clearly, very good fits to the data of SAID~\cite{saiddb} have been obtained.
Comparing with the green dotted curves of our previous analysis (JLMS)~\cite{jlms07}, 
which were obtained by the fit to the data of the $\pi N$ amplitudes 
only up to $F$ wave ($L=3$) and $W=2$ GeV,
there are significant improvements in the  $S_{31}$, $P_{31}$, $D_{35}$ partial waves, 
as seen in Fig.~\ref{fig:pin-amp-3}.
The large improvement in the $P_{31}$ and $D_{35}$ waves in this analysis is 
mainly because we have introduced more bare $N^\ast$ states than JLMS for those partial waves.
These improvements are also perhaps attributable to the change of our fitting strategy by using the data 
below $1.4$ GeV to constrain the parameters of the meson-exchange potentials $v_{M'B',MB}$.
We also find that this procedure prevents the model from generating
undesirable bound states through the strong coupling with 
the bare $N^*$ states in the low-energy region.
We also note that the data of the $G$ and $H$ partial waves with 
the orbital angular momentum $L \geq 4$, which were not included 
in the fit of JLMS, are also fitted very well. 
These high-$L$ partial waves become important 
in fitting the meson production data in the high-$W$ region.

\begin{figure}[t]
\includegraphics[clip,width=0.75\textwidth]{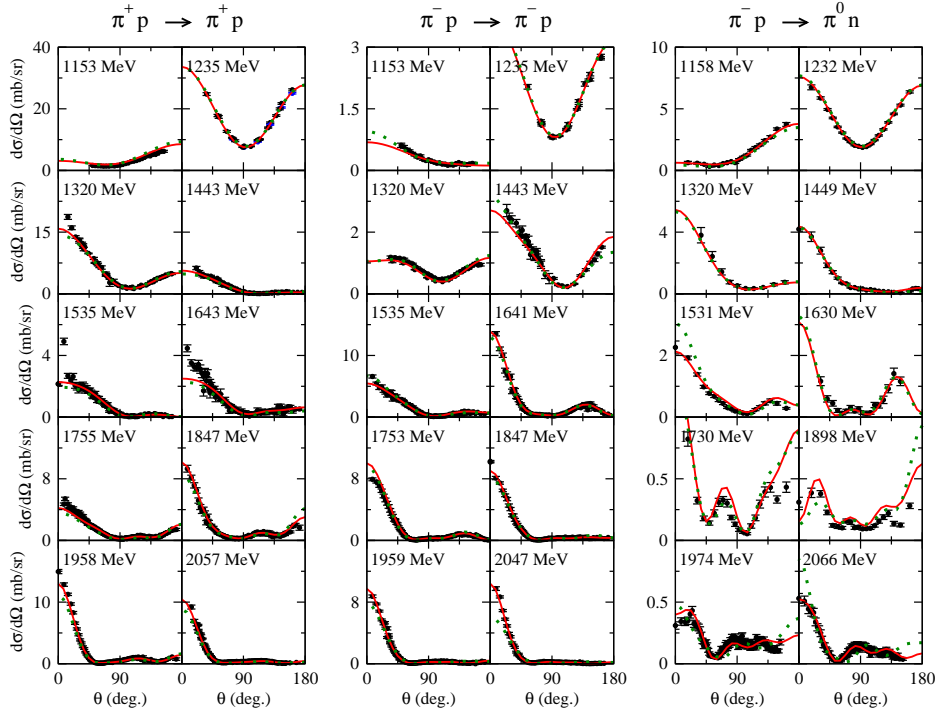}
\caption{(Color online)
Differential cross section $d\sigma/d\Omega$ of $\pi N \to \pi N$.
(Red solid curves) current results; 
(green dotted curves) results from our previous analysis~\cite{jlms07}.
The corresponding values of $W$ are indicated in each panel.
The same applies to the rest of figures.
}
\label{fig:pin-dcs}
\end{figure}

\begin{figure}[t]
\includegraphics[clip,width=0.75\textwidth]{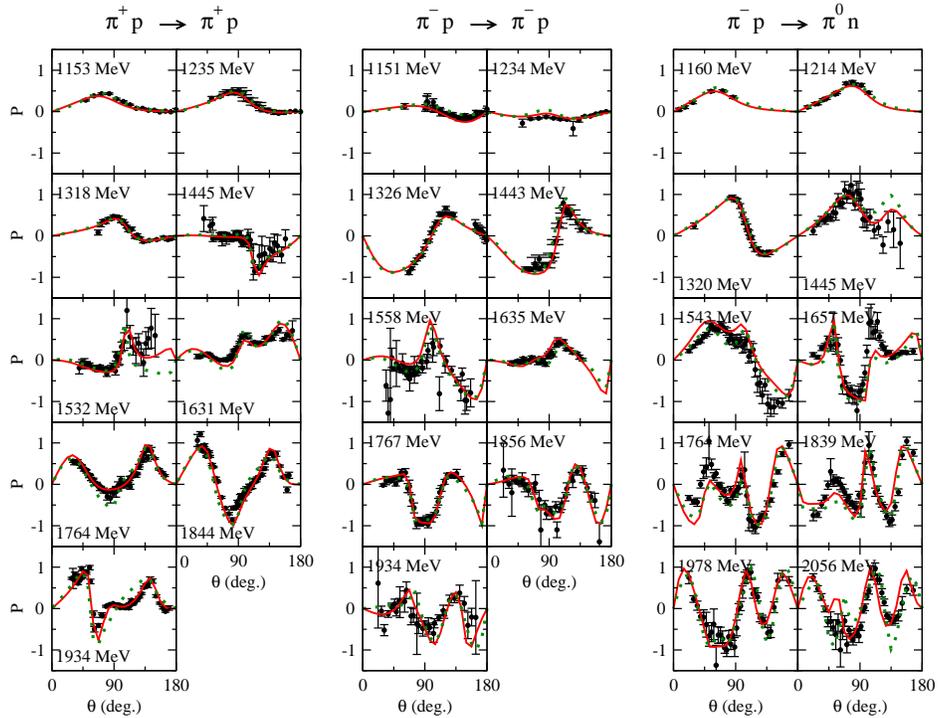}
\caption{(Color online)
Target polarization $P$ of $\pi N \to \pi N$.
(Red solid curves) current results; 
(green dotted curves) results from our previous analysis~\cite{jlms07}.
}
\label{fig:pin-p}
\end{figure}

To test our model directly with the experimental data,
we have also calculated the observables of $\pi N$ scattering. 
Some results of the predicted $\pi N$ elastic scattering observables
are shown in Figs.~\ref{fig:pin-dcs} and~\ref{fig:pin-p}. 
Clearly, good agreements have been obtained. 
Thus our model parameters are consistent with 
the $\sim 30,000$ data points of $\pi N$ scattering compiled in  SAID.
\begin{figure}[t]
\includegraphics[clip,width=0.8\textwidth]{pwa-re3}
\includegraphics[clip,width=0.8\textwidth]{pwa-im3}
\caption{(Color online)
Partial-wave amplitudes of $\pi N$ scattering with isospin $I=3/2$.
Upper (lower) panels are for real (imaginary) parts of the amplitudes.
(Red solid curves) current results; 
(green dotted curves) results from our previous analysis~\cite{jlms07}.
}
\label{fig:pin-amp-3}
\end{figure}

\clearpage

\subsection{$\gamma N\rightarrow \pi N$}

In our previous analysis~\cite{jlmss08},  
we only considered the data of differential cross  section
$d\sigma/d\Omega$ and photon asymmetry $\Sigma$ of the
$\gamma N \rightarrow \pi N$ reactions.
Also, the fits  were performed by
only adjusting the bare $ N^* \to \gamma N$ parameters, and
all other parameters of the model were fixed at the values determined in
the fits to the $\pi N\to \pi N$ data~\cite{jlms07}.
It was found that only the data below $W=1.6$ GeV can be fitted well 
by using such a very restricted procedure.
In this work,  we  allow 
 all of the model parameters to vary to fit simultaneously
the data of
 $\pi N \to \pi N, \eta N, K\Lambda, K\Sigma$ and 
$\gamma N \to \pi N, \eta N, K\Lambda, K\Sigma$.
This change of the fitting strategy gives us more
freedom to fit the data at higher energies above 1.6 GeV
and other polarization observables comprehensively.

\begin{figure}[t]
\includegraphics[clip,width=0.7\textwidth]{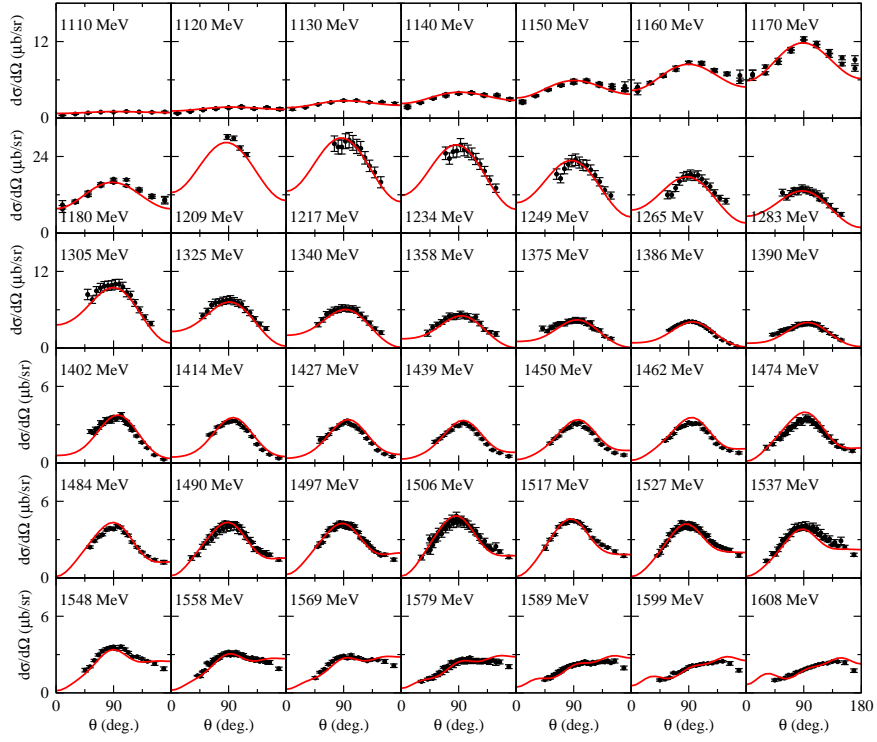}
\caption{(Color online)
$d\sigma/d\Omega$ of $\gamma p\to \pi^0 p$. 
}
\label{fig:gp-pi0p-dcs}
\end{figure}

\begin{figure}[t]
\includegraphics[clip,width=0.7\textwidth]{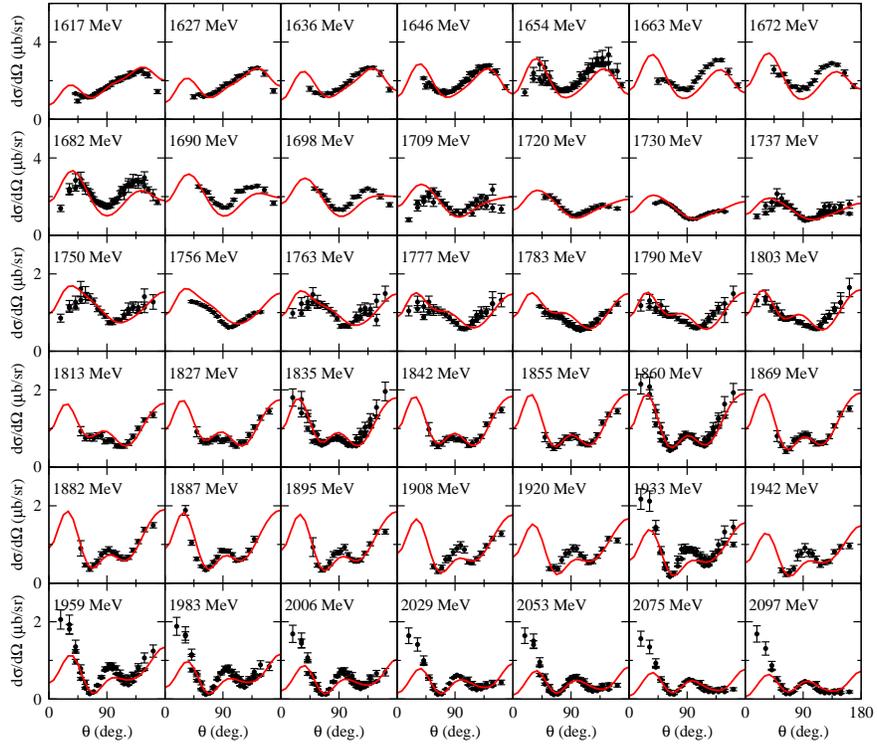}
\caption{(Color online)
$d\sigma/d\Omega$ of $\gamma p\to \pi^0 p$ (continued). 
}
\label{fig:gp-pi0p-dcs-2}
\end{figure}

\begin{figure}[t]
\includegraphics[clip,width=0.7\textwidth]{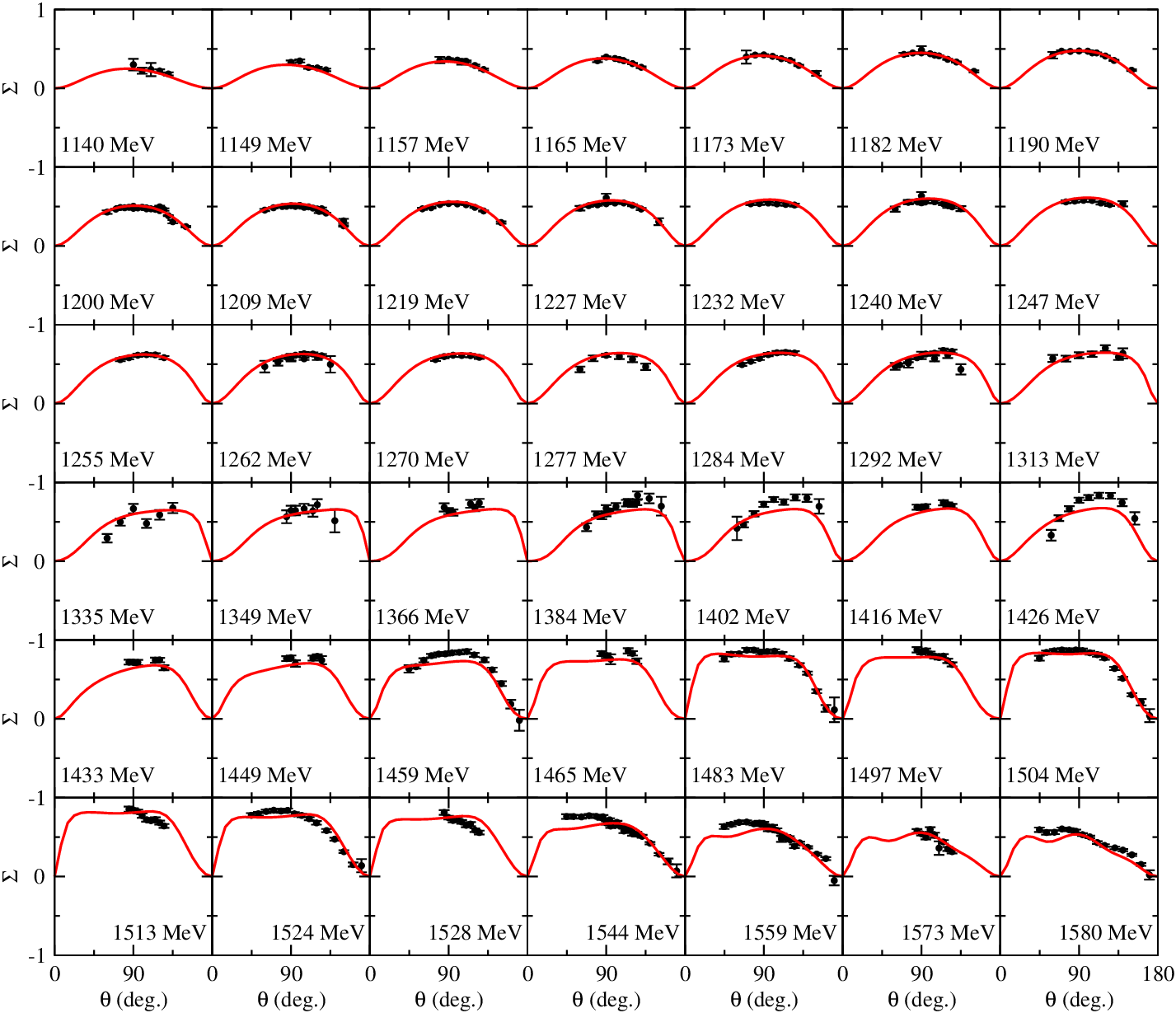}
\caption{(Color online)
$\Sigma$ of $\gamma p\to \pi^0 p$.
}
\label{fig:gp-pi0p-s}
\end{figure}

\begin{figure}[t]
\includegraphics[clip,width=0.7\textwidth]{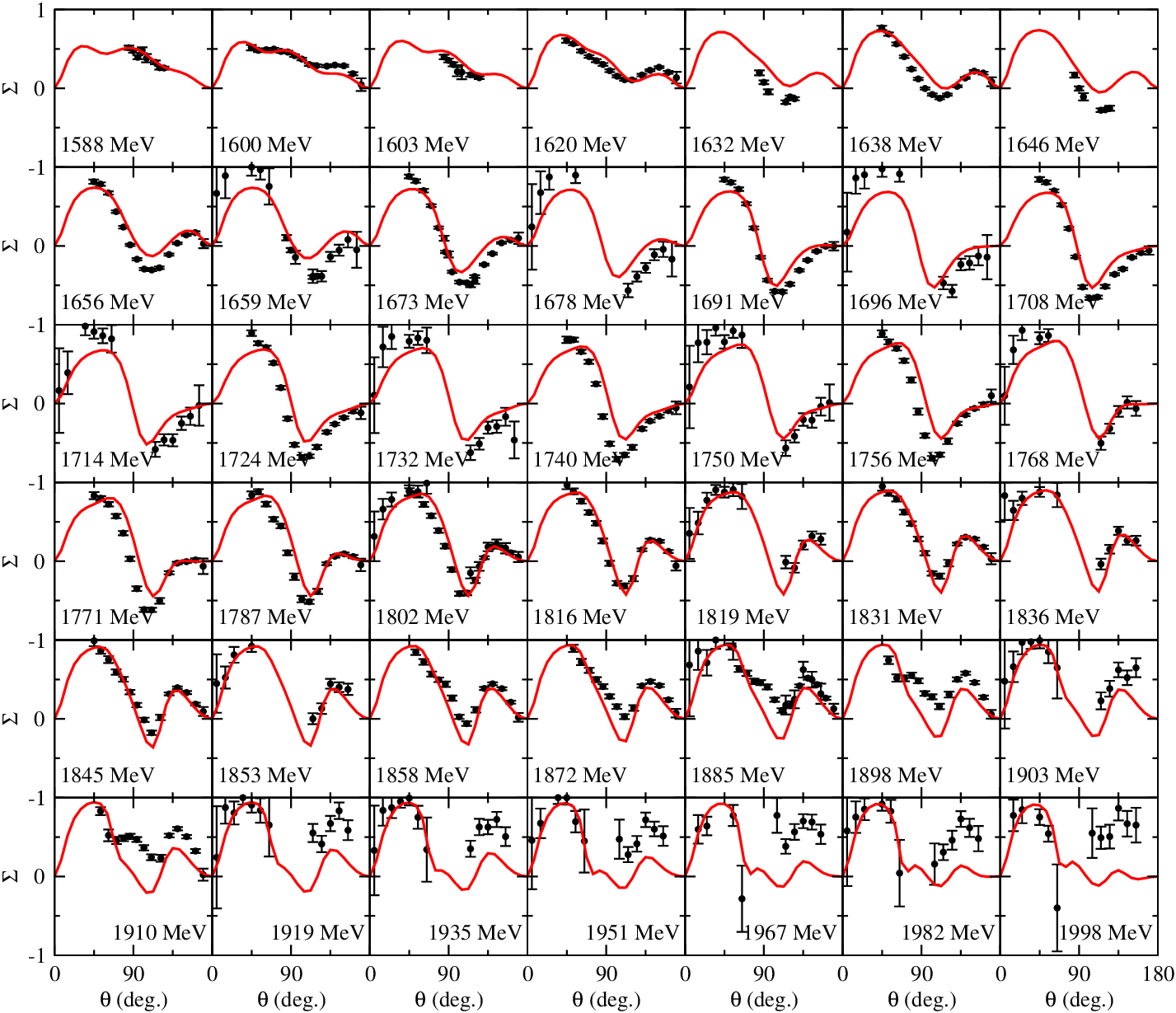}
\caption{(Color online)
$\Sigma$ of $\gamma p\to \pi^0 p$ (continued).
}
\label{fig:gp-pi0p-s-2}
\end{figure}

The $\gamma N \to \pi N$ data included in our fits
are listed in the first two rows of Table~\ref{tab:em-data}.
We see that the data points of
$d\sigma/d\Omega$ and $\Sigma$ of $\gamma p \to \pi^0 p$
are much more than those for the other considered pion
photoproduction observables.
Thus these two data sets have similar large weights
in determining the parameters in our coupled-channels analysis.
In Figs.~\ref{fig:gp-pi0p-dcs}-\ref{fig:gp-pi0p-s-2}, we see that
these rather extensive data below about $W=1.9$ GeV can be fitted very well.
We, however, are not able to account for the forward peaks 
in $d\sigma/d\Omega$ at $W \geq 1.933$ GeV. 
Similar difficulty is also encountered in fitting  the data of $\Sigma$ 
at $W \gtrsim 1.9$ GeV. 
We expect that such data in the forward-angle region
are affected mainly by the $t$-channel processes
rather than the $s$-channel resonance ones.
This might suggest a need for an incorporation of Regge-type processes
at high energies.

\begin{figure}[t]
\includegraphics[clip,width=0.70\textwidth]{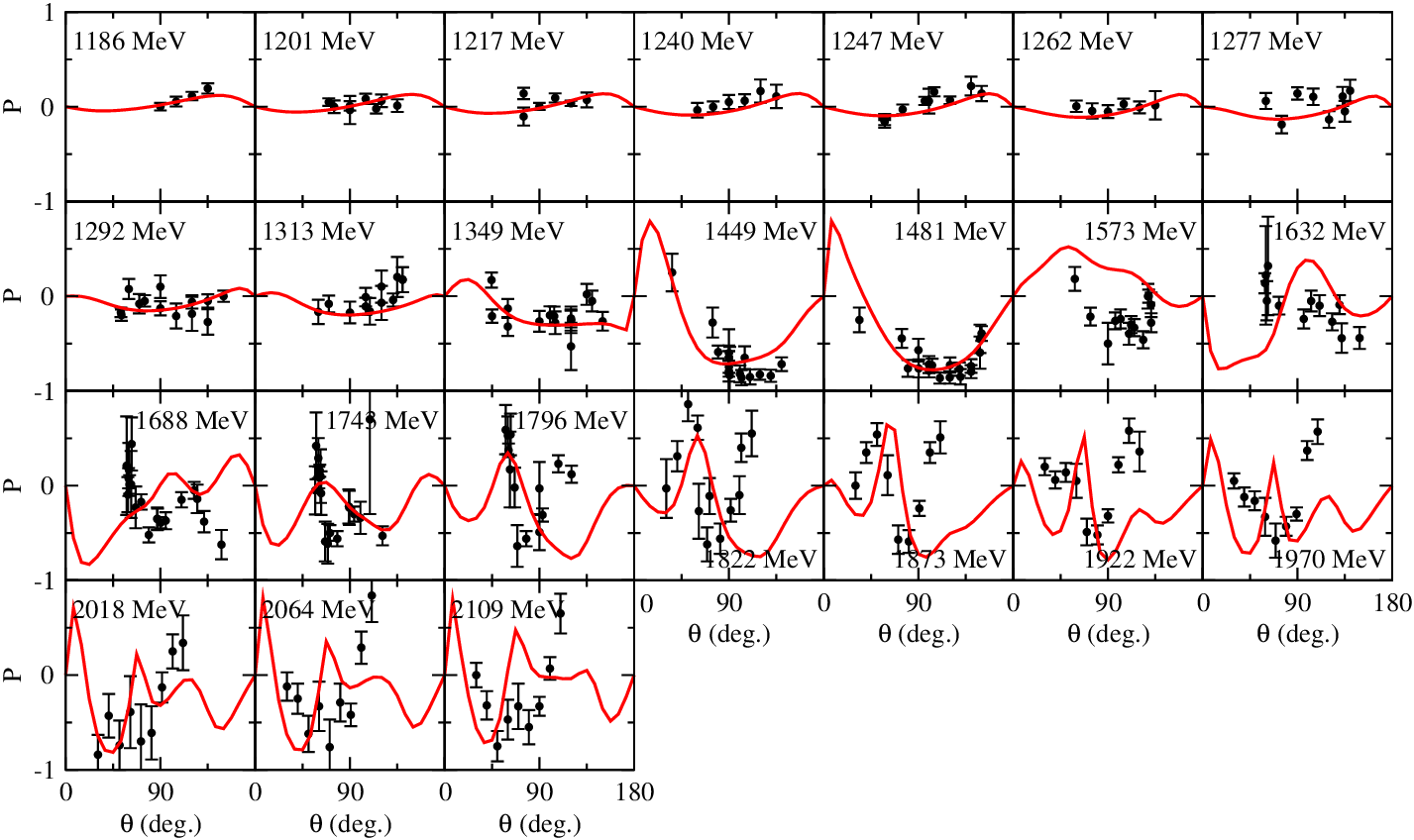}
\caption{(Color online)
$P$ of $\gamma p\to \pi^0 p$.
}
\label{fig:gpp0p-P}
\end{figure}

\begin{figure}[t]
\includegraphics[clip,width=0.70\textwidth]{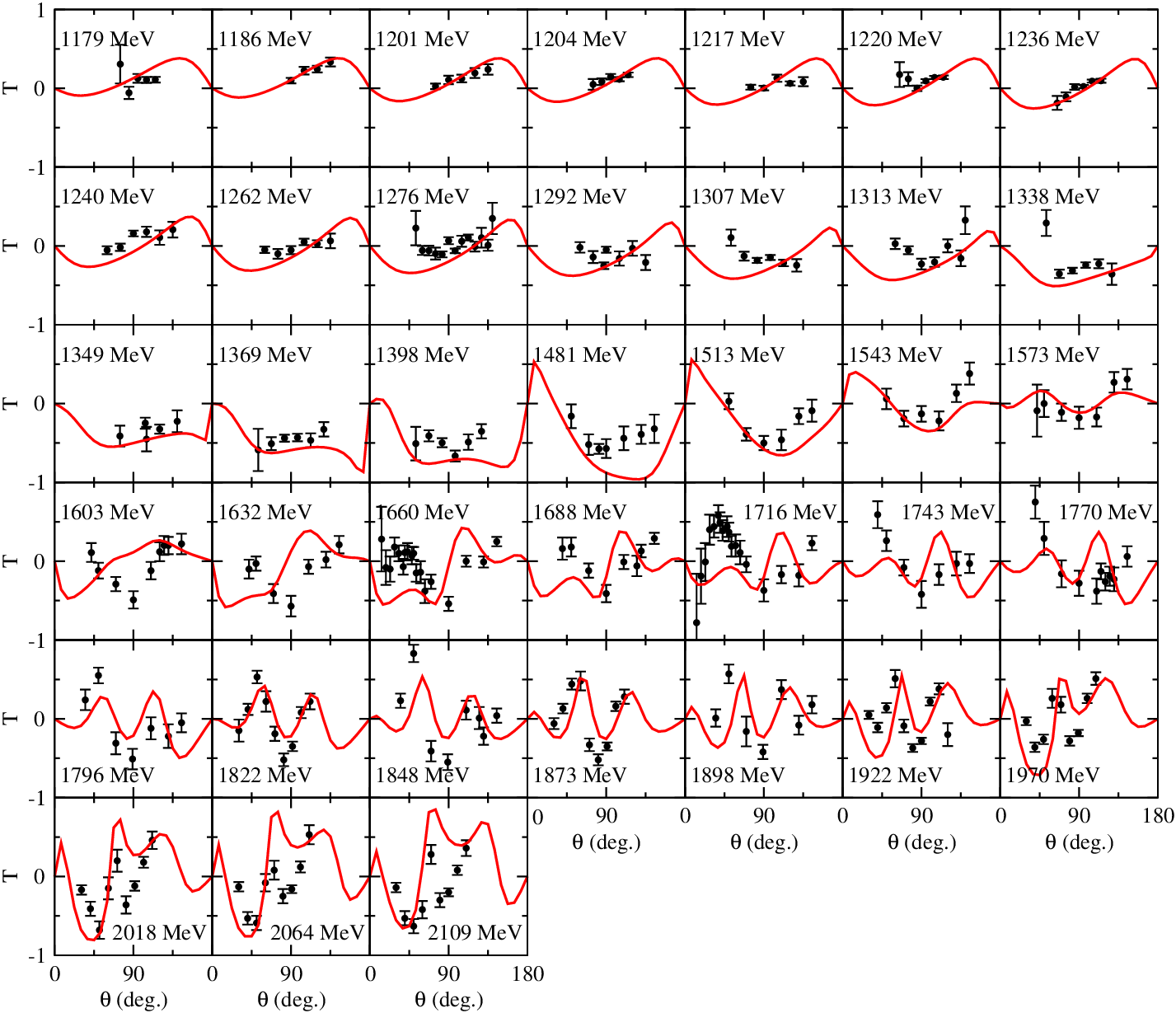}
\caption{(Color online)
$T$ of $\gamma p\to \pi^0 p$.
}
\label{fig:gpp0p-T}
\end{figure}

\begin{figure}[t]
\includegraphics[clip,width=0.70\textwidth]{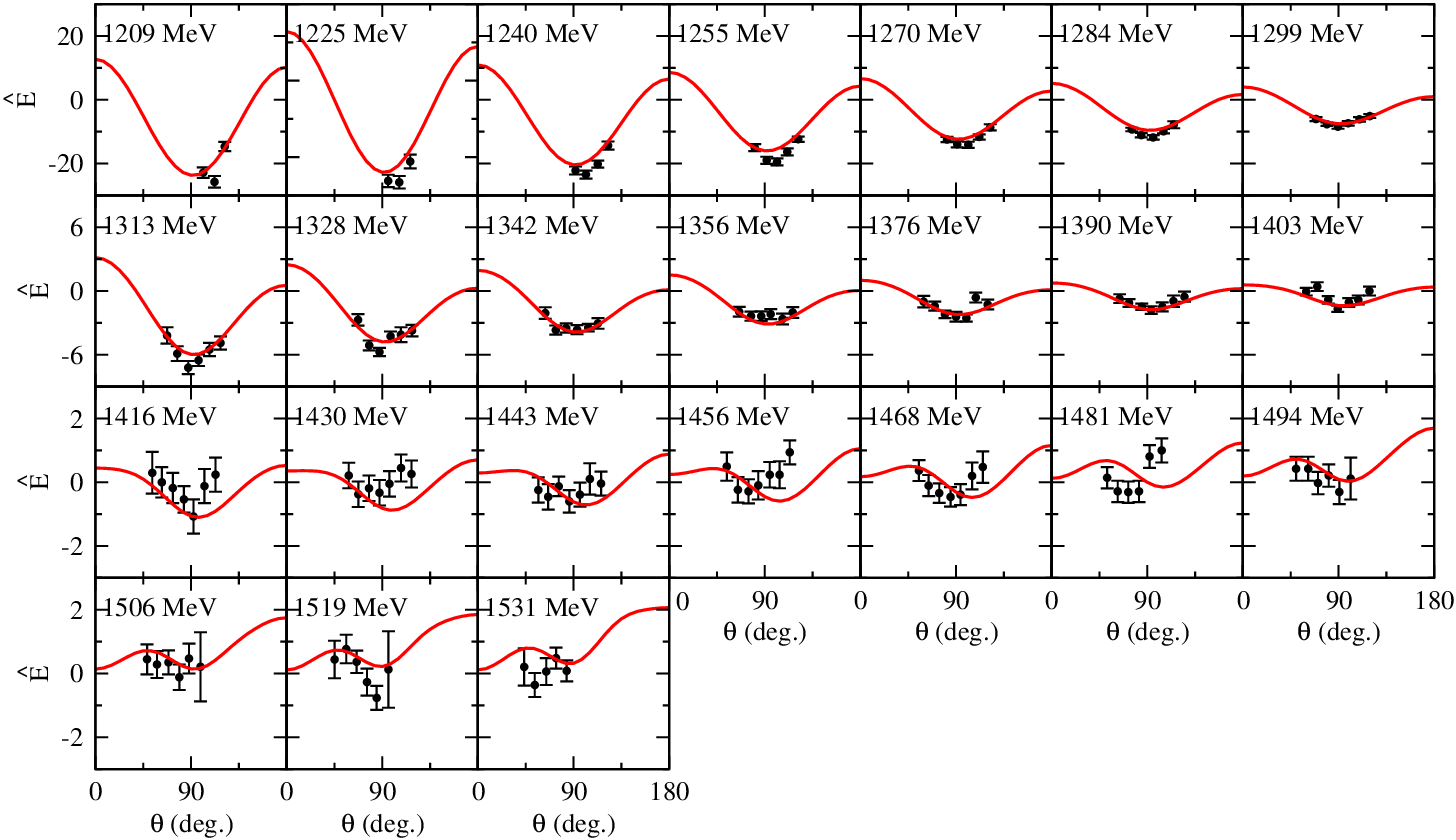}
\caption{(Color online)
$\hat E$ of $\gamma p\to \pi^0 p$.
}
\label{fig:gpp0p-hE}
\end{figure}

\begin{figure}[t]
\includegraphics[clip,width=0.70\textwidth]{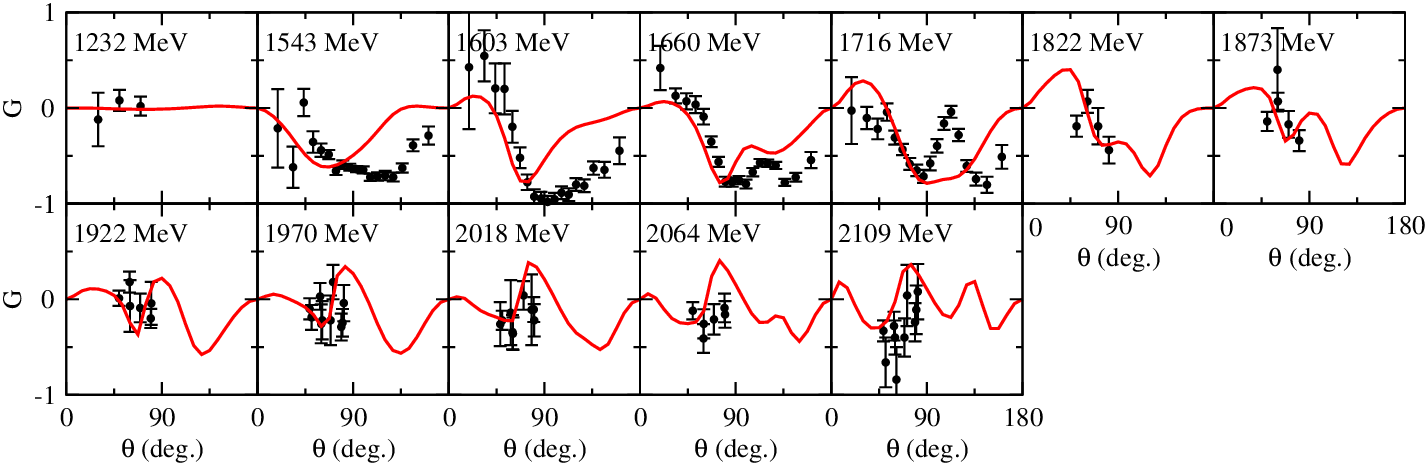}
\includegraphics[clip,width=0.70\textwidth]{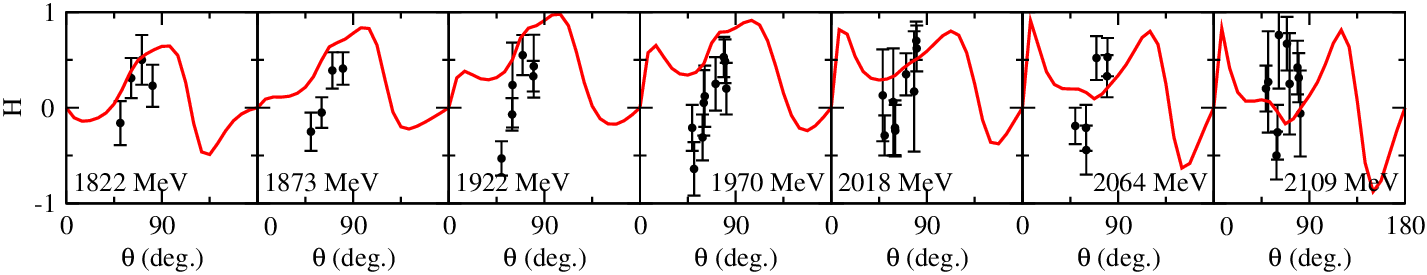}
\caption{(Color online)
$G$ and $H$ of $\gamma p\to \pi^0 p$.
}
\label{fig:gpp0p-GH}
\end{figure}

The fits to the other polarization observables, $P$, $T$, $\hat E$, $G$, and $H$, of 
$\gamma p \to \pi^0 p$ are shown in Figs.~\ref{fig:gpp0p-P}-\ref{fig:gpp0p-GH}. 
[Note that the beam-target polarization $\hat E$ defined in Ref.~\cite{shkl11}
and the quantity $\Delta_{31}$ measured and presented in Ref.~\cite{ehat},
which are used as the data in our fits,
are related with each other by $\hat E = -(1/2)\Delta_{31}$.]
As seen in Table~\ref{tab:em-data}, the numbers of 
these data points are much less than those of $d\sigma/d\Omega$ and $\Sigma$;
it is therefore not easy to improve the fits to these polarization
observables, in particular $\hat E$, $G$, and $H$ 
(Figs.~\ref{fig:gpp0p-hE} and~\ref{fig:gpp0p-GH}).
While we need to improve our fits, more precise data of the polarization
observables are also needed to make further progress.

\begin{figure}[t]
\includegraphics[clip,width=0.70\textwidth]{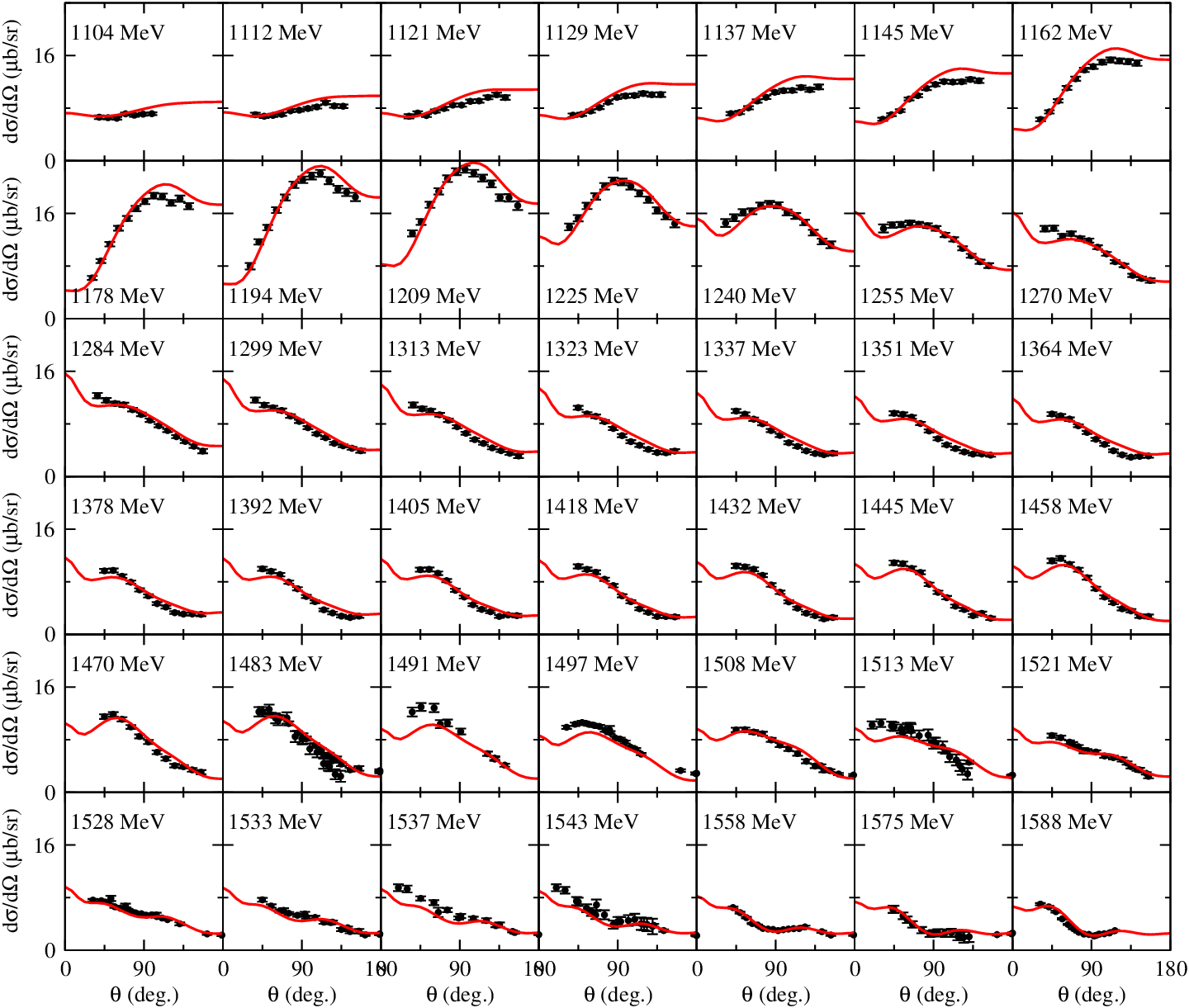}
\caption{(Color online)
$d\sigma/d\Omega$ of $\gamma p\to \pi^+ n$.
}
\label{fig:gp-pipn-dcs}
\end{figure}

\begin{figure}[t]
\includegraphics[clip,width=0.70\textwidth]{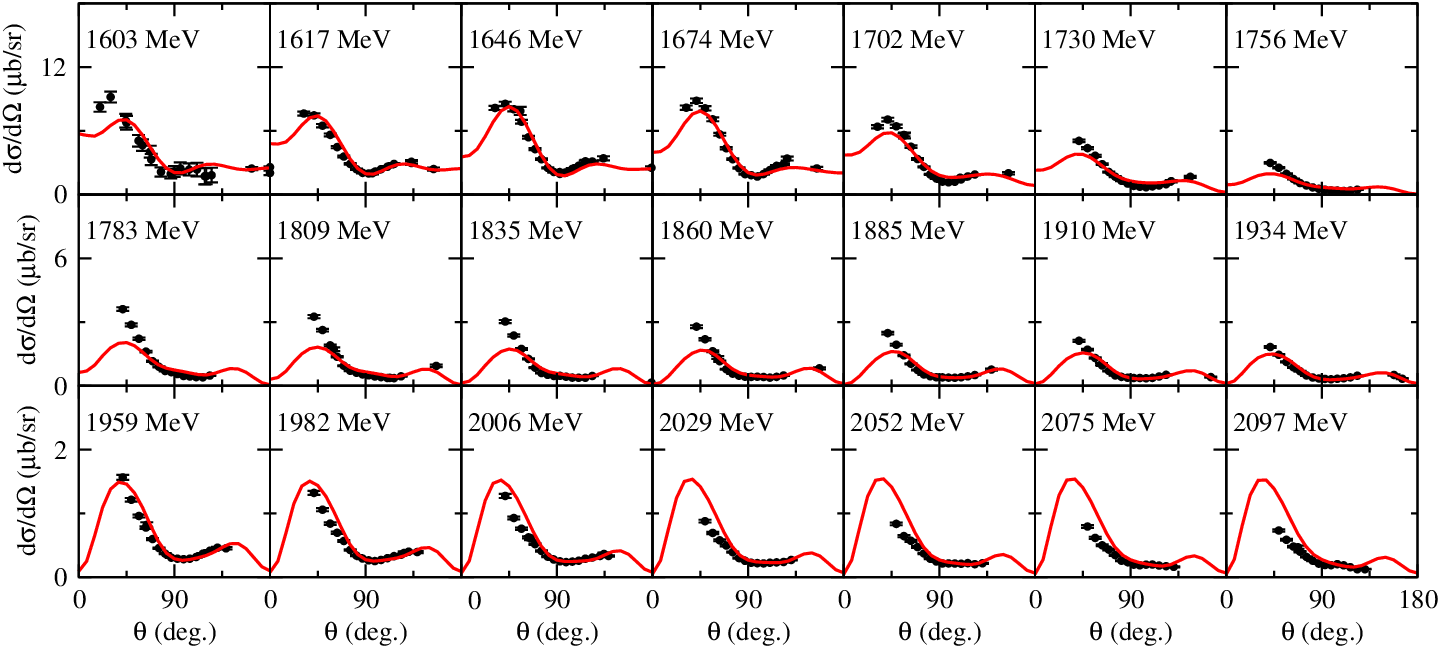}
\caption{(Color online)
$d\sigma/d\Omega$ of $\gamma p\to \pi^+ n$ (continued).
}
\label{fig:gp-pipn-dcs-2}
\end{figure}

\begin{figure}[t]
\includegraphics[clip,width=0.70\textwidth]{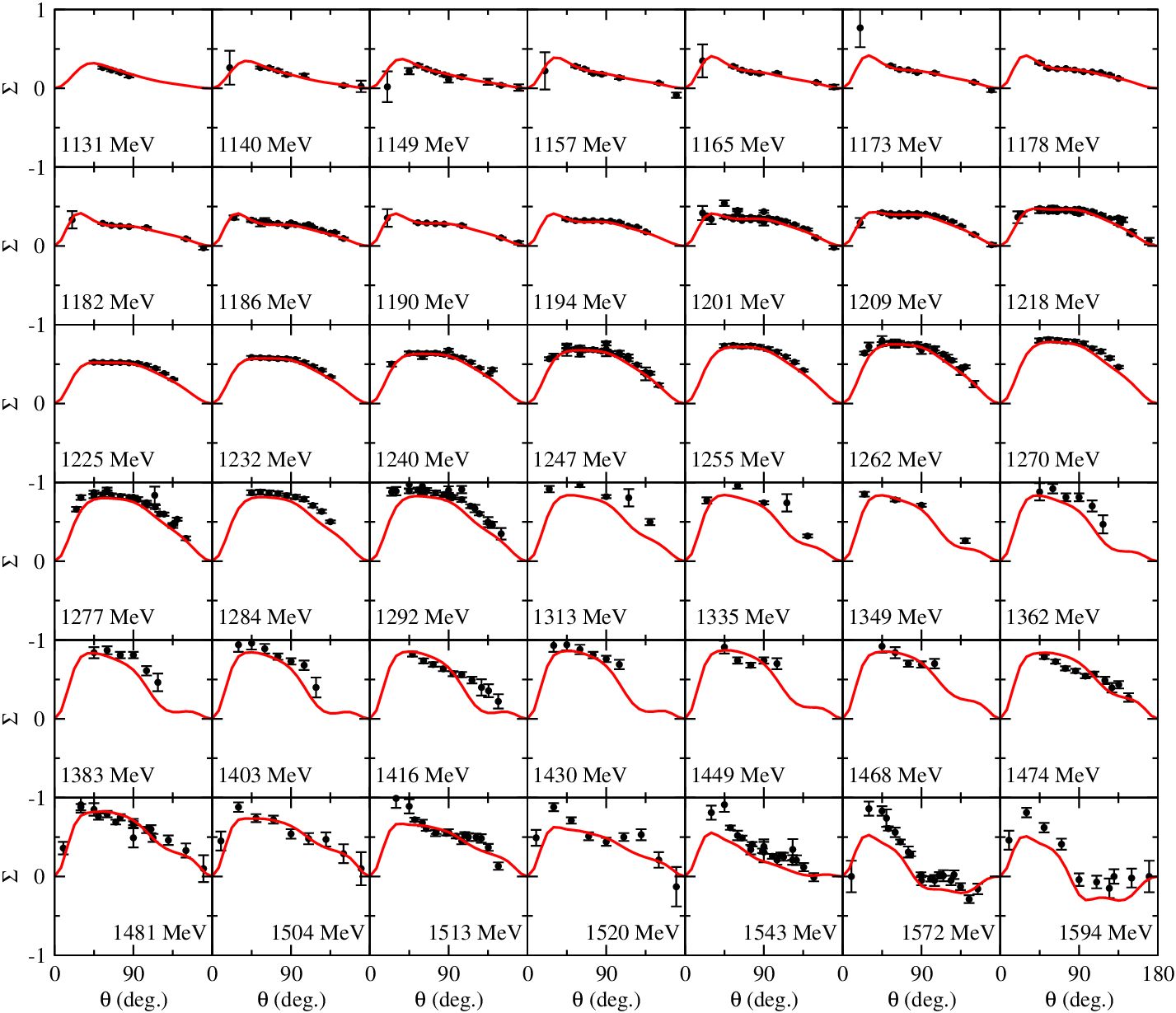}
\caption{(Color online)
$\Sigma$ of $\gamma p\to \pi^+ n$.
}
\label{fig:gp-pipn-s}
\end{figure}

\begin{figure}[t]
\includegraphics[clip,width=0.70\textwidth]{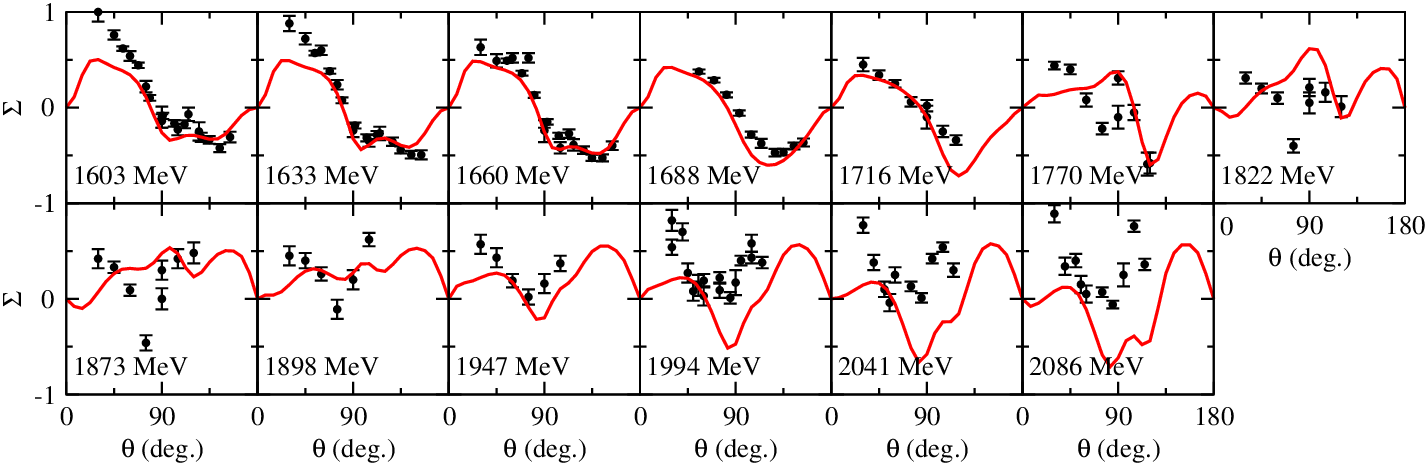}
\caption{(Color online)
$\Sigma$ of $\gamma p\to \pi^+ n$ (continued).
}
\label{fig:gp-pipn-s-2}
\end{figure}

\begin{figure}[t]
\includegraphics[clip,width=0.70\textwidth]{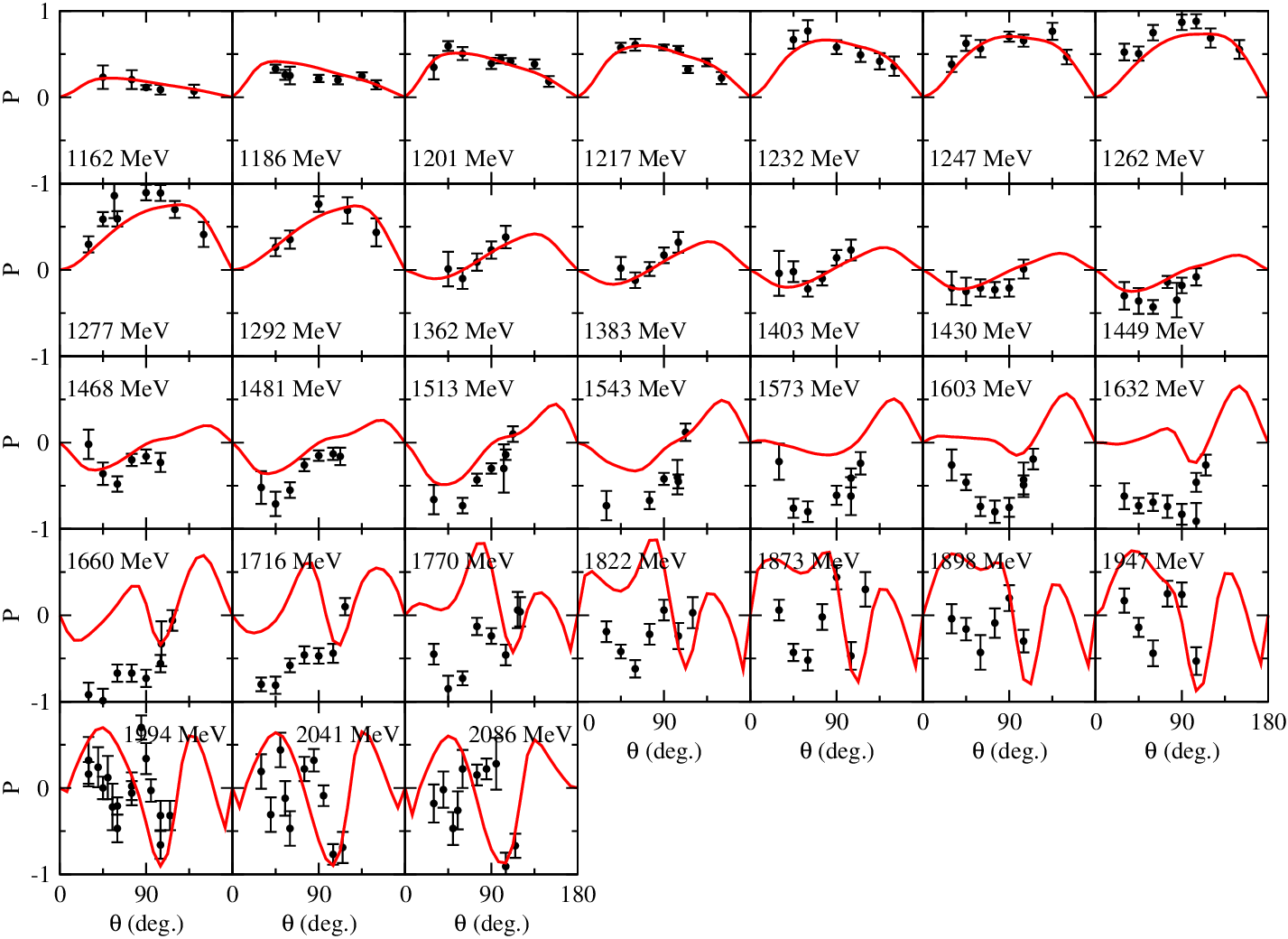}
\caption{(Color online)
$P$ of $\gamma p\to \pi^+ n$.
}
\label{fig:gpppn-P}
\end{figure}

\begin{figure}[t]
\includegraphics[clip,width=0.80\textwidth]{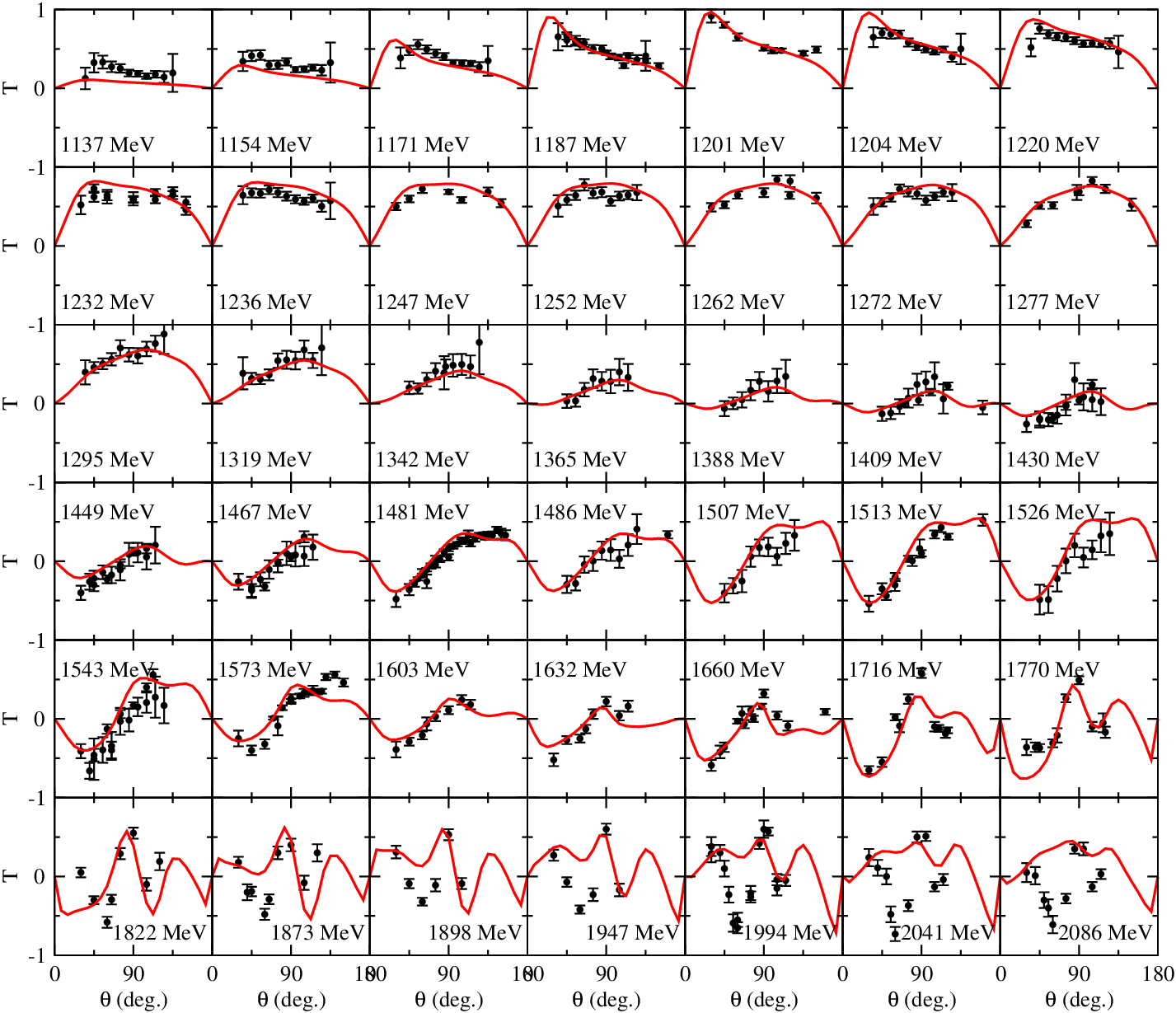}
\caption{(Color online)
$T$ of $\gamma p\to \pi^+ n$.
}
\label{fig:gpppn-T}
\end{figure}

\begin{figure}[t]
\includegraphics[clip,width=0.70\textwidth]{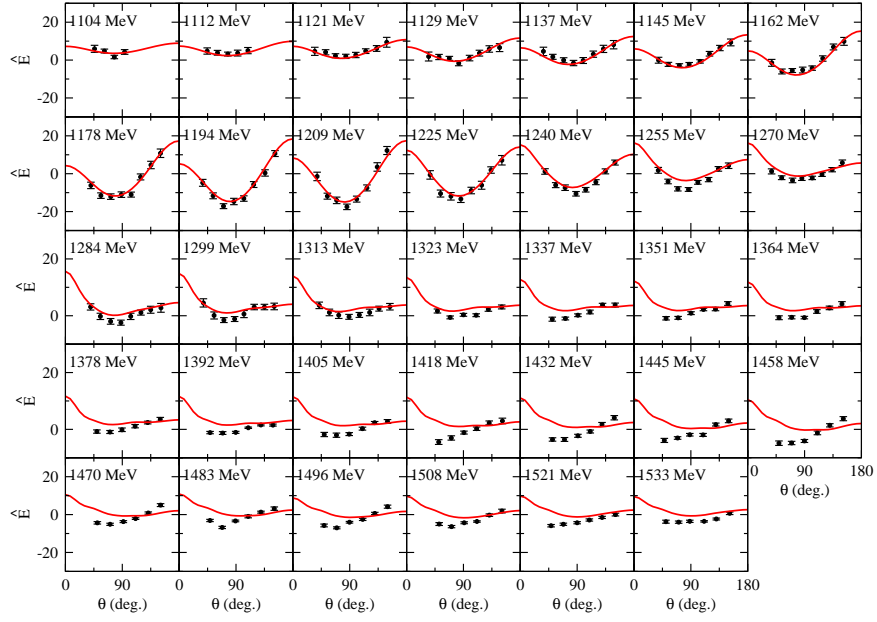}
\caption{(Color online)
$\hat E$ of $\gamma p\to \pi^+ n$.
}
\label{fig:gpppn-hE}
\end{figure}

\begin{figure}[t]
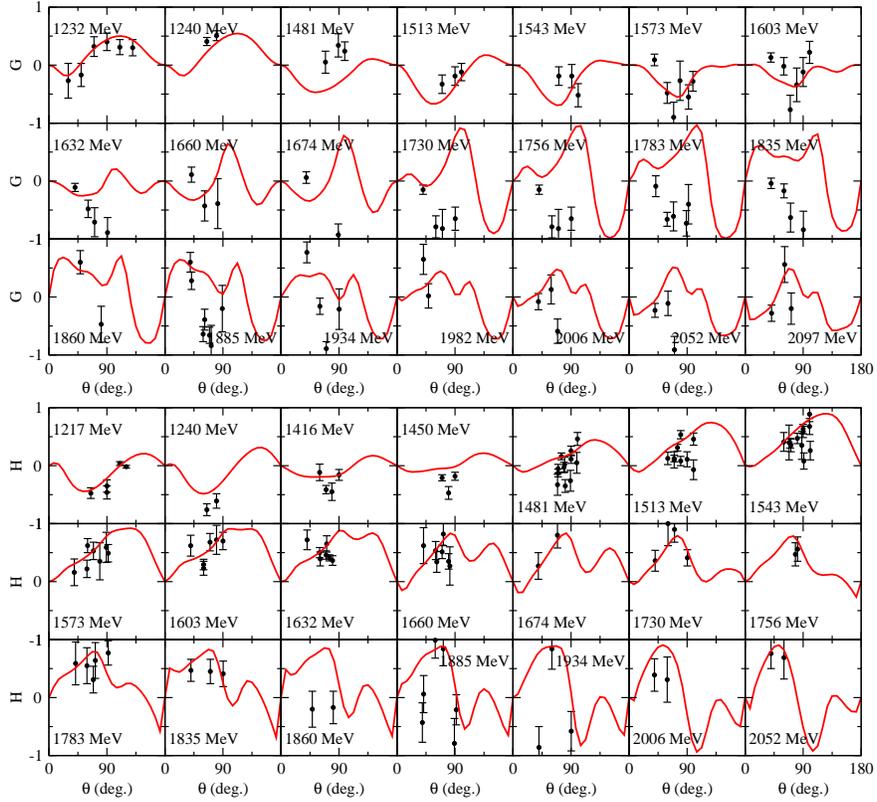

\includegraphics[clip,width=0.70\textwidth]{gpppn-G}
\includegraphics[clip,width=0.70\textwidth]{gpppn-H}
\caption{(Color online)
$G$ and $H$ of $\gamma p\to \pi^+ n$.
}
\label{fig:gpppn-GH}
\end{figure}

Our fits to the data of $d\sigma/d\Omega$ of $\gamma p \to \pi^+n$ are shown in
Figs.~\ref{fig:gp-pipn-dcs} and~\ref{fig:gp-pipn-dcs-2}.
The fits are good in most of the considered energy region.
The data at forward angles in the $W \gtrsim 1.8$ GeV region
cannot be accounted for.
The origin of this difficulty is perhaps related to the similar problem in
$d\sigma/d\Omega$ for $\gamma p \to \pi^0p$.
Our fits to the polarization observables of $\gamma p \to \pi^+n$ become
less accurate as $W$ increases, as seen in Figs.~\ref{fig:gp-pipn-s}-\ref{fig:gpppn-GH}.

Overall, we are able to fit more data of $\gamma p \to \pi^0 p, \pi^+n$
than the JLMS analysis~\cite{jlmss08}.
Not only do we cover the $W \geq 1.6$ GeV region,
our fits to the data in the low-energy region with $W \leq 1.2$ GeV
are much better.
We, however, still cannot fit accurately the data in the $W$ region close to 2.0 GeV.
Here we note that in this high-$W$ region,
the data for the $\eta N, K\Lambda$, and $K\Sigma$ channels
play a very significant role in the analysis through the coupled-channels effects.
Thus, the discrepancies with the data in the high-$W$ region cannot be trivially removed.

\clearpage


\clearpage

\subsection{$\pi^-p  \to \eta n$}

\begin{figure}[h]
\includegraphics[clip,width=0.70\textwidth]{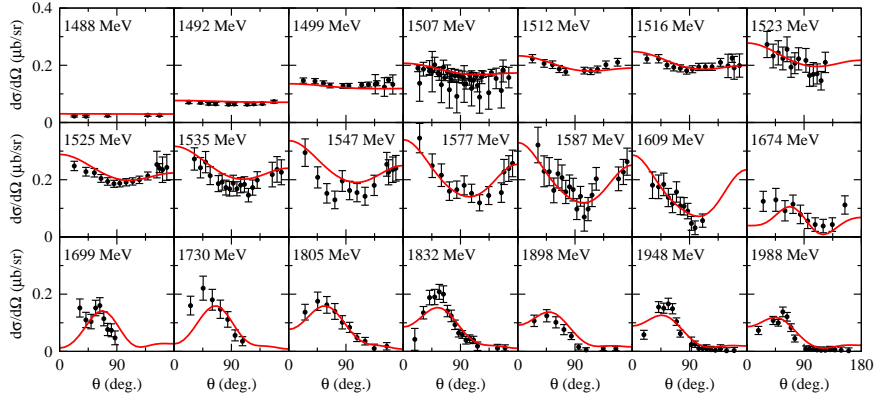}
\caption{(Color online)
$d\sigma/d\Omega$ of $\pi^- p\to \eta n$.
}
\label{fig:pin-etan-dcs}
\end{figure}

The $\pi^-p  \to \eta n$ is simpler than $\pi N$ elastic scattering because
it  depends only on the isospin $I=1/2$ partial waves.
As mentioned in Sec.~\ref{sec:fit-proc}, the data for this reaction
used in  our fits are chosen carefully according to the discussion in Ref.~\cite{djlss08}
because some inconsistency exists among the data sets from different experiments.
In Fig.~\ref{fig:pin-etan-dcs}, we show our fits to the data of
$d\sigma/d\Omega$ data of $\pi^-p  \rightarrow \eta n$.
Clearly, our results reproduce the considered $d\sigma/d\Omega$ data well.

\subsection{$\gamma  p \to \eta p$}

\begin{figure}[t]
\includegraphics[clip,width=0.70\textwidth]{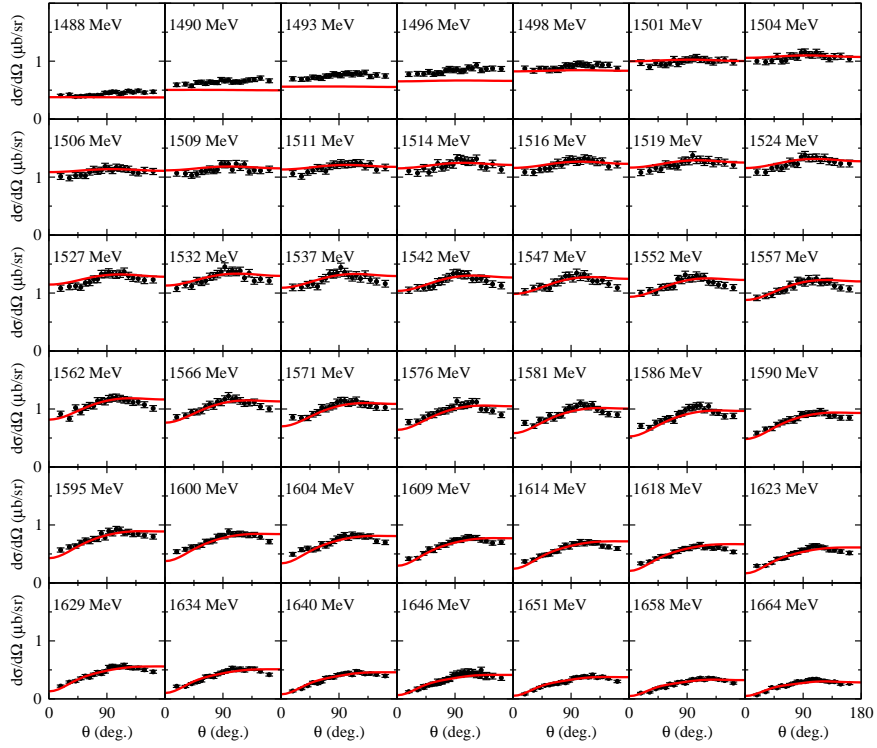}
\caption{(Color online)
$d\sigma/d\Omega$ of $\gamma p\to \eta p$.
}
\label{fig:gn-etan-dcs}
\end{figure}

\begin{figure}
\includegraphics[clip,width=0.70\textwidth]{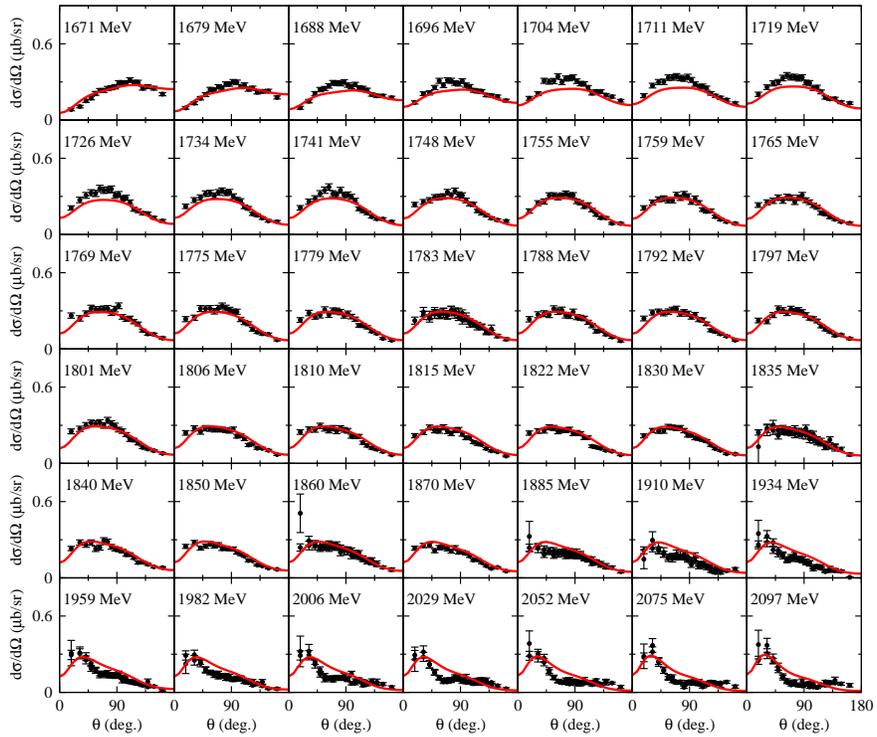}
\caption{(Color online)
$d\sigma/d\Omega$ of $\gamma p\to \eta p$ (continued).
}
\label{fig:gn-etan-dcs-2}
\end{figure}

\begin{figure}
\includegraphics[clip,width=0.70\textwidth]{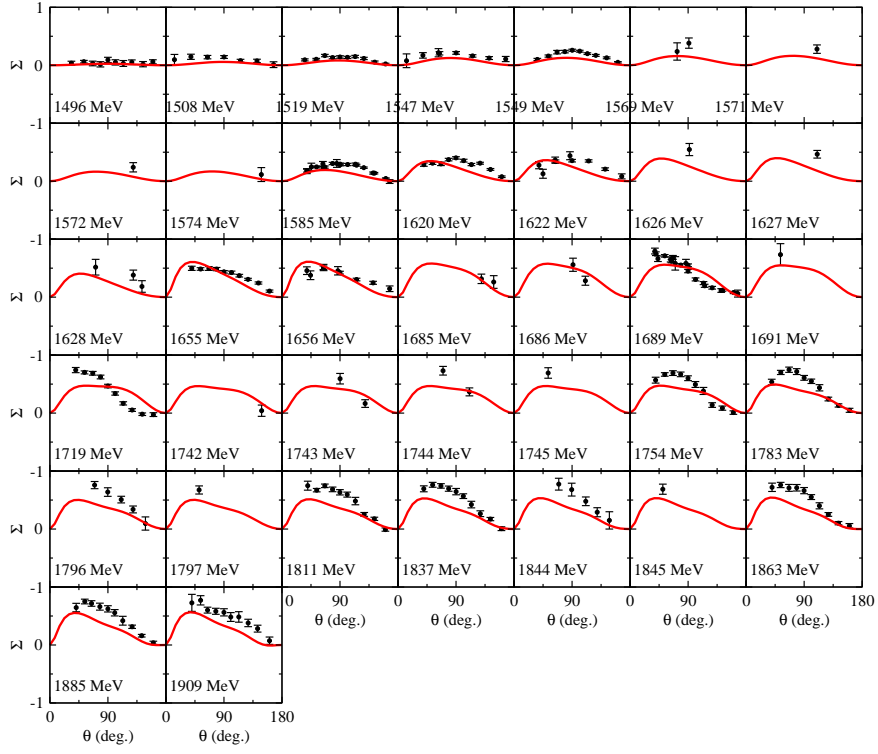}
\caption{(Color online)
$\Sigma$ of $\gamma p\to \eta p$.
}
\label{fig:gn-etan-s}
\end{figure}

\begin{figure}
\includegraphics[clip,width=0.70\textwidth]{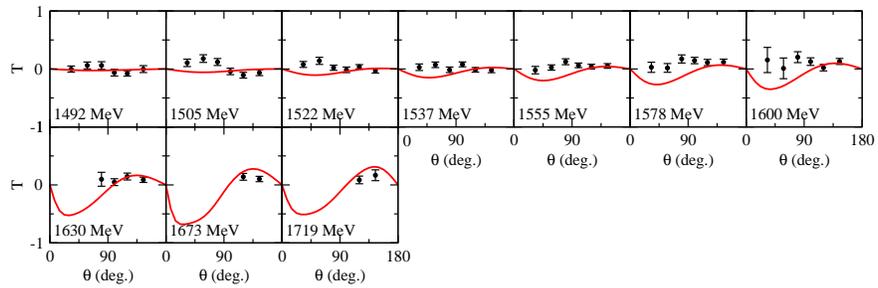}
\caption{(Color online)
$T$ of $\gamma p\to \eta p$.
}
\label{fig:gn-etan-t}
\end{figure}

Our fits to the data of $\gamma  p \to \eta p$ up to $W=2.1$ GeV 
are shown in Figs.~\ref{fig:gn-etan-dcs}-\ref{fig:gn-etan-t}.
Clearly,  both the differential cross sections
$d\sigma/d\Omega$ and the polarization observables $\Sigma$ and
$T$ can be described well in the considered energy region.  
Our results for $\Sigma$ above $W =1.7$ GeV
are  smaller than the data 
in the $0^\circ\lesssim\theta \lesssim 90^\circ$ region,
although the positive values of the data are reproduced to a large extent.

\clearpage

\subsection{$\pi^+ p \to K^+\Sigma^+$}

\begin{figure}[h]
\includegraphics[clip,width=0.67\textwidth]{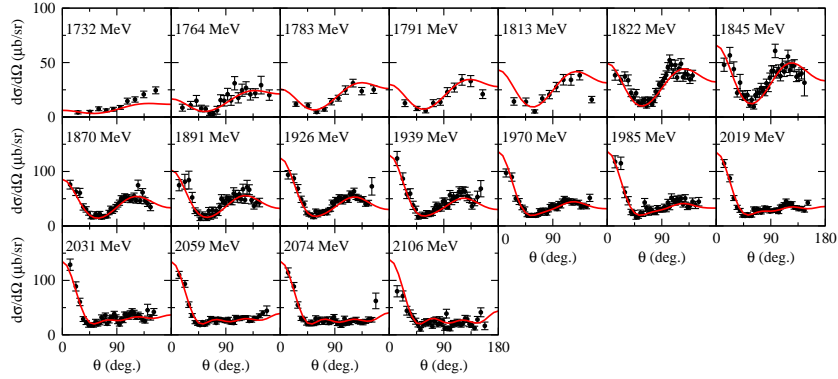}
\caption{(Color online)
$d\sigma/d\Omega$ of $\pi^+ p \to K^+\Sigma^+$.}
\label{fig:pin-pks-dcs}
\end{figure}

\begin{figure}[h]
\includegraphics[clip,width=0.67\textwidth]{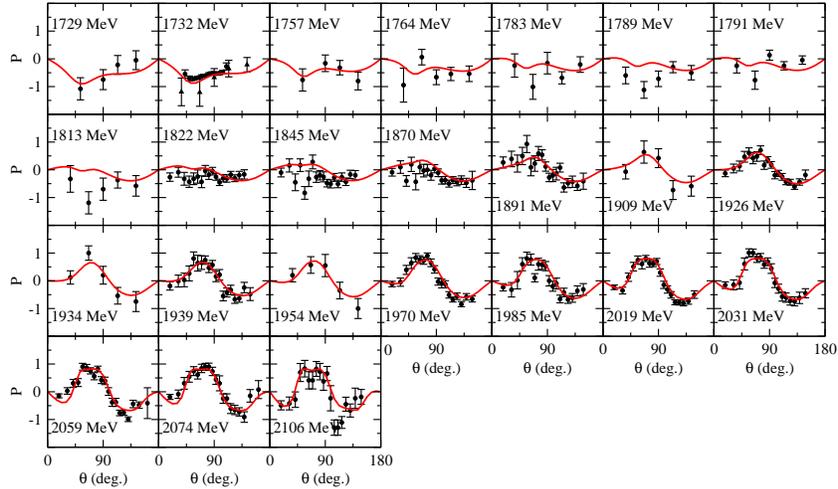}
\caption{(Color online)
$P$ of $\pi^+ p \to K^+\Sigma^+$.}
\label{fig:pin-pks-p}
\end{figure}

\begin{figure}[h]
\includegraphics[clip,width=0.23\textwidth]{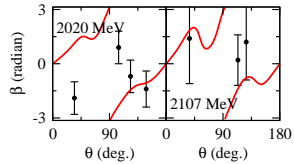}
\caption{(Color online)
$\beta$ of $\pi^+ p \to K^+\Sigma^+$.}
\label{fig:pin-pks-b}
\end{figure}

The $\pi^+ p \to K^+\Sigma^+$ reaction
depends only on the isospin $I=3/2$ partial waves, and
hence the $I=1/2$ $N^*$ states cannot  contribute.
Our fits to both $d\sigma/d\Omega$ and $P$ up to $W=2106$ MeV 
are good, as shown in Figs.~\ref{fig:pin-pks-dcs} and~\ref{fig:pin-pks-p}. 
In Fig.~\ref{fig:pin-pks-b}, we present the spin-rotation $\beta$.
This quantity is modulo $2\pi$, and here we plot it in the range $[-\pi,\pi]$.
At present, almost no data is available for $\beta$ of this reaction 
and these few  data points hardly affect the $\chi^2$-minimization.
We thus are not able to improve our results shown in Fig.~\ref{fig:pin-pks-b}.

\clearpage

\subsection{$\pi^- p \to K^0\Lambda^0$}

\begin{figure}[h]
\includegraphics[clip,width=0.67\textwidth]{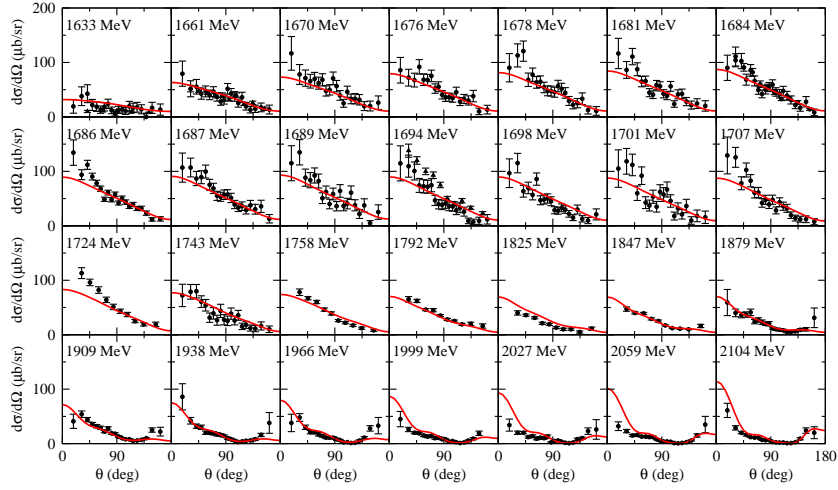}
\caption{(Color online)
$d\sigma/d\Omega$ of $\pi^- p \to K^0\Lambda^0$.}
\label{fig:pin-k0l0-dcs}
\end{figure}

\begin{figure}[h]
\includegraphics[clip,width=0.67\textwidth]{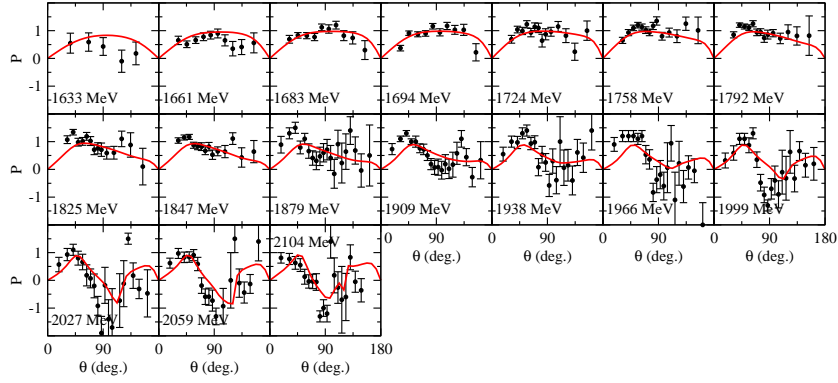}
\caption{(Color online)
$P$ of $\pi^- p \to K^0\Lambda^0$.}
\label{fig:pin-k0l0-p}
\end{figure}

\begin{figure}[h]
\includegraphics[clip,width=0.5\textwidth]{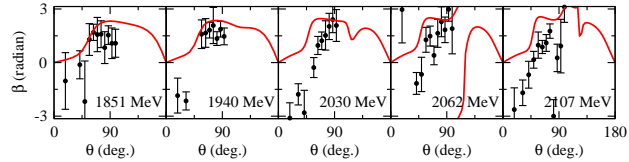}
\caption{(Color online)
$\beta$ of $\pi^- p \to K^0\Lambda^0$.}
\label{fig:pin-k0l0-b}
\end{figure}

In contrast with the $\pi^+ p \rightarrow K^+\Sigma^+$ discussed in the
above section, 
the $\pi^- p \to K^0\Lambda^0$ only involves the isospin $I=1/2$ mechanism.
Our fits to $d\sigma/d\Omega$ and $P$ of this reaction
are shown in Figs.~\ref{fig:pin-k0l0-dcs} and~\ref{fig:pin-k0l0-p},
respectively.
Clearly, our fits are very  good. 
Our results for the spin rotation $\beta$ (Fig.~\ref{fig:pin-k0l0-b}) reproduce the main feature of
the data while these few data points, like those in 
the $\pi^+ p \rightarrow K^+\Sigma^+$ reaction,  have practically no
effect in the $\chi^2$ minimization.
More data for $\beta$ will be necessary to improve the situation.

\clearpage

\subsection{$\pi^- p \to K^0\Sigma^0$}

\begin{figure}[h]
\includegraphics[clip,width=0.70\textwidth]{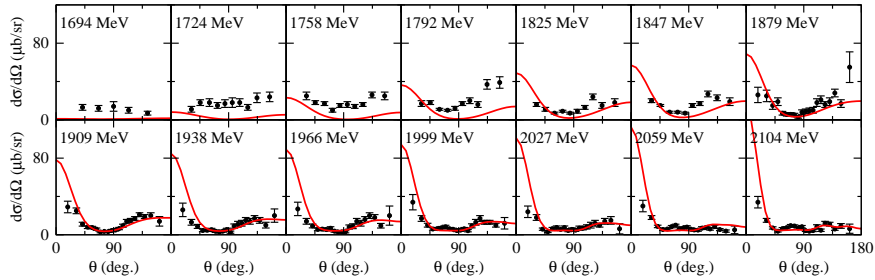}
\caption{(Color online)
$d\sigma/d\Omega$ of $\pi^- p \to K^0\Sigma^0$.}
\label{fig:pin-pks0-dcs}
\end{figure}

\begin{figure}[h]
\includegraphics[clip,width=0.70\textwidth]{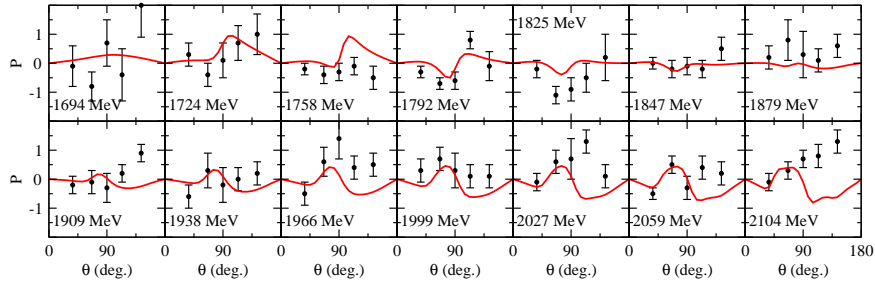}
\caption{(Color online)
$P$ of $\pi^- p \to K^0\Sigma^0$.}
\label{fig:pin-pks0-p}
\end{figure}

The $\pi^- p \to K^0\Sigma^0$ reaction is
more complex than $\pi^+ p \to K^+\Sigma^+$ 
and $\pi^- p \rightarrow K^0\Lambda^0$,
because it involves interfering contributions from
both the isospin $I=1/2$ and $I=3/2$ mechanisms.
Our fits to the data of $d\sigma/d\Omega$ and $P$ of this reaction
are shown in Figs.~\ref{fig:pin-pks0-dcs} and~\ref{fig:pin-pks0-p},
respectively.
We see that our results reproduce the data to a large extent in the considered energy
region, although the $d\sigma/d\Omega$ data in the near-threshold region
are underestimated.

Here we note that all of the $\pi N \to KY$ data used in our fits
are from old measurements more than a quarter of century ago.
Their statistical errors are, in general, quite large compared with
the recent photoproduction data. 
Furthermore, the amount of data is scarce in the kinematical region
relevant to $N^\ast$ above 1.6 GeV,
in particular for the $\pi^- p \to K^0\Sigma^0$ reaction.
An experiment planned at J-PARC~\cite{jparc-p45}
is quite encouraging, where the high-precision data
for $\pi N \to \pi\pi N$ and $\pi N \to K Y$ will be obtained in the wide energy region.

\clearpage

\subsection{$\gamma p \to K^+\Lambda$}

\begin{figure}[b]
\includegraphics[clip,width=0.70\textwidth]{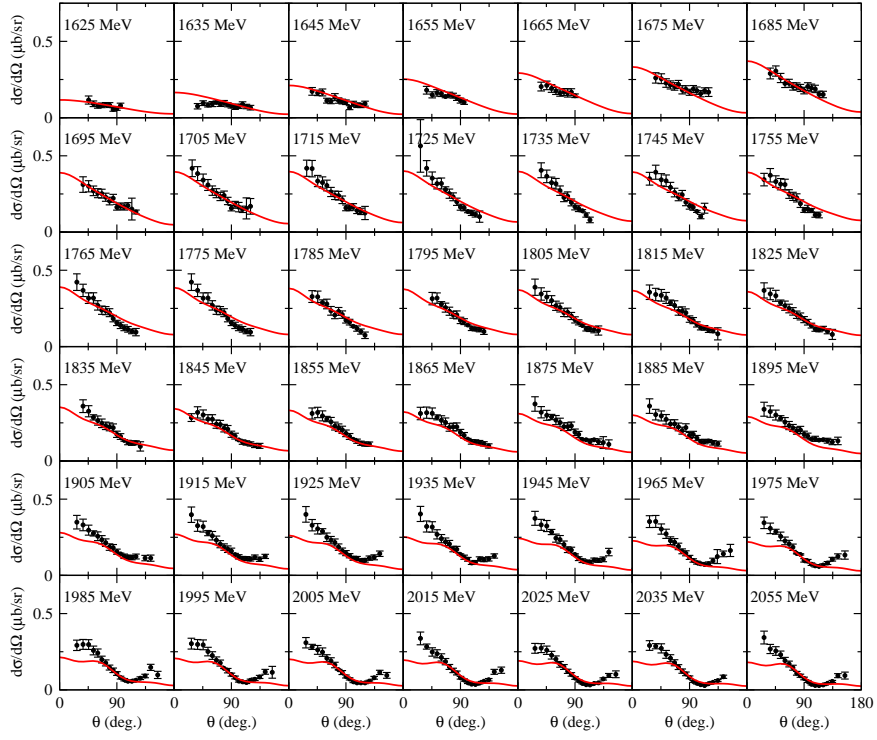}
\caption{(Color online)
$d\sigma/d\Omega$ 
of $\gamma p\to K^+ \Lambda$.}
\label{fig:gp-kl0-dcs}
\end{figure}

\begin{figure}[b]
\includegraphics[clip,width=0.51\textwidth]{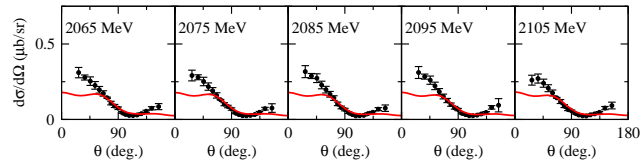}
\caption{(Color online)
$d\sigma/d\Omega$ 
of $\gamma p\to K^+ \Lambda$. (continued)}
\label{fig:gp-kl0-dcs-2}
\end{figure}

Our fits to the  $d\sigma/d\Omega$ data of $\gamma p \to K^+\Lambda$
are  shown in Figs.~\ref{fig:gp-kl0-dcs} and~\ref{fig:gp-kl0-dcs-2}.
The main features of the data up to 
$W = 2106 $ MeV are reproduced reasonably well, in particular in the  
$45^\circ \lesssim \theta \lesssim 135^\circ$ region
where the $s$-channel processes with the isospin $I=1/2$
$N^*$ resonances are found to be important.
However, our results at forward and backward angles underestimate
the data in the  $W \gtrsim 1.9$ GeV high energy region.
This situation is similar to the $\gamma p \to \pi N$ reactions and 
further suggests that the inclusion of additional mechanisms
would be necessary for improving our combined analysis at high energies.

Because of the self-analyzing property of the recoil $\Lambda$ baryon, 
the single and double polarization observables of $\gamma p \to K^+\Lambda$ 
have been extensively measured in recent years.
More data of the polarization observables of $\gamma p \to K^+\Lambda$ 
will soon become available
from JLab and other photon/electron beam facilities.
Actually, all of the 15 possible polarization observables
for this reaction have already been measured at CLAS~\cite{andy-proc}.

\begin{figure}
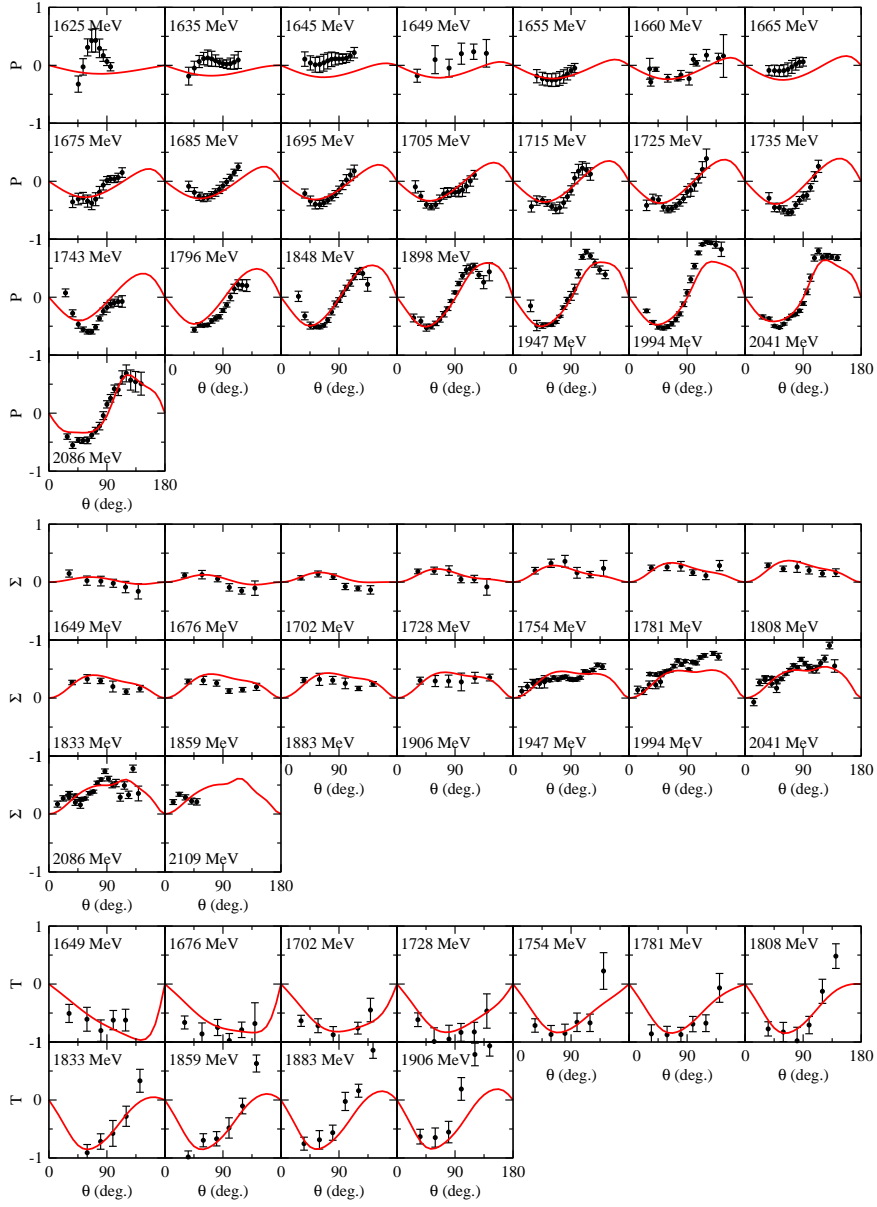

\includegraphics[clip,width=0.7\textwidth]{gpkl-P}
\includegraphics[clip,width=0.7\textwidth]{gpkl-S}
\includegraphics[clip,width=0.7\textwidth]{gpkl-T}
\caption{(Color online)
$P$, $\Sigma$, and $T$ of $\gamma p\to K^+ \Lambda$.}
\label{fig:gp-kl0-p-s-t}
\end{figure}

\begin{figure}
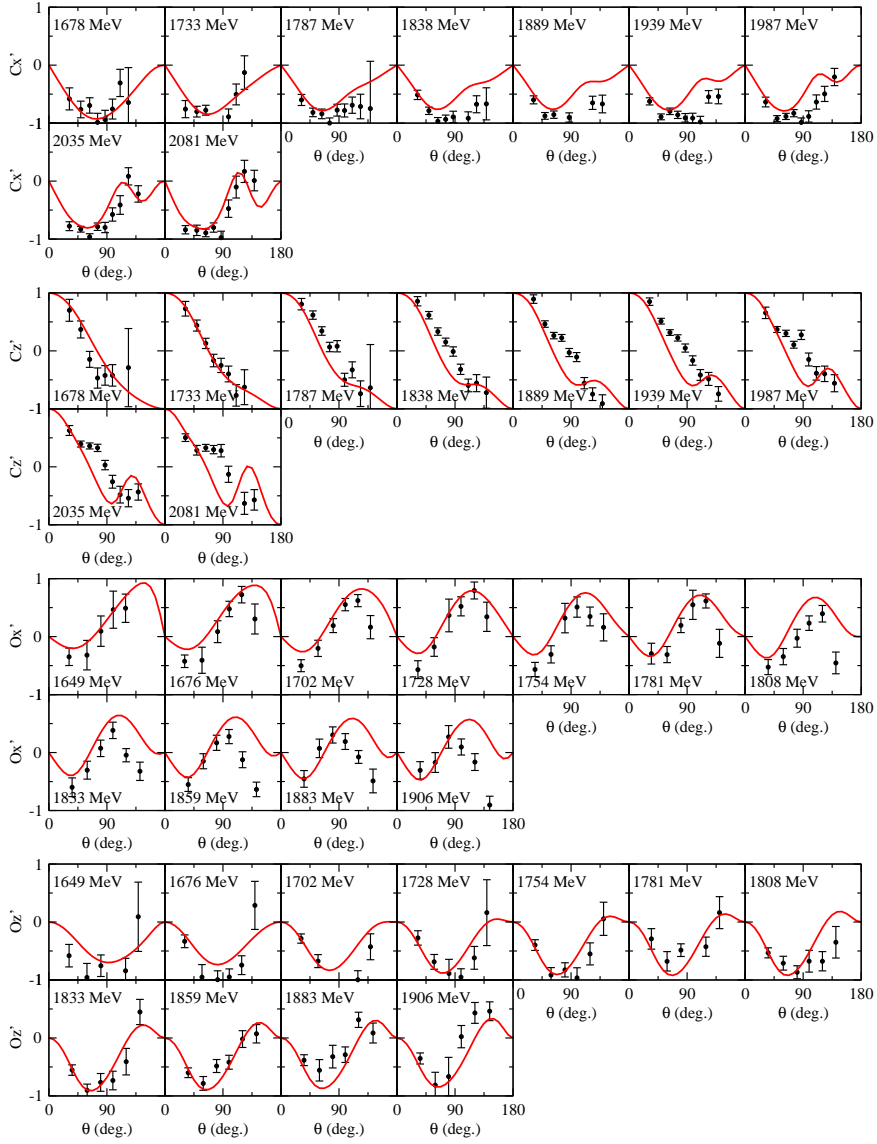

\includegraphics[clip,width=0.7\textwidth]{gpkl-Cx}
\includegraphics[clip,width=0.7\textwidth]{gpkl-Cz}
\includegraphics[clip,width=0.7\textwidth]{gpkl-Ox}
\includegraphics[clip,width=0.7\textwidth]{gpkl-Oz}
\caption{(Color online)
$C_{x'}$, $C_{z'}$, $O_{x'}$, and $O_{z'}$ of $\gamma p\to K^+ \Lambda$.
}
\label{fig:gp-kl0-cx-cz-ox-oz}
\end{figure}

The fits to the available polarization data, as listed in 
Table~\ref{tab:em-data}, are already highly nontrivial 
because they are attributable to delicate interferences between different partial waves.
Nevertheless, as seen in 
Figs.~\ref{fig:gp-kl0-p-s-t} and~\ref{fig:gp-kl0-cx-cz-ox-oz},
our fits can reproduce the main features of these polarization
data to a very large extent. 
Here it is noted that for the double polarizations $C_x$ and $C_z$,
we have plotted them in the \textit{primed} coordinate system
as $C_{x'}$ and $C_{z'}$, by using the relation
\begin{eqnarray}
C_{x'} &=& + C_x \cos\theta - C_z \sin\theta, 
\label{eq:cx-prime} 
\\
C_{z'} &=& + C_x \sin\theta + C_z \cos\theta,
\label{eq:cz-prime}
\end{eqnarray}
as done in the analysis of Ref.~\cite{shkl11}.

\clearpage

\subsection{$\gamma p \to K^+\Sigma^0, K^0\Sigma^+$}

It is more difficult to fit the data of $\gamma p \to K^+\Sigma^0$ and
$\gamma p \to  K^0\Sigma^+$
than that of $\gamma p \to K^+\Lambda^0$  presented in the previous section,
because here the isospin $I=3/2$ $N^*$ resonances also contribute.
In particular, we have found that the  $d\sigma/d\Omega$ data for
both $\gamma p \to K^+\Sigma^0$ and $\gamma p \to K^0\Sigma^+$ can be
fitted well only when the $K\Sigma N$
coupling constant $g_{K\Sigma N}$ is allowed to 
vary far off the SU(3) relation, as given in Eqs.~(\ref{eq:gksn}), in the fits. 
This implies that $\gamma p \to K\Sigma$ reaction could be
an important source of learning about the validity of the SU(3) relation.
This, however, can be done rigorously only when
more complete data for these two photoproduction reactions become available.

\begin{figure}[h]
\includegraphics[clip,width=0.70\textwidth]{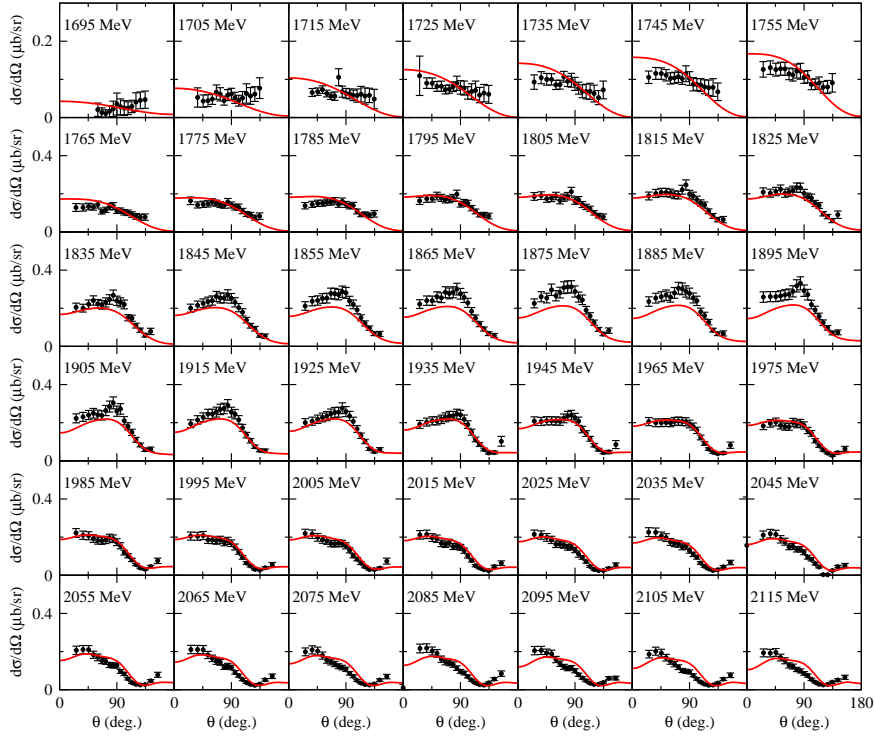}
\caption{(Color online)
$d\sigma/d\Omega$ of $\gamma p\to K^+ \Sigma^0$.}
\label{fig:gpkps0-dcs}
\end{figure}

\begin{figure}[h]
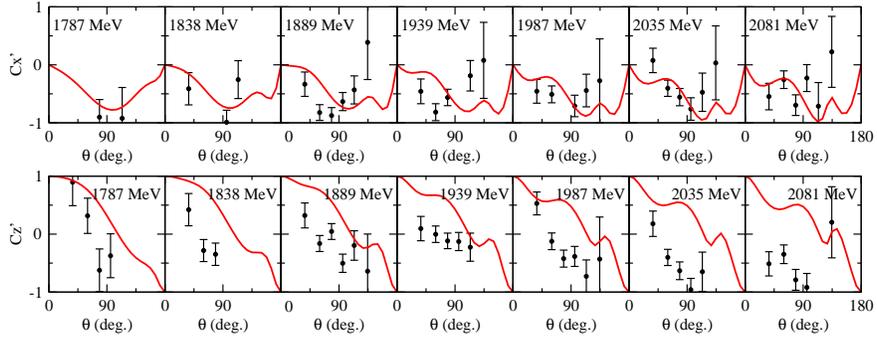

\includegraphics[clip,width=0.70\textwidth]{gpkps0-Cx}
\includegraphics[clip,width=0.70\textwidth]{gpkps0-Cz}
\caption{(Color online)
$C_{x'}$ and $C_{z'}$ of $\gamma p\to K^+ \Sigma^0$.}
\label{fig:gpkps0-cx-cz}
\end{figure}

Our fits to the data of differential cross sections  $d\sigma/d\Omega$ and
polarization observables  $P$, $\Sigma$, $C_{x}$, $C_{z}$
of $\gamma p\to K^+ \Sigma^0$ are shown in 
Figs.~\ref{fig:gpkps0-dcs}-\ref{fig:gpkps0-cx-cz}. 
Again we have plotted $C_{x}$ and $C_{z}$
as $C_{x'}$ and $C_{z'}$ in the \textit{primed} coordinate system
using Eqs.~(\ref{eq:cx-prime}) and~(\ref{eq:cz-prime}).
We see that the fits to the data of differential cross sections 
(Fig.~\ref{fig:gpkps0-dcs}), $P$ and $\Sigma$ (Fig.~\ref{fig:gpkps0-p-s})
are fairly good. However, the fits to the data
of $C_{x}$, $C_{z}$ (Fig.~\ref{fig:gpkps0-cx-cz}) are very qualitative,
in particular in the high-energy region, $W  \gtrsim 1.8$ GeV.
\begin{figure}[h]
\includegraphics[clip,width=0.70\textwidth]{gpkps0-P}
\includegraphics[clip,width=0.70\textwidth]{gpkps0-S}
\caption{(Color online)
$P$ and $\Sigma$ of $\gamma p\to K^+ \Sigma^0$.}
\label{fig:gpkps0-p-s}
\end{figure}

\clearpage

\begin{figure}[h]
\includegraphics[clip,width=0.70\textwidth]{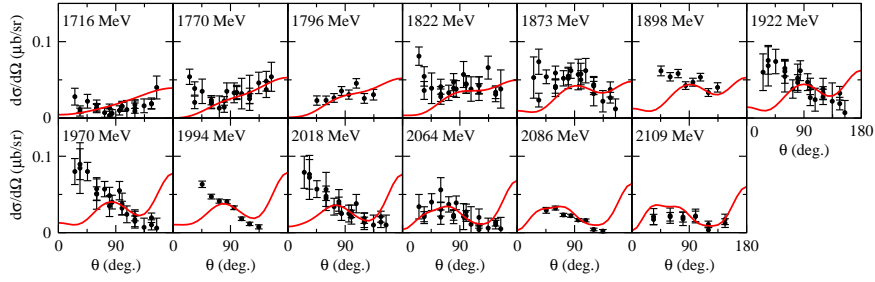}
\caption{(Color online)
$d\sigma/d\Omega$ of $\gamma p\to K^0\Sigma^+$.
}
\label{fig:gpk0sp-dcs}
\end{figure}

\begin{figure}[h]
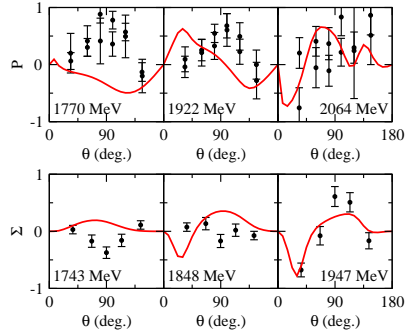

\includegraphics[clip,width=0.32\textwidth]{gpk0sp-P}\\
\includegraphics[clip,width=0.32\textwidth]{gpk0sp-S}
\caption{(Color online)
$P$  and $\Sigma$ of $\gamma p\to K^0\Sigma^+$.
}
\label{fig:gpk0sp-ps}
\end{figure}

We now present our fits to the very limited data of the
$\gamma p\to K^0 \Sigma^+$ reaction in
Figs.~\ref{fig:gpk0sp-dcs} and~\ref{fig:gpk0sp-ps}.
Clearly, our fits to the data of $d\sigma/d\Omega$,
shown in Fig.~\ref{fig:gpk0sp-dcs},
are qualitative only in the $W \gtrsim 1.9$ GeV region.
However, we see in Fig.~\ref{fig:gpk0sp-ps} that
our fits to the data of $P$ and $\Sigma$  are reasonable.

\clearpage

\section{Resonance parameters}
\label{sec:res-para}

Once the model parameters are determined by the simultaneous fits 
to all of the considered data, as presented in Sec.~\ref{sec:results},
we then apply the analytic continuation method of Refs.~\cite{ssl09,ssl10} 
to find the resonance pole positions and their residues.
We expect that the resonances with large widths 
[the widths are defined to be related to the resonance pole position 
$M_R$ by $\Gamma^{\text{tot}}=-2\mathrm{Im}(M_R)$]
have less influence on the physical observables, and thus the extractions of
these resonances from fitting the available reaction data,
as done by all analysis groups, are more model-dependent.
As a result, the extracted information on those resonances is much less reliable.  
We therefore examine only the resonances with the width less than 400 MeV.
Also, as already mentioned in Sec.~\ref{sec:fit-proc}, 
the resonances with $\mathrm{Re}(M_R) > 2$ GeV, found in $F_{17}$, $G$ and $H$ waves, 
are also not presented here.
We will be able to present those high-mass resonances with confidence only after
extending our analysis to higher energy regions and including $\omega N$ 
and $\pi \pi N$ production data.

As defined in Eqs.~(\ref{eq:res-pole})-(\ref{eq:pin-resid}),
we present our results for the resonance pole positions $M_R$
and residues $R_{MB,\pi N}(M_R)$ for $MB= \pi N, \eta N, K\Lambda, K\Sigma$.
Other components of the extracted residues do not correspond to 
the reaction data considered and are not presented. 
The elasticity $\eta_e$ [Eq.~(\ref{eq:elasticity})] and helicity amplitudes 
$A_{\lambda = 1/2, 3/2}$ [Eqs.~(\ref{eq:a32}) and~(\ref{eq:a12})] are also presented.
Our results are given and discussed in the following sections.

\subsection{Resonance pole positions and Residues}

The $N^*$ pole positions $M_R$
extracted from our analysis are presented in Table~\ref{tab:pole}
and compared with those listed by Particle Data Group (PDG).
The $N^*$ states are specified by $J^P(L_{2I2J})$, where $J$ is the spin, 
$P=\pm$ the parity, $I$ the isospin, and  $L$ the orbital angular momentum of
the associated $\pi N$ partial wave.
Here we only list the three- and four-star $N^*$ of PDG because their values are
rather stable in the past two decades and most of their widths are all 
less than 400 MeV that is the limit set in our search. 
We see that we agree well with PDG for the $N^*$ resonances below 2 GeV except
(a) we do not have a third state in $3/2^-(D_{13})$,
(b) we do not have a third state in $3/2^+(P_{33})$, and
(c) we have a second state in $3/2^-(D_{33})$.
In our analysis, a resonance pole with $M_R=(1429,147)$ MeV in $1/2^+(P_{11})$ 
is also found, which turns out to be a ``shadow'' of the first $1/2^+ (P_{11})$ resonance
with respect to the $\pi \Delta$ branch point.
However, this pole is far from the physical region and thus is not listed.
Such a shadow pole is also found in other analyses~(e.g., Ref.~\cite{juelich13-1})
as well as our early analyses~\cite{sjklms10,knls10}.
In Table~\ref{tab:pole}, we also see that the
determined elasticities $\eta_e$ are rather consistent
with the values listed by PDG.
However, some explanation may be required for the result that the elasticity for 
the 1st $3/2^+(P_{33})$ resonance exceeds 100\%.
The elasticity $\eta_e$ is defined by Eq.~(\ref{eq:elasticity}),
where $-2 {\rm Im} (M_R)$ and $2|R_{\pi N, \pi N}|$ are interpreted as 
the total width and the partial width for the decay to the $\pi N$ channel, respectively,
evaluated at \textit{the resonance pole position} in the complex energy plane.
As is well known, the sum of the partial widths defined in this manner does not
agree with the total width defined with the imaginary part of the resonance pole position.

\begin{figure}
\includegraphics[clip,height=0.5\textheight]{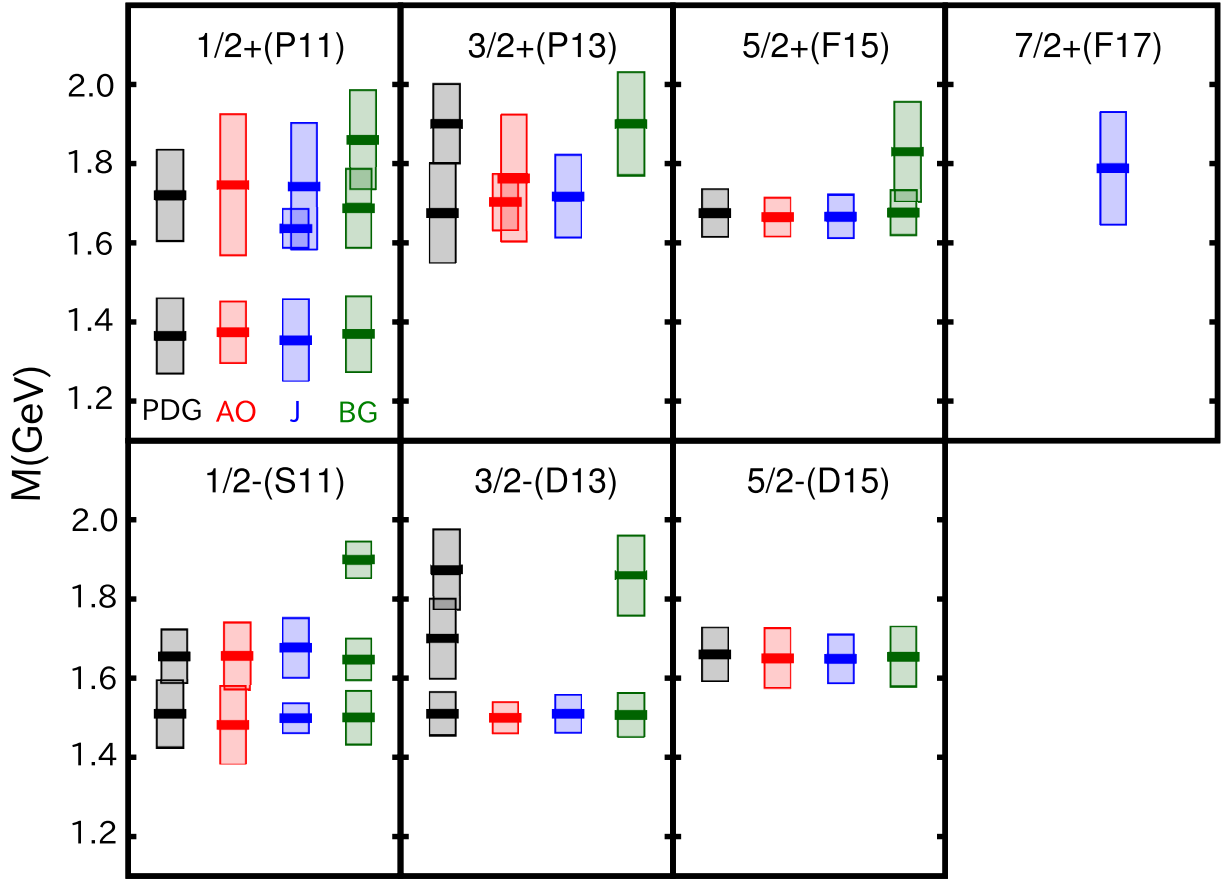}
\caption{(Color online)
$N^*$ spectrum with the isospin $I=1/2$ determined by ANL-Osaka (AO) collaboration.
For each $N^*$ state,
$\mathrm{Re}(M_R)$ together with the $\mathrm{Re}(M_R)\pm\mathrm{Im}(M_R)$ band is plotted.
The results are compared with 4- and 3-star states
listed by the PDG~\cite{pdg} as well as
the results from J\"ulich (J)~(model A in Ref.~\cite{juelich13-1}) 
and Bonn-Gatchina (BG)~\cite{bg2012} groups. 
The spin and parity of states are denoted as $J^P$ with $P=\pm$ and 
the associated $\pi N$ partial wave.
}
\label{fig:spectrum-1}
\end{figure}

\begin{figure}
\includegraphics[clip,height=0.5\textheight]{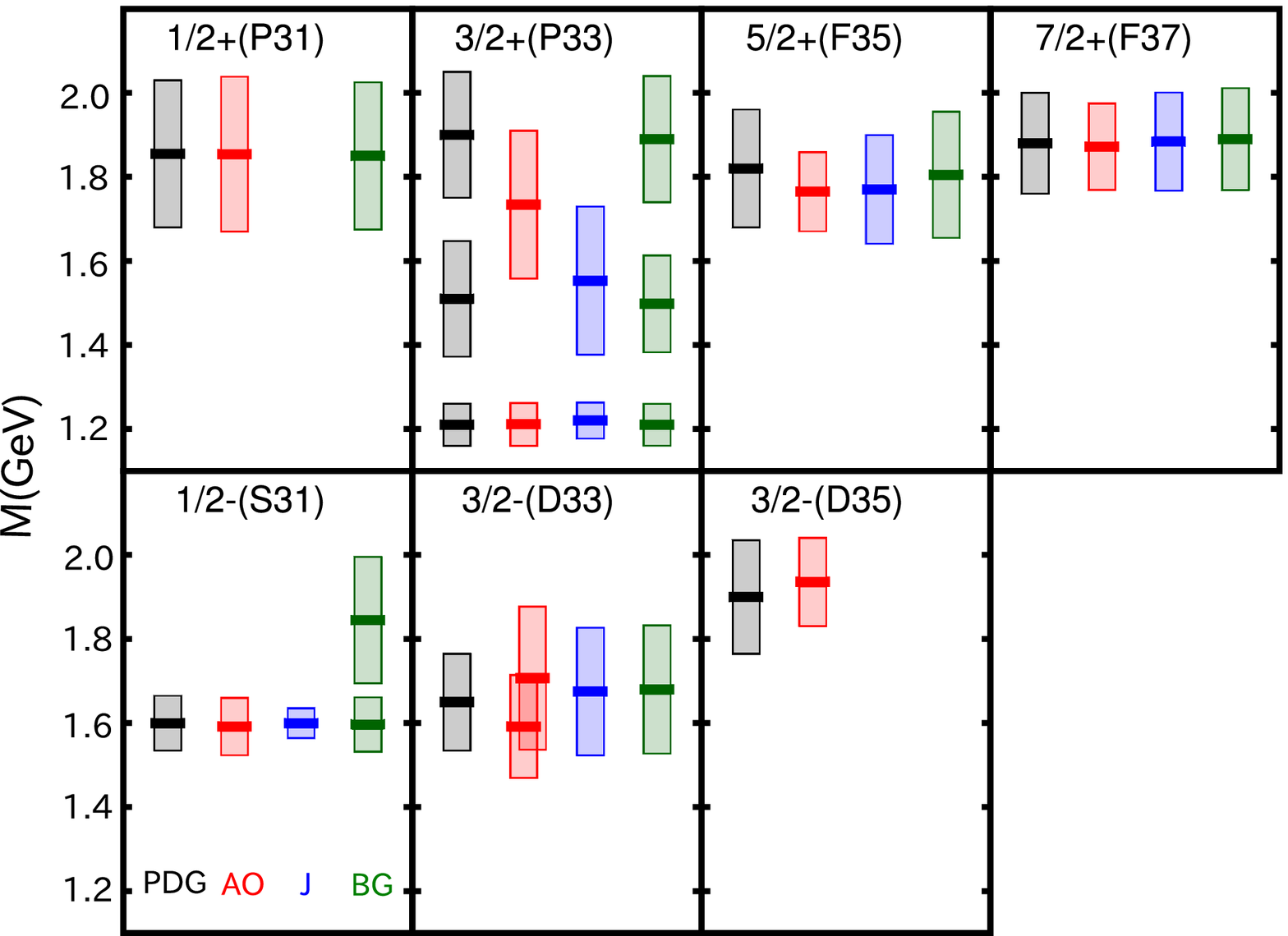}
\caption{(Color online)
$N^*$ spectrum with the isospin $I=3/2$ determined by ANL-Osaka (AO) collaboration.
For each $N^*$ state,
$\mathrm{Re}(M_R)$ together with the $\mathrm{Re}(M_R)\pm\mathrm{Im}(M_R)$ band is plotted.
The results are compared with 4- and 3-star states
listed by the PDG~\cite{pdg} as well as
the results from J\"ulich (J)~(model A in Ref.~\cite{juelich13-1}) 
and Bonn-Gatchina (BG)~\cite{bg2012} groups. 
The spin and parity of states are denoted as $J^P$ with $P=\pm$ and 
the associated $\pi N$ partial wave.
}
\label{fig:spectrum-3}
\end{figure}

In Figs.~\ref{fig:spectrum-1} and~\ref{fig:spectrum-3}, 
we compare our (ANL-Osaka) values (AO, 2nd column) of the $N^*$ spectrum with 
those from PDG (PDG, first column) and the analyses by 
the J\"ulich~\cite{juelich13-1} (J, 3rd column) and the Bonn-Gatchina~\cite{bg2012} 
(BG, 4th column) groups.
In the figures only the resonances with the total widths less than 400 MeV 
are compared.
It is noted, however, that 
some broad $N^*$ resonances with 
$\left(\mathrm{Re}(M_R), -2\mathrm{Im}(M_R)\right)=(1787, 575)$ MeV 
in $1/2^+(P_{31})$, $(1727, 866)$ MeV in $3/2^+(P_{33}) $, and
$(1776, 646)$ MeV in $5/2^-(D_{35}) $
have also been reported by the J\"ulich group,
while
$\left(\mathrm{Re}(M_R), -2\mathrm{Im}(M_R)\right)=(1660, 450)$ MeV 
in $3/2^+(P_{13}) $,
$(1770, 420)$ MeV in $3/2^-(D_{13}) $, and
$(1990, 450)$ MeV in $3/2^-(D_{33}) $ have been reported 
by the Bonn-Gatchina group.

We see in Figs.~\ref{fig:spectrum-1} and~\ref{fig:spectrum-3} that
the first $N^*$ resonances from four results in each $(J^P,I)$ state agree well except:
(a) the $3/2^+(P_{13})$ from the Bonn-Gatchina analysis is about 200 MeV higher than the others, 
(b) the J\"ulich analysis does not have $1/2^+(P_{31})$,
(c) only the J\"ulich analysis has $7/2^+(F_{17})$ below 2 GeV, and
(d) the J\"ulich and Bonn-Gatchina analyses do not have $5/2^-(D_{35})$. 
For higher mass states, the number of states and their positions from four results 
do not agree well. 
Here we mention that the PDG values are mainly from the earlier results of 
$\pi N$ elastic scattering data. 
The photoproduction data of $\gamma p \rightarrow \pi N, \eta N, K\Lambda, K\Sigma$
have not been included in the J\"ulich analysis,
while the data included in the AO and Bonn-Gatchina analyses are not too different.
As discussed in the Introduction, the $K$-matrix model used in the 
Bonn-Gatchina analysis is very different from the DCC model used in the AO 
and J\"ulich analyses.
In the Bonn-Gatchina analysis, like other $K$-matrix model analyses, 
the parameters in different partial waves are varied independently in the fits. 
However, the parameters of meson-exchange mechanisms in 
the AO and J\"ulich analyses can affect all partial waves and channels.
Thus, the Bonn-Gatchina analysis is more flexible and efficient in fitting the data;
in particular, in the area where the data have fluctuating structure with large uncertainties.  

Thus, the differences between four results seen in Figs.~\ref{fig:spectrum-1} 
and~\ref{fig:spectrum-3}
could be attributed to the differences in the employed analysis methods
and the data included in the analysis.
The much more divergent results for the second and third states could also be 
attributed to the fact that
in the  high $ W \gtrsim 1.6$ GeV region,
the available data are far from complete for determining
the partial wave amplitudes model independently. The difficulties
in extracting the partial wave amplitudes model independently
from the data, even they are complete, have been investigated
recently in Ref.~\cite{shkl11}.

In Table~\ref{tab:residue-mb}, we list the extracted residues $R_{MB,\pi N}$
for $MB=\pi N, \eta N, K\Lambda, K\Sigma$. 
In Table~\ref{tab:residues}, we compare the extracted residues $R_{\pi N,\pi N}$  
with those from the Bonn-Gatchina and J\"ulich analyses.
Three analyses agree very well for the well-established $\Delta(1211)~3/2^+ (P_{33})$.
For some states, three results agree qualitatively.
However, the differences between three results can be very large for several states.
This is perhaps attributable to the fact that the residues
are more sensitive to the functional forms of the amplitudes which are 
very different between different analyses.
More detailed investigations of
this issue are needed for advancing the field, but are beyond the scope
of this paper.

\begin{table}
\caption{\label{tab:pole} 
$N^\ast$ pole mass $M_R$ and $\pi N$ elasticity $\eta_e$ extracted in this work.
$M_R$ is listed as $\left(\mathrm{Re}(M_R),-\mathrm{Im}(M_R)\right)$ in the unit
of MeV.
As a reference, we also list the PDG values of the $N^\ast$ states for which 
either 4- or 3-star status is assigned~\cite{pdg}.
The $N^\ast$ states for which the asterisk (*) is marked locate in the complex energy plane
slightly off the closest sheet to the physical real energy axis, yet are still expected
to visibly affect the physical observables. 
}
\begin{ruledtabular}
\begin{tabular}{lcccrr}
&$J^P(L_{2I2J})$  &$M_R$ & $M_R$ (PDG)& $\eta_e$ & $\eta_e$ (PDG)\\
\hline
$N$-baryons&&&&\\
&$1/2^-(S_{11})$&(1482,~~98)$^*$&(1490-1530,~~45-125)      &  64\%&35-55\% \\
&               &(1656,~~85)&(1640-1670,~~50-~85)       &  62\%&50-90\% \\
&$1/2^+(P_{11})$&(1374,~~76)&(1350-1380,~~80-110)      &  48\%&55-75\% \\
&               &(1746,~177)& (1670-1770,~~40-190)      &  11\%&5-20\% \\
&$3/2^+(P_{13})$&(1703,~~70)&(1660-1690,~~75-200)      &  11\%&9-14\% \\
&               &(1763,~159)&(1870-1930,~~70-150)      &  18\%& $\sim$10\%\\
&$3/2^-(D_{13})$&(1501,~~39)&(1505-1515,~~52-~60)       &  67\%&55-65\% \\
&               &(1702,~141)$^*$&(1650-1750,~~50-150)  &  1\%& 7-17\%\\
                     &               &           &(1800-1950,~~75-125)  &     & 2-22\%\\
&$5/2^-(D_{15})$&(1650,~~75)&(1655-1665,~~62-~75)       &  37\%&35-45\% \\
&$5/2^+(F_{15})$&(1665,~~49)&(1665-1680,~~55-~68)       &  69\%&65-70\% \\
$\Delta$-baryons&&&&\\
&$1/2^-(S_{31})$&(1592,~~68)&(1590-1610,~60-~70)       &  29\%&20-30\% \\
&               &(1702,~193)$^*$&   &    10\%     & \\
&$1/2^+(P_{31})$&(1854,~184)&(1830-1880,~100-250)&  12\%& 15-30\%\\
&$3/2^+(P_{33})$&(1211,~~51)&(1209-1211,~~49-~51)       & 105\%&100\% \\
&             &(1734,~176)&(1460-1560,~100-175)     &   5\%&10-25\%\\
&             &           &(1850-1950,~100-200)     &      & 5-20\%\\
&$3/2^-(D_{33})$&(1592,~122)&(1620-1680,~~80-150)      &  15\%&10-20\% \\
&              &(1707,~170)&                             &   7\%& \\
&$5/2^-(D_{35})$&(1936,~105)&(1840-1960,~~88-180)      &   2\%&5-15\% \\
&$5/2^+(F_{35})$&(1765,~~94)&(1805-1835,~132-150)     &  12\%&9-15\% \\
&$7/2^+(F_{37})$&(1872,~103)&(1870-1890,~110-130)     &  45\%&35-45\% \\
\end{tabular}
\end{ruledtabular}
\end{table} 
\begin{table}
\caption{\label{tab:residue-mb} 
Residues for $\pi N\to N^*\to MB$ amplitudes ($R_{MB,\pi N}$) at the $N^\ast$ resonance pole position.
The listed values of $R_{MB,\pi N}$ are in the unit of MeV.
Each resonance is specified by its quantum numbers and the real part of the pole mass
$\mathrm{Re}(M_{R})$.
}
\begin{ruledtabular}
\begin{tabular}{lrrrrrrrr}
Particle $J^P(L_{2I2J})$ & 
\multicolumn{2}{c}{$R_{\pi N,\pi N}$} &
\multicolumn{2}{c}{$R_{\eta N,\pi N}$} &
\multicolumn{2}{c}{$R_{K\Lambda,\pi N}$} &
\multicolumn{2}{c}{$R_{K\Sigma,\pi N}$} \\
& Re & Im & Re & Im & Re & Im & Re & Im \\
\hline
$N(1482)~1/2^-(S_{11})$&$ 45.$&$-43.$&-&-&-&-&-&-\\
$N(1656)~1/2^-(S_{11})$&$ 18.$&$-49.$&$-31.$&$-21.$&$ -2.$&$-10.$&-&-\\
$N(1374)~1/2^+(P_{11})$&$ 13.$&$-34.$&-&-&-&-&-&-\\
$N(1746)~1/2^+(P_{11})$&$ 20.$&$  1.$&$  9.$&$ -9.$&$  4.$&$-10.$&$  2.$&$ 12.$\\
$N(1703)~3/2^+(P_{13})$&$  8.$&$ -0.$&$  1.$&$ -0.$&$  0.$&$ -0.$&$  0.$&$  0.$\\
$N(1763)~3/2^+(P_{13})$&$ -8.$&$-28.$&$ -1.$&$  0.$&$  0.$&$ -8.$&$  1.$&$ -5.$\\
$N(1500)~3/2^-(D_{13})$&$ 25.$&$ -5.$&$ -0.$&$ -1.$&-&-&-&-\\
$N(1702)~3/2^-(D_{13})$&$ -0.$&$  2.$&$ -1.$&$  0.$&$ -1.$&$  1.$&$ -0.$&$ -1.$\\
$N(1650)~5/2^-(D_{15})$&$ 24.$&$-15.$&$ -6.$&$  5.$&$  0.$&$ -0.$&-&-\\
$N(1665)~5/2^+(F_{15})$&$ 32.$&$-12.$&$ -1.$&$  1.$&$  0.$&$ -0.$&-&-\\
\\
$\Delta(1592)~1/2^-(S_{31})$&$ -7.$&$-18.$&-&-&-&-&-&-\\
$\Delta(1702)~1/2^-(S_{31})$&$  8.$&$ 18.$&-&-&-&-&$ 13.$&$ -9.$\\
$\Delta(1854)~1/2^+(P_{31})$&$-12.$&$-19.$&-&-&-&-&$-18.$&$-30.$\\
$\Delta(1211)~3/2^+(P_{33})$&$ 37.$&$-39.$&-&-&-&-&-&-\\
$\Delta(1734)~3/2^+(P_{33})$&$ -4.$&$- 7.$&-&-&-&-&$ -0.$&$ -1.$\\
$\Delta(1592)~3/2^-(D_{33})$&$  8.$&$-16.$&-&-&-&-&-&-\\
$\Delta(1707)~3/2^+(D_{33})$&$  8.$&$  9.$&-&-&-&-&$  2.$&$ -3.$\\
$\Delta(1936)~5/2^-(D_{35})$&$  2.$&$ -1.$&-&-&-&-&$  2.$&$ -3.$\\
$\Delta(1765)~5/2^+(F_{35})$&$  5.$&$-10.$&-&-&-&-&$ -1.$&$ -1.$\\
$\Delta(1872)~7/2^+(F_{37})$&$ 38.$&$-27.$&-&-&-&-&$  0.$&$ -1.$
\end{tabular}
\end{ruledtabular}
\end{table}
\begin{table}
\caption{\label{tab:residues}
Comparison of residue $R_{\pi N,\pi N} = R e^{i\phi}$ of $\pi N\to\pi N$ amplitude
between existing multichannel analyses.
}
\begin{ruledtabular}
\begin{tabular}{lrrrrrr}
Particle $J^P(L_{2I2J})$
& \multicolumn{2}{c}{ANL-Osaka}
& \multicolumn{2}{c}{Bonn-Gatchina~\cite{bg2012}}
& \multicolumn{2}{c}{J\"ulich (model A)~\cite{juelich13-1}}\\
&$R$&$\phi$&$R$&$\phi$&$R$&$\phi$\\\hline
$N(1482)~1/2^-(S_{11})$& $ 63$& $ -44$& $31\pm 4$   &$-(29\pm 5)$  & $16$&$ -36$\\
$N(1656)~1/2^-(S_{11})$& $ 53$& $ -70$& $24\pm 3$   &$-(75\pm 12)$ & $46$&$ -42$\\
$N(1374)~1/2^+(P_{11})$& $ 37$& $ -69$& $48\pm 3$   &$-(78\pm 4)$  & $58$&$-104$\\
$N(1746)~1/2^+(P_{11})$& $ 20$& $   3$& $6\pm  4$   &$(120\pm 70)$ & $ 4$&$ -30$\\
$N(1703)~3/2^+(P_{13})$& $  8$& $  -3$& $22\pm 8$   &$-(115\pm 30)$& $7$ &$-73$\\
$N(1763)~3/2^+(P_{13})$& $ 29$& $-106$& -            & -           &-    &-\\
$N(1500)~3/2^-(D_{13})$& $ 26$& $ -11$& $36\pm 3$   &$-(14\pm 3)$  & $32$&$-11$\\
$N(1702)~3/2^-(D_{13})$& $  2$& $ 104$& $50\pm 40$  &$-(100\pm 40)$&-    &-    \\
$N(1650)~5/2^-(D_{15})$& $ 28$& $ -31$& $28\pm 1$   &$-(26\pm 4)$  & $24$&$-19$\\
$N(1665)~5/2^+(F_{15})$& $ 34$& $ -20$& $43\pm 4$   &$-(2\pm 10)$  & $36$&$-24$\\
\\
$\Delta(1592)~1/2^-(S_{31})$& $ 20$& $-111$& $18\pm 2$   &$-(100\pm 5)$  & $17$&$-106$\\
$\Delta(1702)~1/2^-(S_{31})$& $ 19$& $  65$& $10\pm 3$   &$-(125\pm 20)$ &-    &-     \\
$\Delta(1854)~1/2^+(P_{31})$& $ 23$& $-123$& $24\pm 6$   &$-(145\pm 30)$ & $54$&$-140$\\
$\Delta(1211)~3/2^+(P_{33})$& $ 53$& $ -47$& $51.6\pm 0.6$&$-(46\pm 1)$   & $44$&$-35$\\
$\Delta(1734)~3/2^+(P_{33})$& $  8$& $-118$& $11\pm 6$   &$-(160\pm 33)$ & $20$&$-158$\\
$\Delta(1592)~3/2^-(D_{33})$& $ 18$& $ -62$&  -           &-             & -   &-\\
$\Delta(1707)~3/2^-(D_{33})$& $ 11$& $  49$&  $42\pm 7$   &$-(3\pm 15)$   & $24$&$-9$\\
$\Delta(1936)~5/2^-(D_{35})$& $  2$& $ -32$&  -           &-             & $18$&$-159$\\
$\Delta(1765)~5/2^+(F_{35})$& $ 11$& $ -62$&  $20\pm 2$   &$-(44\pm 5)$   & $17$&$-59$\\
$\Delta(1872)~7/2^+(F_{37})$& $ 46$& $ -35$&  $58\pm 2$   &$-(24\pm 3)$   & $58$&$-25$\\
\end{tabular}
\end{ruledtabular}
\end{table}

\subsection{$\gamma N\to N^*$ helicity amplitudes}

To determine the  $\gamma N \to N^*$ helicity amplitude,
we first extract the residue $R_{\pi N,\gamma N}(M_R)$
of the $\gamma N \to \pi N$ amplitude at the resonance pole position $M_R$.
By using Eq.~(\ref{eq:res-r}) of the factorized form of $R_{M'B',MB}(M_R)$,
the dressed vertex $\bar{\Gamma}_{\gamma N}(M_R)$ can be determined
from the extracted $R_{\pi N,\gamma N}(M_R)$ and
the $R_{\pi N, \pi N}(M_R)$ listed in Table~\ref{tab:residue-mb}. 
The $\gamma N \to N^*$ helicity amplitudes can then be
calculated from the resulting $\bar{\Gamma}_{\gamma N}$ by
using Eqs.~(\ref{eq:a32})-(\ref{eq:coef-c}).

Our results are listed in Table~\ref{tab:helicity}
and are compared with those extracted by the Bonn-Gatchina group.
We see that two results agree very well for the $\Delta(1211)$. 
For the $N^\ast$ resonances with $L \leq 2$ and $\mathrm{Re}(M_R) < 1.7$ GeV,
$N(1482)~1/2^-(S_{11})$, $N(1656)~1/2^-(S_{11})$, $N(1374)~1/2^+(P_{11})$,
$N(1500)~3/2^-(D_{13})$, $N(1650)~5/2^-(D_{15})$, 
$\Delta(1592)~1/2^-(S_{31}) $, and $\Delta(1872)~7/2^+(F_{37}) $,
some qualitative agreements between the two analyses can be seen.
However, it is difficult to compare the results for
other resonance states. 
Here we note that the phase of $\gamma N \to N^*$ chosen 
by different analysis groups can be different, as discussed in Ref.~\cite{inna}.
This adds other complications in comparing the results listed in Table~\ref{tab:helicity}.
Our choice of the phase has been given below Eq.~(\ref{eq:pin-resid}).
We do not compare our results with those from Refs.~\cite{maid07,gwu-said},
because their results are from the Breit-Wigner parametrization which 
cannot be related model independently to the residues 
at resonance pole, as discussed in Ref.~\cite{ssl10}.
\begin{table}
\caption{\label{tab:helicity}
Helicity amplitudes for $\gamma p\to N^*$. 
The values are presented in the unit of $10^{-3}\ {\rm GeV}^{-1/2}$.
As a comparison, we also list those from Bonn-Gatchina group~\cite{bg2012}.
}
\begin{ruledtabular}
\begin{tabular}{lrrrrrrrr}
Particle $J^P(L_{2I2J})$
& \multicolumn{4}{c}{ANL-Osaka}& \multicolumn{4}{c}{Bonn-Gatchina~\cite{bg2012}}\\
\cline{2-9}
& \multicolumn{2}{c}{$A_{3/2}$} & \multicolumn{2}{c}{$A_{1/2}$} & \multicolumn{2}{c}{$A_{3/2}$} & \multicolumn{2}{c}{$A_{1/2}$}\\
&Re&Im&Re&Im&Re&Im&Re&Im\\
\hline
$N(1482)~1/2^-(S_{11})$&-      &-      &$ 159.$&$  24.$&-       &-       &$ 115.1$&$  14.1$\\
$N(1656)~1/2^-(S_{11})$&-      &-      &$  29.$&$ -28.$&-       &-       &$  32.6$&$  -5.2$\\
$N(1374)~1/2^+(P_{11})$&-      &-      &$  49.$&$ -10.$&-       &-       &$ -34.7$&$  27.1$\\
$N(1746)~1/2^+(P_{11})$&-      &-      &$ -24.$&$  83.$&-       &-       &$  54.2$&$  -9.6$\\
$N(1703)~3/2^+(P_{13})$&$ -70.$&$   8.$&$ 234.$&$   8.$&$  63.4$&$ 135.9$&$ 110.0$&$   0.0$\\
$N(1763)~3/2^+(P_{13})$&$ -44.$&$   1.$&$ 126.$&$ -72.$&-       &-       &-       &-       \\
$N(1500)~3/2^-(D_{13})$&$ -93.$&$ -11.$&$  38.$&$   2.$&$ 131.9$&$   4.6$&$ -21.0$&$   0.0$\\
$N(1702)~3/2^-(D_{13})$&$ -40.$&$  36.$&$ -11.$&$ -23.$&$  -37.$&$   0.0$&$   3.8$&$  43.8$\\
$N(1650)~5/2^-(D_{15})$&$  30.$&$ -13.$&$   5.$&$  -2.$&$  24.6$&$  -8.5$&$  23.1$&$  -6.6$\\
$N(1665)~5/2^+(F_{15})$&$ -38.$&$  -2.$&$  53.$&$  -5.$&$ 133.9$&$  -4.7$&$ -11.8$&$   5.5$\\
\\
$\Delta(1592)~1/2^-(S_{31})$&-      &-      &$ 113.$&$  -2.$&-       &-       &$  51.4$&$  -8.1$\\
$\Delta(1702)~1/2^-(S_{31})$&-      &-      &$  35.$&$   3.$&-       &-       &$  29.5$&$  51.1$\\
$\Delta(1854)~1/2^+(P_{31})$&-      &-      &$ -51.$&$   9.$&-       &-       &$  17.6$&$  14.8$\\
$\Delta(1211)~3/2^+(P_{33})$&$-257.$&$  12.$&$-129.$&$  34.$&$-250.9$&$  39.7$&$-123.9$&$  42.6$\\
$\Delta(1734)~3/2^+(P_{33})$&$ -18.$&$-135.$&$ -23.$&$ -68.$&$ -39.6$&$  10.6$&$ -34.1$&$  40.6$\\
$\Delta(1592)~3/2^-(D_{33})$&$ -89.$&$ -76.$&$-123.$&$ -38.$&-       &-       &-       &-       \\
$\Delta(1707)~3/2^-(D_{33})$&$  32.$&$-121.$&$  20.$&$ -56.$&$ 120.2$&$ 120.2$&$ 109.3$&$ 130.2$\\
$\Delta(1936)~5/2^-(D_{35})$&$  34.$&$  -9.$&$  50.$&$ -19.$&-       &-       &-       &-       \\
$\Delta(1765)~5/2^+(F_{35})$&$   0.$&$ -18.$&$  -1.$&$  -8.$&$ -50.0$&$   0.0$&$  23.0$&$  -9.8$\\
$\Delta(1872)~7/2^+(F_{37})$&$ -76.$&$  -2.$&$ -61.$&$  10.$&$ -95.3$&$  11.7$&$ -71.5$&$   8.8$
\end{tabular}
\end{ruledtabular}
\end{table}

\section{Summary and Future Developments}
\label{sec:summary}

We have extended the DCC model developed in
Refs.~\cite{msl07,jlms07,jlmss08,jklmss09,djlss08,kjlms09-1,kjlms09-2,ssl09,sjklms10,ssl10,knls10}
to include the $K\Lambda$ and $K\Sigma$ channels and have completed a combined analysis of
the data of $\pi N, \gamma N \rightarrow \pi N, \eta N, K\Lambda, K\Sigma$ reactions.
The pole positions and residues of nucleon resonances with masses below 2 GeV and total
widths less than 400 MeV have been extracted.
From the extracted residues, we have determined 
the $N^* \to \gamma N, \pi N, \eta N, K\Lambda, K\Sigma$ transition
amplitudes at the resonance positions. 
How this information can be related to the results from the hadron models and lattice QCD 
calculations is an important challenge 
in advancing our understanding of the structure of the nucleon and
its excited states. 
While some progress in this direction has been made~\cite{sl,dse13}
for the $\Delta(1232)\to \gamma N$ transitions, much more work is needed.

The  $N^*$ masses extracted from our analysis, PDG, J\"ulich analysis, 
and Bonn-Gatchina analysis agree well only in the low-mass region. 
In the higher mass region the differences among four results are rather large.
This could be mainly attributed to the fact that in the high $ W \gtrsim 1.7$ GeV region,
the available data of $\pi N$ and $\gamma N$ reactions are far from 
complete to determine the partial-wave-amplitudes model independently.  
The difficulties in extracting the partial-wave amplitudes model-independently
from the data, even when they are complete, have been investigated 
in Refs.~\cite{tabakin,shkl11,cmu-b,maid-said}. 
Of course, the differences in the analysis
methods and the data included in each analysis could also lead to
large disagreements, as discussed in Sec.~\ref{sec:res-para}. 
It is necessary to clarify the situation.

The extracted residues of $\pi N \to \pi N$ partial-wave amplitudes 
are compared with Bonn-Gatchina analysis and J\"ulich analysis.
Except for the well-established $\Delta(1232)$ resonance, the three analyses agree only 
qualitatively even for the cases that their pole positions are close.
A similar situation is also found in comparing the extracted $N^* \to
\gamma N$ helicity amplitudes with  those from the Bonn-Gatchina analysis.
This is perhaps attributable to the fact that the residues
are more sensitive to the model or parametrization used in the analyses.
It is necessary to clarify this issue for advancing the field.

To improve our analysis in the higher mass region, we need to include 
the data of $\pi N, \gamma N \to \pi\pi N$ reactions that dominate 
the $\pi N$ and $\gamma N$ reaction cross sections at $W \gtrsim 1.6$ GeV. 
To proceed, we need detailed data for these two processes.
For $\gamma N \to \pi \pi N$, the precise data of invariant mass distributions 
and some polarization observables 
are becoming available from facilities such as JLab, Mainz, Bonn, SPring-8, and ELPH at Tohoku University.
For $\pi N \to \pi \pi N$, however,
what is available is only the old data of the total cross sections with rather 
large uncertainties and very limited invariant mass distributions 
in the energy region above $W = 1.6$ GeV~\cite{ksu,kjlms09-1}.
The situation can be improved greatly when the data from the  new
experiments on $\pi N\rightarrow \pi\pi N$ reactions at J-PARC~\cite{jparc-p45} 
become available in the near future.
To pin down the reaction mechanisms associated 
with the $\pi\Delta$, $\rho N$, and $\sigma N$ channels, 
the Dalitz plot data of the $\pi\pi N$ distributions
will be desirable.
The importance of fitting the Dalitz plot data has been illustrated recently
in Refs.~\cite{knls11,knls12} for the three-pion decays of heavy mesons.  
We also need to extend the analysis to include the $\omega N$ channel which 
has significant contributions to the $\pi N$ and $\gamma N$ reaction cross sections 
at $W \gtrsim 1.7$ GeV. 
It will be useful if more extensive data of $\pi N \rightarrow \omega N$
can also be obtained at J-PARC.

Our DCC model developed in this work can be readily applied
to the electron- and neutrino-induced meson production reactions.
The application to the electroproduction reactions, which
corresponds to the extension of our early analysis of $p(e,e'\pi)N$~\cite{jklmss09},
is crucial for determining the $Q^2$ dependence of $N$-$N^*$ electromagnetic
transition form factors. 
This analysis is a key to understanding the quark-gluon substructure
of the $N^\ast$ states~\cite{lee-review} and will be closely related to the $N^\ast$
program at JLab after the 12-GeV upgrade~\cite{emnn}.
Also, precise knowledge of the neutrino-nucleon/nucleus reactions in the GeV-energy
region is expected to be very important for determining 
the leptonic $CP$-phase and neutrino-mass hierarchy from the neutrino-oscillation 
measurements through the accelerator and atmospheric experiments~\cite{neuint}.
By following the procedure of Refs.~\cite{sul03,msl05}, the DCC model presented in this work
can be extended straightforwardly to describe the neutrino-induced reactions
in the nucleon resonance region.
A first attempt to study the neutrino-induced reactions within the current
DCC model has been done in Ref.~\cite{knls-neutrino}. 
A further extensive study of the neutrino-induced reactions is ongoing
and will be presented elsewhere.

\begin{acknowledgments}
The authors thank  J.~Durand, B.~Juli\'a-D\'iaz, A.~Matsuyama, B.~Saghai, L.~C.~Smith,
N.~Suzuki, and K. Tsushima for their collaborations at EBAC, and
would also like to thank A.~W.~Thomas for his strong support
and his many constructive discussions.
We thank the support from JLab to the development of the first stage
of this analysis as reported in Ref.~\cite{kl12}.
This work was supported by the JSPS KAKENHI Grant 
No. 25800149 (H.K.) and No. 24540273 (T.S.),
and by the U.S. Department of Energy, Office of Nuclear Physics Division,
under Contract No. DE-AC02-06CH11357.
H.K. acknowledges the support of the HPCI Strategic Program
(Field 5 ``The Origin of Matter and the Universe'') of
Ministry of Education, Culture, Sports, Science and Technology (MEXT) of Japan.
S.X.N. is the Yukawa Fellow and his work is supported in part by Yukawa Memorial Foundation,
the Yukawa International Program for Quark-hadron Sciences (YIPQS),
and by Grants-in-Aid for the global COE program
``The Next Generation of Physics, Spun from Universality and Emergence'' from MEXT.
This research used resources of the National Energy Research Scientific Computing Center,
which is supported by the Office of Science of the U.S. Department of Energy
under Contract No. DE-AC02-05CH11231, and resources provided on ``Fusion,''
a 320-node computing cluster operated by the Laboratory Computing Resource Center
at Argonne National Laboratory. 
\end{acknowledgments}

\appendix

\section{Self-energies in meson-baryon Green functions}
\label{app:self}

In this appendix, we give an explicit expression of the self-energy $\Sigma_{MB}(k;W)$
appearing in the meson-baryon Green's function [Eq.~(\ref{eq:prop-unstab})] for 
the unstable channels $MB=\pi\Delta, \rho N,\sigma N$.

As for the $\pi\Delta$ and $\rho N$ channels, the self-energies are explicitly given by
\begin{eqnarray}
\Sigma_{\pi \Delta}(k;W) &=& \frac{m_\Delta}{E_\Delta(k)}
\int_{C_3} q^2 dq \frac{ M_{\pi N}(q)}{[M^2_{\pi N}(q) + k^2]^{1/2}}
\frac{\left|f_{\Delta \to \pi N}(q)\right|^2}
{W-E_\pi(k) -[M^2_{\pi N}(q) + k^2]^{1/2} + i\epsilon},
\label{eq:self-pidelta}
\\
\Sigma_{\rho N}(k;W) &=& \frac{m_\rho}{E_\rho(k)}
\int_{C_3} q^2 dq \frac{ M_{\pi\pi}(q)}{[M^2_{\pi \pi}(q) + k^2]^{1/2}}
\frac{\left|f_{\rho \to \pi \pi}(q)\right|^2}
{W-E_N(k) -[M^2_{\pi \pi}(q) + k^2]^{1/2} + i\epsilon},
\label{eq:self-rhon}
\end{eqnarray}
where $m_\Delta=1280$ MeV, $m_\rho = 812$ MeV, $M_{\pi N}(q) = E_\pi (q) + E_N(q)$, 
and $M_{\pi \pi}(q) = E_\pi (q) + E_\pi(q)$.
The form factors $f_{\Delta\to\pi N}(q)$ and $f_{\rho \to \pi \pi}(q)$ are for describing
the $\Delta\to\pi N$ and $\rho \to \pi \pi$ decays in the $\Delta$ and $\rho$ rest frames,
respectively.
Those are parametrized as~\cite{betz-lee,jlms07}
\begin{eqnarray}
f_{\Delta \to \pi N}(q) &=& -i\frac{(0.98)}{[2(m_N+m_\pi)]^{1/2}}
\left(\frac{q}{m_\pi}\right)
\left(\frac{1}{1+[q/(358~\text{MeV})]^2}\right)^2 ,
\\
f_{\rho \to \pi \pi}(q) &=& \frac{(0.6684)}{\sqrt{m_\pi}}
\left(\frac{q}{(461~\text{MeV})}\right)
\left(\frac{1}{1+[q/(461~\text{MeV})]^2}\right)^2 .
\end{eqnarray}

To construct the $\sigma$ self-energy in the $\sigma N$ Green's function, $\Sigma_{\sigma N}(k;W)$, 
we first consider a $\pi \pi$ scattering model for the isospin $I=0$ and $s$ wave,
described by the following separable potential
in the $\pi\pi$ center-of-mass system:
\begin{equation}
v(p',p;E)=
g(p')\frac{1}{E-m_\sigma+i\varepsilon}g(p) +h_0h(p')h(p).
\label{eq:pipi-pot}
\end{equation}
Here $p$ and $p'$ are the magnitude of the initial and final momentum.
$E$ is the total scattering energy in the $\pi\pi$ system, 
and the form factors are parametrized as
\begin{eqnarray}
g(p)&=&\frac{g_0}{\sqrt{m_\pi}}\frac{1}{1+(cp)^2},
\\
h(p)&=&\frac{1}{m_\pi}\frac{1}{1+(dp)^2}.
\label{eq:pipi-pot2}
\end{eqnarray}
With this potential, one can solve
the Lippmann-Schwinger equation to obtain the $\pi\pi$ scattering amplitude $T$.
The solution is expressed as
\begin{equation}
T=t+t^R,
\end{equation}
where $t$ is the ``non-resonant'' part of the amplitude given by
\begin{eqnarray}
t(p',p;E)&=&h(p')\tau(E)h(p),
\\
\tau(E) &=& \frac{h_0}{1-h_0\langle hG_{\pi\pi}h\rangle(E)},
\\
\langle hG_{\pi\pi}h\rangle(E) &=& 
\int dq q^2h(q)\frac{1}{E-M_{\pi\pi}(q)+i\varepsilon}h(q),
\end{eqnarray}
while $t_R$ the ``resonant'' part is given by
\begin{eqnarray}
t^R(p',p;E)&=&\frac{\bar\Gamma_{\sigma\to\pi\pi}(p';E)\bar\Gamma_{\pi\pi\to\sigma}(p;E)}
{E-m_\sigma - \Sigma_\sigma (E)}.
\end{eqnarray}
The $\sigma$ self-energy $\Sigma_\sigma(E)$ and the dressed $\sigma\to\pi\pi$
vertex $\bar \Gamma_{\sigma\to\pi\pi}(p';E)$ are then given by
\begin{eqnarray}
\Sigma_\sigma(E) &=& \langle gG_{\pi\pi}g\rangle(E) +\tau(E)[\langle gG_{\pi\pi}h\rangle(E)]^2,
\\
\bar \Gamma_{\sigma\to\pi\pi}(p';E) &=& g(p') + 
h(p') \tau(E) \langle hG_{\pi\pi}g\rangle (E),
\end{eqnarray}
and $\bar \Gamma_{\pi\pi\to\sigma}(p;E) = \bar \Gamma_{\sigma\to\pi\pi}(p;E)$.
The parameters in 
Eqs.~(\ref{eq:pipi-pot})-(\ref{eq:pipi-pot2}) are determined by fitting
the phase shift of the $I=0$ and $s$ wave $\pi\pi$ scattering up to $E=1$ GeV,
and the resulting values are $m^0_\sigma = 700.0$ MeV, 
$g_0 = 1.638$, $h_0 = 0.556$, $c=1.02$ fm, and $d=0.514$ fm.

With the above $\pi\pi$ model, the $\sigma$ self-energy in the $\sigma N$ Green's function, 
$\Sigma_{\sigma N}(k;W)$ in Eq.~(\ref{eq:prop-unstab}), is obtained by 
\begin{eqnarray}
\Sigma_{\sigma N}(k;W) &=& \langle gG_{\pi\pi}g\rangle(k;W) 
+\tau(k;W)[\langle gG_{\pi\pi}h\rangle(k;W)]^2,
\end{eqnarray}
with
\begin{eqnarray}
\tau(k;W) &=& \frac{h_0}{1-h_0\langle h G_{\pi\pi} h\rangle (k;W)},
\\
\langle hG_{\pi\pi}h\rangle(k;W) &=&
\int_{C_3} dq q^2 \frac{M_{\pi\pi}(q)}{ [M_{\pi\pi}^2(q) + k^2]^{1/2}} 
\nonumber\\
&& \qquad\qquad\qquad \times
\frac{h(q)^2}{W - E_N(k) - [M_{\pi\pi}^2(q) + k^2]^{1/2}+i\varepsilon},
\\
\langle gG_{\pi\pi}g\rangle(k;W) &=& 
\frac{m_\sigma}{E_\sigma(k)}
\int_{C_3} dq q^2 \frac{M_{\pi\pi}(q)}{ [M_{\pi\pi}^2(q) + k^2]^{1/2}} 
\nonumber\\
&& \qquad\qquad\qquad \times
\frac{g(q)^2}{W - E_N(k) - [M_{\pi\pi}^2(q) + k^2]^{1/2}+i\varepsilon},
\\
\langle gG_{\pi\pi}h\rangle(k;W) &=&
\sqrt{\frac{m_\sigma}{E_\sigma(k)}}
\int_{C_3} dq q^2 \frac{M_{\pi\pi}(q)}{ [M_{\pi\pi}^2(q) + k^2]^{1/2}} 
\nonumber\\
&& \qquad\qquad\qquad \times
\frac{g(q)h(q)}{W - E_N(k) - [M_{\pi\pi}^2(q) + k^2]^{1/2}+i\varepsilon}.
\end{eqnarray}
The momentum integral path $C_3$ is appropriately deformed 
when we perform the analytic continuation of the scattering amplitudes.
With the self-energies defined above, the branch points of the $\pi\Delta$,
$\rho N$, and $\sigma N$ Green's functions are those as listed 
in Table~\ref{tab:mbgreen-branch}.
Although two branch points are found in the $\sigma N$ Green's function,
the second one at $(1032.3-i247.7) + m_N$ MeV hardly affects
the resonance properties shown in this work because of the large imaginary
part.

\section{Model Lagrangian}
\label{app:lag}

In this appendix, we present a set of Lagrangians for deriving the meson-exchange
potentials $v_{M'B',MB}$ and $v_{M'B',\gamma N}$.
The details of the $Z^{(E)}_{M'B',MB}(k',k;W)$ term in Eq.~(\ref{eq:veff-mbmb})
can be found in Ref.~\cite{msl07} and is not shown here.

It is necessary to define notations associated with isospin quantum numbers.
For the isospin $I=1/2$ hadrons, we use the following field operators:
\begin{eqnarray}
N & = & 
\left( \begin{array}{c}
                 p\\
                 n
                \end{array}
                 \right)\ , \\
\Xi & = & 
\left( \begin{array}{c}
                 \Xi^-\\
                 \Xi^0
                \end{array}
                 \right)\ , \\
K & = & 
\left( \begin{array}{c}
                 K^+\\
                 K^0
                \end{array}
                 \right)\ , \\
K_c & = & 
\left( \begin{array}{c}
                 \bar{K}^0\\
                - K^-
                \end{array}
                 \right)\ . 
\end{eqnarray}
For the isovector ($I=1$) pion field operator, we use the usual notations
\begin{eqnarray}
\vec\pi &=& (\pi^1, \pi^2, \pi^3)\,,
\end{eqnarray}
with the isospin triplet
\begin{eqnarray}
\pi=\left( \begin{array}{c}
                 \pi^+\\
                 \pi^0\\
                 \pi^-
                \end{array}
                 \right)\ ,
\end{eqnarray}
where
\begin{eqnarray}
\pi^{\pm} &=& \mp\frac{1}{\sqrt{2}}(\pi^1\pm i \pi^2) \,,
\\
\pi^0 &=& \pi^3 \,.
\end{eqnarray}
The same definition is also applied to other $I=1$ hadron operators.
For $\Delta$ with $I=3/2$, we define
\begin{eqnarray}
\Delta & = &
\left( \begin{array}{c}
                 \Delta^{++}\\
                 \Delta^{+}\\
                 \Delta^{0}\\
                 \Delta^{-}
                \end{array}
                 \right)\,.
\end{eqnarray}

For the doublet states, the isospin operator $\vec{\tau}$, 
with the spherical components
$\tau^{\pm}=\mp(\tau_1\pm i \tau_2)/\sqrt{2}$ and $\tau_0=\tau_3$,
is defined by the matrix elements
\begin{eqnarray}
\bra{s\, m_s} \tau_m \ket{s\, m'_s}
&=&
\inp{s\,1\, m'_s\, m}{s\, m_s}
\bra{s|} \tau \ket{|s} /\sqrt{2s+1} ,
\end{eqnarray}
where 
$\inp{j_1\,j_2\,m_1\,m_2}{J\,M}$ is the usual Clebsch-Gordon coefficient,
$m=\pm 1,0$, and the reduced matrix element is
$\bra{\frac{1}{2}|} \tau \ket{|\frac{1}{2}} =\sqrt{6}$. 
The isospin operators $\vec{T}_\Delta$ for the $\Delta \to \Delta$ and
$\vec{T}$ for the $N \to \Delta$ transitions are defined by 
\begin{eqnarray}
\bra{j_\Delta m_s} T_{\Delta, m} \ket{j_\Delta m'_s} 
&=&
\inp{j_\Delta 1 m'_s m}{j_\Delta  m_s}
\bra{j_\Delta|} T_\Delta \ket{|j_\Delta} / \sqrt{2j_\Delta+1} ,
\\
\bra{j m_s} T_m \ket{j' m'_s}
&=&
\inp{j' 1 m'_s m}{jm_s}
\bra{j|} T \ket{|j'} / \sqrt{2j+1} ,
\end{eqnarray}
where $m=\pm 1,0$ is the spherical component of 
$\vec{T}_\Delta$ and $\vec{T}$, $j_\Delta=3/2$,
$j,j' = 1/2$ or $3/2$, and the reduced matrix elements are
\begin{eqnarray}
\bra{{\textstyle\frac{3}{2}}|} T_\Delta \ket{|{\textstyle\frac{3}{2}}} &=& \sqrt{15}, 
\\
\bra{{\textstyle\frac{3}{2}}|}T\ket{|{\textstyle\frac{1}{2}}}
&=&
- \bra{{\textstyle\frac{1}{2}}|} T \ket{|\textstyle{\frac{3}{2}}} = 2 .
\end{eqnarray}

With the above definitions of isospin components, 
we list all Lagrangian used in our calculations. 
We follow the conventions of Bjoken and Drell~\cite{bj}
in defining the metric tensor $g_{\mu\nu}$ and 
the Dirac matrices $\gamma_\mu$ and $\gamma_5$.
Also, we set $\epsilon^{0123} = +1$.

\subsection{Hadronic Interactions}

\subsubsection{$PBB'$ interaction}
\label{sec:pbb}

The interaction Lagrangian between a pseudoscalar-octet meson ($P$)
and spin-$1/2$ octet baryons ($B$, $B'$) is given by
\begin{equation}
L_{PBB'} = 
-\frac{f_{PBB'}}{m_P} \bar B \gamma_\mu \gamma_5 B' \partial^\mu P 
\times L^{\mathrm{iso}}_{PBB'} + [\mathrm{h.c.~for}~B\not=B'] \,.
\end{equation}
Here $L^{\mathrm{iso}}_{PBB'}$ is the isospin structure of the interactions given by
\begin{eqnarray}
L^{\mathrm{iso}}_{\pi NN} &=& 
(N^\dagger \vec{\tau} N) \cdot \vec{\pi} \,,
\label{eq:pNN}
\\
L^{\mathrm{iso}}_{\pi\Xi\Xi} &=& 
(\Xi^\dagger \vec{\tau} \Xi) \cdot \vec{\pi} \,,
\\ 
L^{\mathrm{iso}}_{\pi\Lambda\Sigma} &=& 
\Lambda^\dag (\vec{\Sigma}\cdot \vec{\pi}) \,,
\\
L^{\mathrm{iso}}_{\pi\Sigma\Sigma} &=& 
i [\vec{\Sigma}^\dagger \times \vec{\Sigma}]\cdot \vec{\pi} \,,
\\
L^{\mathrm{iso}}_{K\Sigma N} &=& 
\vec{\Sigma}^\dag  \cdot (K^\dag\vec{\tau}N) \,,
\\
L^{\mathrm{iso}}_{K\Xi\Sigma} &=& 
(\Xi^\dagger\vec{\tau}K) \cdot \vec{\Sigma} \,,
\\
L^{\mathrm{iso}}_{K\Lambda N} &=& 
\Lambda^\dag (K^\dag N)  \,,
\\
L^{\mathrm{iso}}_{K\Xi\Lambda} &=& 
(\Xi^\dagger K_c)\Lambda \,,
\\
L^{\mathrm{iso}}_{\eta NN} &=& 
(N^\dagger N) \eta \,,
\\
L^{\mathrm{iso}}_{\eta\Xi\Xi} &=& 
(\Xi^\dagger \Xi) \eta \,,
\\
L^{\mathrm{iso}}_{\eta\Lambda\Lambda} &=& 
\Lambda^\dagger \Lambda \eta \,,
\\
L^{\mathrm{iso}}_{\eta\Sigma\Sigma} &=& 
(\vec{\Sigma}^\dagger \cdot \vec{\Sigma}) \eta \,.
\label{eq:eSS}
\end{eqnarray}
The coupling constants of $PBB'$ are fixed by the SU(3) relations 
and the $\pi NN$ coupling constant:
\begin{eqnarray}
g_{\pi\Xi\Xi} &=& (1-2\alpha) g_{\pi NN}  \,,
\label{eq:gPXX}
\\
g_{\pi\Lambda\Sigma} &=& \frac{2}{\sqrt{3}}\alpha g_{\pi NN}  \,,
\\
g_{\pi\Sigma\Sigma} &=& 2(-1+\alpha) g_{\pi NN}  \,,
\\
g_{K\Sigma N} &=& (-1+2\alpha) g_{\pi NN}  \,,
\label{eq:gksn}
\\
g_{K\Xi\Sigma} &=& (-1) g_{\pi NN}  \,,
\\
g_{K\Lambda N} &=& \frac{1}{\sqrt{3}}(-3+2\alpha) g_{\pi NN}  \,,
\\
g_{K\Xi\Lambda} &=& \frac{1}{\sqrt{3}}(+3-4\alpha) g_{\pi NN}  \,,
\\
g_{\eta NN} &=& \frac{1}{\sqrt{3}}(+3-4\alpha) g_{\pi NN}  \,,
\label{eq:genn}
\\
g_{\eta\Lambda\Lambda} &=& \frac{1}{\sqrt{3}}(-2\alpha) g_{\pi NN}  \,,
\\
g_{\eta\Sigma\Sigma} &=& \frac{1}{\sqrt{3}}(+2\alpha) g_{\pi NN}  \,,
\label{eq:gESS}
\end{eqnarray}
where $g_{PBB'}= f_{PBB'}/m_P$, $\alpha = 0.635$, and $f_{\pi NN} = \sqrt{4\pi\times 0.08}$.
In this work, however, we do not follow the SU(3) relations~(\ref{eq:gksn}) and~(\ref{eq:genn})
for the $\eta NN$ and $K\Sigma N$ couplings, 
but vary $f_{\eta NN}$ and $f_{K\Sigma N}$ freely in the fit.

\subsubsection{$VBB'$ interaction}
\label{sec:vbb}

The interaction Lagrangian between a vector-octet meson ($V$)
and spin-$1/2$ octet baryons ($B$, $B'$) is given by
\begin{equation}
L_{VBB'} = 
+g_{VBB'} \bar B 
\left[
\vec{\slas{V}} -\frac{\kappa_{VBB'}}{m_B+m_{B'}}\sigma_{\mu\nu}(\partial^\nu \vec V^\mu)
\right] 
B' 
\times L^{\mathrm{iso}}_{VBB'} + [\mathrm{h.c.~for}~B\not=B'] \,.
\end{equation}
The isospin structure of the interactions ($L^{\mathrm{iso}}_{VBB'}$) 
involving $\rho$, $K^\ast$, $K^\ast_c$, and $\omega_8$ is
given by the replacement of
$\pi\to \rho$, $K\to K^\ast$, $K_c\to K^\ast_c$, and $\eta\to \omega_8$ in
Eqs.~(\ref{eq:pNN})-(\ref{eq:eSS}).
Here $\omega_8$ is the eighth component of the octet representation 
of the vector mesons. 
Assuming the ideal mixing, it is related to the physical $\omega$ and $\phi$ mesons as
\begin{equation}
\omega_8 = \frac{1}{\sqrt{3}}\omega - \sqrt{\frac{2}{3}}\phi \,.
\end{equation}
In this work, only $g_{\rho NN}$, $\kappa_{\rho NN}$, $g_{\omega NN}$, and $\kappa_{\omega NN}$
are free parameters for $VBB'$ coupling constants determined by the global fit.
As for the other $VBB'$ interactions, $g_{VBB'}$ is fixed by the value of $g_{\rho NN}$ 
and the corresponding SU(3) relations to Eqs.~(\ref{eq:gPXX})-(\ref{eq:gESS}) 
with the replacement of $\pi\to \rho$, $K\to K^\ast$, $K_c\to K^\ast_c$, and $\eta\to \omega_8$,
while $\kappa_{VBB'}$ is fixed with $\kappa_{VBB'}/(m_B+m_{B'})\equiv\kappa_{\rho NN}/(2m_N)$.

\subsubsection{$SBB'$ interaction}

The interaction Lagrangian between a scalar meson ($S$)
and spin-$1/2$ octet baryons ($B$, $B'$) used in this work is:
\begin{equation}
L_{SBB'} = 
+g_{SBB'} \bar B B' S
\times L^{\mathrm{iso}}_{SBB'} + [\mathrm{h.c.~for}~B\not=B'] \,.
\end{equation}
The isospin structure of the interactions ($L^{\mathrm{iso}}_{SBB'}$) 
is given by
\begin{eqnarray}
L^{\mathrm{iso}}_{\sigma NN} &=& (N^\dag N) \sigma\,,
\\
L^{\mathrm{iso}}_{f_0 NN} &=& (N^\dag N) f_0\,,
\\
L^{\mathrm{iso}}_{\kappa \Lambda N} &=& \Lambda^\dag (\kappa^\dag N) \,,
\\
L^{\mathrm{iso}}_{\kappa \Sigma N} &=& \vec{\Sigma}^\dag \cdot (\kappa^\dag \vec{\tau} N)\,.
\end{eqnarray}
Other $SBB'$ interactions are not considered in this work.

\subsubsection{$ABB'$ interaction}

As for the interaction between a axial-vector meson ($A$)
and spin-$1/2$ octet baryons ($B$, $B'$), we consider
only $a_1 NN$ interaction given by
\begin{equation}
L_{a_1 NN} = +g_{a_1 NN}\bar N \gamma^\mu\gamma_5 \vec \tau N \cdot \vec a_{1\mu} \,.
\end{equation}

\subsubsection{$PVBB'$ and $VVBB'$ interaction}

The following contact terms are also included in the calculation,
\begin{eqnarray}
L_{\rho\pi NN} &=&+\frac{f_{\pi NN}}{m_\pi}g_{\rho NN}
\bar{\psi}_N\gamma_\mu \gamma_5 \vec{\tau}\psi_N
\cdot \left[\vec{\rho^\mu} \times \vec \pi\right] \,, 
\label{eq:Lrpnn}
\\
L_{\rho\rho NN} &=& -\frac{\kappa_\rho g_{\rho NN}^2}{8m_N}
\bar{\psi}_N\sigma^{\mu\nu}\vec{\tau} \psi_N 
\cdot \left[ \vec{\rho_\mu} \times \vec{\rho_\nu} \right] \,.
\label{eq:Lrrnn}
\end{eqnarray}
Note that these contact terms are derived from applying 
$[\partial^\mu \to \partial^\mu - g_{\rho NN} \vec{\rho}^\mu \times ]$ 
to $L_{\pi NN}$ and $L_{\rho NN}$, respectively.
However, in this work we replace 
$f_{\pi NN}g_{\rho NN}$ [$\kappa_\rho g_{\rho NN}^2/8$]
in Eq.~(\ref{eq:Lrpnn}) [Eq.~(\ref{eq:Lrrnn})]
with a new parameter $c_{\rho\pi NN}$ [$c_{\rho\rho NN}$]
and vary the new parameter in the fit.

\subsubsection{$PBD$ interaction}

The interaction Lagrangian involving a pseudoscalar-octet meson ($P$),
a spin-$1/2$ octet baryon ($B$), and a spin-$3/2$ decuplet baryon ($D$)
is given by
\begin{equation}
L_{PBD} = -\frac{f_{PBD}}{m_P} \bar B D^\mu \partial_\mu P 
\times L^{\mathrm{iso}}_{PBD} + [\mathrm{h.c.~for}~B\not=B'] \,, 
\end{equation}
The isospin structure of the interactions ($L^{\mathrm{iso}}_{PBD}$) 
is given by
\begin{eqnarray}
L^{\mathrm{iso}}_{\pi N\Delta} &=& (N^\dag \vec T \Delta) \cdot \vec \pi \,,
\\
L^{\mathrm{iso}}_{\pi \Lambda\Sigma^\ast} &=& \Lambda^\dag (\vec \Sigma^* \cdot \vec \pi) \,,
\\
L^{\mathrm{iso}}_{\pi \Sigma\Sigma^\ast} &=& i[\vec\Sigma^\dag \times \vec \Sigma^\ast]\cdot \vec \pi \,,
\\
L^{\mathrm{iso}}_{KN\Sigma^*} &=& 
(N^\dag \vec \tau K) \cdot \vec \Sigma^* \,.
\end{eqnarray}

\subsubsection{$VBD$ interaction}

As for the interaction involving a vector-octet meson ($V$),
a spin-$1/2$ octet baryon ($B$), and a spin-$3/2$ decuplet baryon ($D$),
we consider only $\rho N \Delta$ interaction:
\begin{equation}
L_{\rho N \Delta} = 
-i\frac{f_{\rho N\Delta}}{m_\rho}\bar \Delta^\mu
\gamma^\nu \gamma_5 
\left[
\partial_\mu \vec{\rho}_\nu-\partial_\nu\vec{\rho}_\mu
\right]
\cdot \vec{T} 
N
+\mathrm{h.c.} \,. 
\end{equation}

\subsubsection{$PDD'$ interaction}

As for the interaction between a pseudoscalar-octet meson ($P$)
and spin-$3/2$ decuplet baryons ($D, D'$), 
we consider only the $\pi\Delta\Delta$ interaction:
\begin{equation}
L_{\pi\Delta\Delta} = 
+\frac{f_{\pi\Delta\Delta}}{m_\pi}\bar \Delta_\mu
\gamma^\nu \gamma_5 \vec{T}_\Delta \Delta^\mu
\cdot \partial_\nu\vec{\pi} \,, 
\end{equation}

\subsubsection{$VDD'$ interaction}

As for the interaction between a vector-octet meson ($V$)
and spin-$3/2$ decuplet baryons ($D, D'$), 
we consider only the $\rho\Delta\Delta$ interaction:
\begin{equation}
L_{\rho\Delta\Delta} = 
+g_{\rho \Delta\Delta} \bar \Delta_\alpha
\left[
\vec{\slas{\rho}} - \frac{\kappa_{\rho\Delta\Delta}}{2m_\Delta}\sigma_{\mu\nu}(\partial^\nu \vec{\rho^\mu})
\right]
\cdot \vec{T}_\Delta \Delta^\alpha \,.
\end{equation}

\subsubsection{$VPP'$ interaction}

The interaction Lagrangian between a vector-octet meson ($V$)
and pseudoscalar-octet mesons ($P$, $P'$) used in this work is given by 
\begin{eqnarray}
L_{\rho\pi\pi} &=&+g_{\rho\pi\pi} 
\left[ \vec\pi \times \partial_\mu \vec\pi \right]
\cdot \vec{\rho}^\mu \,,
\\
L_{\rho K K} &=&+i g_{\rho K K}
(K^\dagger \vec{\tau} \partial_\mu K) \cdot \vec{\rho}^\mu
+\mathrm{h.c.} \,,
\\
L_{K^*K\pi} &=&+i g_{K^*K\pi} 
\left[
  (K^{* \mu \dagger}\vec{\tau} K) \cdot\partial_\mu\vec\pi
- (K^{* \mu \dagger}\vec{\tau} \partial_\mu K) \cdot\vec\pi
\right]
+\mathrm{h.c.} \,,
\\
L_{\omega_8 KK} &=&+i g_{\omega_8 KK}
\left[
K^\dagger(\partial_\mu K) - (\partial_\mu K^\dagger) K
\right] \omega_8^\mu \,,
\\
L_{K^* K \eta} &=&+i g_{K^* K \eta} 
K^{* \mu \dagger} \left[ K (\partial_\mu \eta) - (\partial_\mu K)\eta \right]
+\mathrm{h.c.} \,.
\end{eqnarray}
In this work, the coupling constant $g_{VPP'}$ is fixed by the SU(3) relation,
\begin{eqnarray}
g_{\rho KK} & = & \frac{1}{2}g_{\rho \pi \pi} \, ,\\
g_{K^*  K\pi} & = & \frac{1}{2}g_{\rho \pi \pi} \, ,\\
g_{\omega KK} & = & \frac{1}{\sqrt{3}}g_{\omega_8 KK} = \frac{1}{2}g_{\rho \pi\pi} \, ,\\
g_{\phi  KK} & = & -\sqrt{\frac{2}{3}}g_{\omega_8 KK} =  -\frac{1}{\sqrt{2}}g_{\rho \pi\pi} \, ,\\
g_{K* K\eta} & = & \frac{\sqrt{3}}{2}g_{\rho \pi \pi} \, ,
\end{eqnarray}
so that only $g_{\rho \pi\pi}$ is a free parameter for the $VPP'$ couplings.

\subsubsection{$SPP'$ interaction}

The interaction Lagrangian between a scalar meson ($S$)
and pseudoscalar-octet mesons ($P$, $P'$) used in this work is given by 
\begin{eqnarray}
L_{\sigma\pi\pi}&=& 
-\frac{g_{\sigma\pi\pi}}{2m_\pi} 
(\partial^\mu \vec\pi) \cdot (\partial_\mu \vec\pi) \sigma 
+\frac{\tilde g_{\sigma\pi\pi} m_\pi^2 }{2f_\pi} 
\vec\pi \cdot \vec\pi \sigma \,,
\\
L_{f_0\pi\pi}&=& 
- \frac{g_{f_0\pi\pi}}{2m_\pi} 
(\partial^\mu \vec\pi) \cdot (\partial_\mu \vec\pi) f_0 \,.
\\
L_{\kappa K\pi}&=& 
- \frac{g_{\kappa K\pi}}{m_\pi} 
(\partial_\mu K^\dag) \vec\tau \kappa 
\cdot (\partial^\mu \vec\pi) 
+\mathrm{h.c.} \,,
\\
L_{\kappa K\eta}&=& 
- \frac{g_{\kappa K\eta}}{m_\eta} 
(\partial_\mu K^\dag) \kappa (\partial^\mu \eta) 
+\mathrm{h.c.} \,.
\end{eqnarray}

\subsubsection{$VV'P$ interaction}

As for the interaction between a pseudoscalar-octet meson ($P$)
and vector-octet mesons ($V$, $V'$), in this work 
we consider only $\omega\rho\pi$ interaction:
\begin{eqnarray}
L_{\omega\rho\pi} &=& 
-\frac{g_{\omega\rho\pi}}{m_\omega}
\epsilon_{\mu\alpha\lambda\nu}
(\partial^\alpha\vec{\rho^\mu})\cdot
(\partial^\lambda \vec\pi) \omega^\nu \,.
\end{eqnarray}

\subsection{Electromagnetic Interactions}

The electromagnetic interactions are obtained from the 
usual non-interacting Lagrangian and the above
hadronic Lagrangian by using the minimum substitution
$\partial_\mu \rightarrow \partial_\mu - ie A_\mu$.
The resulting Lagrangians divided by $-e$  ($e=\sqrt{4\pi/137}$) are listed below.

\subsubsection{$\gamma BB'$ interaction}

The interaction Lagrangian between a photon and spin-$1/2$ octet baryons
is given by
\begin{eqnarray}
L_{\gamma BB'} &=& 
\bar B
\left[ 
\hat{e}_{BB'}(Q^2) \sla{A} - \frac{\hat{\kappa}_{BB'}(Q^2)}{2m_N}\sigma^{\mu\nu} (\partial_\nu A_\mu)
\right]
B' 
+ [\mathrm{h.c.~for}~B\not=B'] \,.
\end{eqnarray}
Here we have defined
\begin{eqnarray}
\hat{e}_{NN} & = & \frac{F_{1S}+ F_{1V}\tau^3}{2} \,, 
\\
\hat{\kappa}_{NN}& =&  \frac{F_{2S} + F_{2V}\tau^3}{2} \,,
\\
\hat{e}_{\Lambda\Lambda} & = & 0 \,, 
\\
\hat{\kappa}_{\Lambda\Lambda}& =& -0.61 \,,
\\
\hat{e}_{\Sigma\Sigma} & = & T^3_\Sigma \,, 
\\
\hat{\kappa}_{\Sigma\Sigma}& =& \kappa_{\mathrm{s}}^\Sigma + \kappa_{\mathrm{v}}^\Sigma T^3_\Sigma \,,
\\
\hat{e}_{\Lambda\Sigma} & \sim & (0,0,0) \,, 
\\
\hat{\kappa}_{\Lambda\Sigma}& =& (0,-1.61,0)\,.
\end{eqnarray}
Here $F_{1S}= F_{1V}=1$; $F_{2S}=\mu_p + \mu_n -1 \sim -0.12 $;
$F_{2V}=\mu_p - \mu_n -1 \sim 3.7 $;
$\kappa_{\mathrm{s}}^\Sigma =0.65$; $\kappa_{\mathrm{v}}^\Sigma =0.81$;
and $T_\Sigma^3=\mathrm{diag}(1,0,-1)$.

\subsubsection{$\gamma PBB'$ and $\gamma VBB'$ interaction}

The following four-point interactions are obtained
by applying the minimum substitution $\partial_\mu \rightarrow \partial_\mu - ie A_\mu$
to the $PBB'$ and $VBB'$ interactions defined in Appendices~\ref{sec:pbb} and~\ref{sec:vbb}:

\begin{eqnarray}
L_{\gamma \pi N N} &=&+\frac{f_{\pi NN}}{m_\pi}
\left[ 
(\bar N\sla{A} \gamma_5 \vec{\tau} N)\times \vec \pi
\right]_3 \,, 
\\
L_{\gamma KN\Lambda} &=&+i \frac{f_{K N\Lambda}}{m_K}
\left[
\bar p\sla{A}\gamma_5 K^+ \Lambda
 - \bar\Lambda\sla{A}\gamma_5 K^- p
\right] \,,
\\
L_{\gamma KN\Sigma} &=&+i \frac{f_{K N\Sigma}}{m_K}
\left[ 
 \bar n\sla{A}\gamma_5 K^+\Sigma^{1+i2}
+\bar p\sla{A}\gamma_5 K^+ \Sigma^3
\right. 
\nonumber\\
&&
\left. 
- \bar \Sigma^{1-i2}\sla{A}\gamma_5 K^- n
- \bar \Sigma^{3}\sla{A}\gamma_5 K^- p
\right] \,, 
\\
L_{\gamma \rho N N} &=&+\frac{g_{\rho NN}\kappa_\rho}{2m_N}
\left[
\left(\bar N \frac{\vec{\tau}}{2}\sigma^{\nu\mu} N \right) \times \vec{\rho_\nu}
\right]_3 
A_\mu \,.
\end{eqnarray}
Here the symbol $[ \vec A \times \vec B ]_3$ in the above equations
means the third component of the outer product of the isospin vectors
$\vec A$ and $\vec B$.
In this work, we multiply 
the $\gamma \pi NN$, $\gamma KN\Lambda$, and $\gamma KN\Sigma$ interactions
by additional phenomenological factors 
$c_{\gamma \pi NN}$, $c_{\gamma KN\Lambda}$, and $c_{\gamma KN\Sigma}$,
respectively, and we treat those factors as parameters in the fit.

\subsubsection{$\gamma BD$ interaction}

The following Lagrangian is used for the interactions involving a photon ($\gamma$),
spin-$1/2$ octet baryon ($B$), and a spin-$3/2$ decuplet baryon ($D$):
\begin{eqnarray}
L_{\gamma N \Delta} &=& 
-i \bar \Delta^\mu \Gamma^{\mathrm{em},\Delta N}_{\mu\nu} T^3 NA^\nu 
+\mathrm{h.c.} \,,  
\\
L_{\gamma \Sigma \Sigma^\ast} &=& 
-i (\bar\Sigma^*)^\mu \Gamma^{\mathrm{em},\Sigma^*\Sigma}_{\mu\nu} \Sigma A^\nu 
+\mathrm{h.c.} \,.  
\end{eqnarray}
The matrix element of $\gamma BD$ vertex is explicitly given by
\begin{eqnarray}
\bra{D(p_D)} \Gamma^{\mathrm{em},DB}_{\mu\nu} \ket{B(p_B)}
&=&
\frac{m_D+m_B}{2m_B} \frac{1}{(m_D+m_B)^2-q^2}
\nonumber\\
&&\times
\left[ 
(G^{DB}_M-G^{DB}_E) 3\epsilon_{\mu\nu\alpha\beta}P^\alpha q^\beta
\right.
\nonumber\\
&& \qquad
+ G^{DB}_E i\gamma_5 \frac{12}{(m_D-m_B)^2-q^2} 
\epsilon_{\mu\lambda\alpha\beta} P^\alpha q^\beta {\epsilon^\lambda}_{\nu\gamma\delta} p_D^\gamma q^\delta
\nonumber\\
&& \qquad
\left.
+ G^{DB}_C i\gamma_5 \frac{6}{(m_D-m_B)^2-q^2} q_\mu (q^2 P_\nu - q\cdot P q_\nu)
\right] \,,
\label{eq:vgDB}
\end{eqnarray}
with $P=(p_D+p_B)/2$ and $q=p_D-p_B$. 
Note that the index $\mu$ of $\Gamma^{\mathrm{em},DB}_{\mu\nu}$ contracts with the 
$D$ field and $\nu$ with the photon field.
The $\gamma N\Delta$ coupling strength
$G^{\Delta N}_M = 1.85$, $G^{\Delta N}_E = 0.025$, and $G_C^{\Delta N}=-0.238$ are taken
from the SL model~\cite{sl}. 
However, for the $\gamma \Sigma\Sigma^*$ interactions,
in this work we set 
$G_{E,C}^{(\Sigma^*)^0\Sigma^0}= G_{E,C}^{(\Sigma^*)^\mp\Sigma^\pm}=0$, 
and only $G_{M}^{(\Sigma^*)^0\Sigma^0}$ and $G_{M}^{(\Sigma^*)^\pm\Sigma^\pm}$ 
are treated as free parameters varied freely in the fit.

\subsubsection{$\gamma PBD$ interaction}

The following four-point interaction is given
by applying the minimum substitution $\partial_\mu \rightarrow \partial_\mu - ie A_\mu$
to $L_{\pi N\Delta}$:
\begin{eqnarray}
L_{\gamma \pi N \Delta} &=&+\frac{f_{\pi N\Delta}}{m_\pi}
\left[
(\bar \Delta^\mu \vec{T}{N}) \times \vec\pi
\right]_3
A_\mu 
+\mathrm{h.c.} \,. 
\label{eq-Lgnd} 
\end{eqnarray}

\subsubsection{$\gamma DD'$ interaction}

As for the interaction involving a photon ($\gamma$)
and spin-$3/2$ decuplet baryons ($D$, $D'$),
we consider only the $\gamma \Delta\Delta$ interaction,
\begin{eqnarray}
L_{\gamma\Delta\Delta} &=& 
+\bar \Delta^\eta
\left( T^3_\Delta+\frac{1}{2} \right)
\left[
- \gamma_\mu g_{\eta \nu} + (g_{\mu\eta} \gamma_\nu + g_{\mu\nu}
\gamma_\eta) + \frac{1}{3}\gamma_\eta \gamma_\mu \gamma_\nu
\right] \Delta^\nu A^\mu \ .
\end{eqnarray}

\subsubsection{$\gamma PP'$ interaction}

As for the interaction involving a photon ($\gamma$)
and pseudoscalar octet mesons ($P$, $P'$),
in this work only the following $\gamma \pi\pi$ and $\gamma KK$ interactions are considered:
\begin{eqnarray}
L_{\gamma \pi\pi} &=& 
+\left[ \vec\pi \times (\partial^\mu \vec\pi) \right]_3 A_\mu \,, 
\\
L_{\gamma KK} &=& 
+i \left[ K^-\partial^\mu K^+ - (\partial^\mu K^-)K^+ \right] A_\mu \,.
\end{eqnarray}

\subsubsection{$\gamma VP$ interaction}

The following Lagrangian is considered for interactions 
involving a photon ($\gamma$), a vector octet mesons ($V$),
and a pseudoscalar octet mesons ($P$),
\begin{eqnarray}
L_{\gamma\rho\pi} &=& 
+\frac{g_{\gamma\rho\pi}}{m_\pi} \epsilon_{\alpha\beta\gamma\delta}
(\partial^\alpha A^\beta) (\partial^\gamma \vec{\rho^\delta}) \cdot \vec\pi \,, 
\\
L_{\gamma\omega\pi} &=&
+\frac{g_{\gamma\omega\pi}}{m_\pi} \epsilon_{\alpha\beta\gamma\delta}
(\partial^\alpha A^\beta) (\partial^\gamma \omega^\delta) \pi^3 \,, 
\\
L_{\gamma\rho\eta} &=&
+ \frac{g_{\gamma\rho\eta}}{m_\rho}
\epsilon^{\mu\nu\alpha\beta}
(\partial_\mu\rho_\nu^3)(\partial_\alpha A_\beta) \eta \,, 
\\
L_{\gamma\omega\eta} &=&
+\frac{g_{\gamma\omega\eta}}{m_\omega} \epsilon_{\alpha\beta\gamma\delta}
(\partial^\alpha A^\beta) (\partial^\gamma \omega^\delta) \eta \,, 
\\
L_{\gamma K^*K} &=& 
+\frac{g^0_{\gamma K^*K}}{m_K}\epsilon_{\alpha\beta\gamma\delta}
\left[  
 \bar K^0 (\partial^\gamma K^{*0, \delta})
+K^0 (\partial^\gamma \bar{K}^{*0, \delta})
\right]
\partial^\alpha A^\beta
\\
&& + 
\frac{g^c_{\gamma K^*K}}{m_K}\epsilon_{\alpha\beta\gamma\delta}
\left[    
 K^+ (\partial^\gamma K^{*-, \delta})
+K^- (\partial^\gamma K^{*+, \delta})
\right]
\partial^\alpha A^\beta \, .
\end{eqnarray}
In this work, we treat 
$g_{\gamma\rho\pi}$, $g_{\gamma\omega\pi}$, $g_{\gamma\rho\eta}$, and $g_{\gamma\omega\eta}$
as free parameters in the fit, while we use the fixed values for 
$\gamma K^*K$ couplings, i.e., $g^0_{\gamma K^*K}/m_K =-0.388$ GeV$^{-1}$
and $g^c_{\gamma K^*K}=0.254$ GeV$^{-1}$~\cite{onl}.

\subsubsection{$\gamma VPP'$ interaction}

The following four-point interaction is given
by applying the minimum substitution $\partial_\mu \rightarrow \partial_\mu - ie A_\mu$
to $L_{\rho\pi\pi}$:
\begin{eqnarray}
L_{\gamma \rho\pi\pi} &=& -g_{\rho\pi\pi}
\left[ 
(\vec{\rho^\mu} \times \vec\pi)\times \vec\pi
\right]_3
A_\mu \,. 
\end{eqnarray}

\subsubsection{$\gamma AP$ interaction}

As for the interaction involving a photon ($\gamma$),
an axial-vector meson ($A$), and a pseudoscalar meson ($P$),
in this paper only the $\gamma a_1 \pi$ interaction is considered:
\begin{eqnarray}
L_{\gamma a_1\pi} &=&+\frac{1}{m_{a_1}}
(\partial^\mu A^\nu-\partial^\nu A^\mu)
\nonumber\\
&& \times
\left\{
  2[(\partial_\mu \vec\pi)\times \vec a_{1\nu}]_3 
- 2[(\partial_\nu \vec \pi)\times \vec a_{1\mu}]_3
+[\vec \pi \times (\partial_\mu \vec a_{1\nu}-\partial_\nu \vec a_{1\mu})]_3
\right\} \,.
\end{eqnarray}

\subsubsection{$\gamma VV'$ interaction}

As for the interaction involving a photon ($\gamma$)
and vector octet mesons ($V$, $V'$),
in this paper only the $\gamma \rho\rho$ interaction is considered:
\begin{eqnarray}
L_{\gamma \rho\rho} &=& 
+\left[ 
(\partial^\mu\vec{\rho^\nu} -\partial^\nu\vec{\rho^\mu}) \times \vec{\rho}_\nu
\right]_3
A_\mu \,.
\end{eqnarray}

\section{Matrix elements of meson-baryon potentials}
\label{app:pot}

It is convenient to get the partial-wave matrix elements of 
the meson-exchange potential $v_{M'B',MB}$ 
by first evaluating $v_{M'B',MB}$ in helicity representation and then
transforming them into the usual $\ket{(LS)J T }$ representation with
$J$, $T$, $L$, and $S$  denoting the total angular momentum,
isospin, orbital angular momentum, and spin quantum numbers, respectively.
For each meson-baryon ($MB$) state, we use $k$($p$) to denote
the momentum of $M$($B$). In the center-of-mass frame,
we thus have $\vec{p}=-\vec{k}$.
Following the Jacob-Wick formulation~\cite{jw},
the partial-wave matrix elements of $v_{M'B',MB}$ can be written as 
\begin{eqnarray}
{\it v}^{JT}_{L' S' M'B', L S MB}(k',k,W) 
&=& \sum_{ \lambda'_M\lambda'_B\lambda_M\lambda_B } 
\frac{\sqrt{(2L+1)(2L'+1)}}{2J+1} \nonumber \\
& & \times 
\inp{ j'_M j'_B \lambda'_M  -\lambda'_B }{ S' S'_z } 
\inp{ L'S'0S'_z }{ J S'_z } 
\nonumber \\
& & \times 
\inp{ j_M j_B \lambda_M  -\lambda_B }{ S S_z }
\inp{ LS0S_z }{ J S_z} 
\nonumber \\
& & \times 
\bra{ J,k' \lambda'_M -\lambda'_B } v_{M'B',MB} \ket{ J,k \lambda_M -\lambda_B } \,,
\label{eq:c-1}
\end{eqnarray}
where $j_M$ and $j_B$ are the spins of the meson  and baryon, 
respectively, and $\lambda_M$ and $\lambda_B$ are their helicities, and
\begin{eqnarray}
& &
\bra{ J,k' \lambda'_M -\lambda'_B } v_{M'B',MB} \ket{ J,k \lambda_M -\lambda_B } 
\nonumber \\
& & = 2\pi \int_{-1}^{+1} d(\cos\theta)
d^{J}_{\lambda_M-\lambda_B,\lambda'_M-\lambda'_B}(\theta)
\nonumber \\
& & \times 
\bra{ M'(\vec{k}', s'_M \lambda'_M) B'(-\vec{k}', s'_B,-\lambda'_B) } 
v_{M'B',MB}
\ket{ M(\vec{k}, s_M \lambda_M) B(-\vec{k}, s_B, -\lambda_B) }\,.
\label{eq:c-2}
\end{eqnarray}
Here we have chosen the coordinates such that
\begin{eqnarray}
\vec{k}' &=& (k'\sin\theta ,0, k'\cos\theta )\,, \\
\vec{k} &=& (0 ,0, k )\,,
\end{eqnarray}
and the helicity eigenstates are defined by
\begin{eqnarray}
\hat{k}\cdot \vec{s}_M \ket{ M( \vec{k}, s_M \lambda_M) } &=& \lambda_M
\ket{ M( \vec{k}, s_M \lambda_M) }  \,,
\end{eqnarray}
\begin{eqnarray}
[-\hat{k}\cdot \vec{s}_B] \ket{ B( -\vec{k}, s_B \lambda_B) } 
&=& \lambda_B \ket{ B( -\vec{k}, s_B \lambda_B) } \,.
\label{eq:helicity-2}
\end{eqnarray}
Note the ``$-$'' sign in Eq.~(\ref{eq:helicity-2}).

To evaluate the matrix elements in the right hand side of Eq.~(\ref{eq:c-2}),
we define (suppressing the helicity and isospin indices)
\begin{eqnarray}
\bra{ k^\prime(j),p^\prime } v_{M'B',MB} \ket{ k(i), p} 
&=& \frac{1}{(2\pi)^3}
\sqrt{\frac{m_B^\prime}{E_{B^\prime}(p^\prime)}}
\frac{1}{\sqrt{2E_{M^\prime}(k^\prime)}}
\sqrt{\frac{m_B}{E_{B}(p)}}
\frac{1}{\sqrt{2E_{M}(k)}} \nonumber \\
& & \times \bar{u}_{B'}(\vec{p'}) \bar{V}(n) u_B(\vec{p}) \ .
\label{eq:pwa-pot-diag}
\end{eqnarray}
Here the label $n$ indicates the considered $MB \to M'B'$ transition
as specified in Table~\ref{tab:pot}; $i, j$ are the isospin indices of the mesons.
We also use the notation $q= k^\prime-k$ or $q=p-p^\prime$ in this appendix.
The expressions of each term in $\bar V(n)$ are given in the following sections.
\begin{table}
\caption{\label{tab:pot}
Label $n$ for $\bar V(n)$ in Eq.~(\ref{eq:pwa-pot-diag}).
}
\begin{ruledtabular}
\begin{tabular}{lrrrrrrr}
Channel     &$\pi N$&$\eta N$&$\sigma N$&$\rho N$&$\pi \Delta$&$K\Lambda$&$K\Sigma$ \\
\hline
$\pi N$     &1      &2       &4         &7       &11          &16        &19        \\
$\eta N$    &       &3       &5         &8       &12          &17        &20        \\
$\sigma N$  &       &        &6         &9       &13          &-         &-         \\
$\rho N$    &       &        &          &10      &14          &-         &-         \\
$\pi \Delta$&       &        &          &        &15          &-         &-         \\
$K\Lambda$  &       &        &          &        &            &18        &21        \\
$K\Sigma$   &       &        &          &        &            &          &22
\end{tabular}
\end{ruledtabular}
\end{table}

\subsection{$\pi(k,i)+ N(p) \to \pi(k^\prime,j)+ N(p^\prime)$}

\begin{eqnarray}
\bar{V}(1) = \bar{V}^1_a +\bar{V}^1_b+\bar{V}^1_c+\bar{V}^1_d+\bar{V}^1_e+\bar{V}^1_f\ ,
\label{eq:v1}
\end{eqnarray}
with
\begin{eqnarray}
\bar{V}^1_a &=& \left( \frac{f_{\pi NN}}{m_\pi} \right)^2 \slas{k}^\prime
\gamma_5 \tau^j S_N(p+k) \slas{k}\gamma_5 \tau^i\ , 
\label{eq:v1a}
\\
\bar{V}^1_b &=& \left( \frac{f_{\pi NN}}{m_\pi} \right)^2 \slas{k}
\gamma_5 \tau^i S_N(p-k^\prime) \slas{k}^\prime\gamma_5 \tau^j \ , 
\\
\bar{V}^1_c &=& \left( \frac{f_{\pi N\Delta}}{m_\pi} \right)^2 {k}_\alpha
(T^\dagger)^i S^{\alpha\beta}_\Delta(p-k^\prime){k}^\prime_{\beta} T^j \ , 
\\
\bar{V}^1_d &=& 
-C_{V1}\frac{g_{\rho NN}g_{\rho\pi\pi}}{m_\rho^2}
\left[
(\slas{k}+\slas{k}^\prime)+
C_{V2}\frac{\kappa_\rho}{4m_N}\{(\slas{k}+\slas{k}^\prime)\slas{q}
-\slas{q}(\slas{k}+\slas{k}^\prime)\}
\right] 
i\epsilon_{jil}
\frac{\tau^l}{2}
\nonumber \\ 
&&
+ g_{\rho NN}g_{\rho\pi\pi} \frac{q^2}{m^2_\rho(q^2-m^2_\rho)} 
\left[
(\slas{k}+\slas{k}^\prime)+
\frac{\kappa_\rho}{4m_N}\{(\slas{k}+\slas{k}^\prime)\slas{q}
-\slas{q}(\slas{k}+\slas{k}^\prime)\}
\right] 
i\epsilon_{jil}
\frac{\tau^l}{2} \ , 
\\
\bar{V}^1_e &=& 
+C_S \frac{k\cdot k^\prime}{m_\pi} \delta_{ij} 
\nonumber\\
&&
-
\left(
g_{\sigma NN} g_{\sigma \pi\pi} \frac{q^2}{m_\sigma^2(q^2-m_\sigma^2)} 
+g_{f_0 NN} g_{f_0 \pi\pi} \frac{q^2}{m_{f_0}^2(q^2-m_{f_0}^2)} 
\right)
\frac{k\cdot k^\prime}{m_\pi} \delta_{ij} \ ,
\\
\bar{V}^1_f &=& 
-\frac{g_{\sigma NN} \tilde g_{\sigma \pi\pi}m_\pi^2}{f_\pi} \frac{1}{q^2-m_\sigma^2} 
\delta_{ij} \ .
\end{eqnarray}
Here one can see that parameters $C_{V1}$, $C_{V2}$, and $C_S$, which are somewhat
unusual and require some explanation,
appear in the meson-exchange potentials $\bar V_d^1$ and $\bar V_e^1$.
Let us explain this by taking $\rho$-meson-exchange potential $\bar V_d^1$
as an example.
It is first noticed that if one takes $C_{V1} = C_{V2}=1$,
$\bar V_d^1$ reduces to the familiar $\rho$-meson exchange $\pi N\to\pi N$ potential 
derived from $L_{\rho\pi\pi}$ and $L_{\rho NN}$.
In effective theories, the propagator of heavy mesons $M$ is expanded as
\begin{equation}
\frac{1}{q^2-m_M^2} = -\frac{1}{m_M^2} + {\cal O}\left(\frac{q^2}{m_M^4}\right),
\end{equation}
and the effect of the heavy-particle exchanges is absorbed into contact terms of the 
effective theories.
We have taken into account contributions of higher excited $\rho$ meson states 
phenomenologically by making $C_{V1}$ and $C_{V2}$ as free parameters 
that can be different from $C_{V1} = 1$
and $C_{V2}=1$, respectively.
Furthermore, we attach different cutoff factors
for the first and second terms of $\bar V_d^1$.
A similar prescription has been applied also to the scalar-meson-exchange
potential $\bar V_e^1$. 
In addition, in this work we include a phenomenological contact potential 
$\bar V^1 = c_{S31}$ in getting a good fit to $S_{31}$ partial wave.
We do not need this for other partial waves.

\subsection{$\pi(k,i)+ N (p)\to \eta(k^\prime)+ N (p^\prime)$ }

\begin{eqnarray}
\bar{V}(2) = \bar{V}^2_a +\bar{V}^2_b\ ,
\end{eqnarray}
with 
\begin{eqnarray}
\bar{V}^2_a &=& \frac{f_{\pi NN}f_{\eta NN}}{m_\pi m_\eta} \slas{k}^\prime
\gamma_5  S_N(p+k) \slas{k}\gamma_5 \tau^i\ , \\
\bar{V}^2_b &=& \frac{f_{\pi NN}f_{\eta NN}}{m_\pi m_\eta} \slas{k}
\gamma_5 \tau^i S_N(p-k^\prime) \slas{k}^\prime\gamma_5 \ .
\end{eqnarray}

\subsection{$\eta(k)+ N(p) \to \eta(k^\prime)+ N(p^\prime)$ }

\begin{eqnarray}
\bar{V}(3) = \bar{V}^3_a +\bar{V}^3_b\ ,
\end{eqnarray}
with
\begin{eqnarray}
\bar{V}^3_a&=&\left(\frac{f_{\eta NN}}{m_\eta}\right)^2\slas{k}^\prime
\gamma_5 S_N(p+k) \slas{k}\gamma_5 \ , 
\\
\bar{V}^3_b&=&\left(\frac{f_{\eta NN}}{m_\eta}\right)^2\slas{k}
\gamma_5 S_N(p-k^\prime) \slas{k}^\prime\gamma_5 \ .
\end{eqnarray}

\subsection{$\pi(k,i)+ N (p)\to \sigma(k^\prime)+ N (p^\prime)$ }

\begin{eqnarray}
\bar{V}(4) = \bar{V}^4_a +\bar{V}^4_b+\bar{V}^4_c\ ,
\end{eqnarray}
with
\begin{eqnarray}
\bar{V}^4_a&=& i g_{\sigma NN}\frac{f_{\pi NN}}{m_\pi}S_N(p+k)\slas{k}
\gamma_5\tau^i \ , \\
\bar{V}^4_b&=& i g_{\sigma NN}\frac{f_{\pi NN}}{m_\pi}\slas{k}
\gamma_5 S_N(p-k^\prime) \tau^i \ , \\
\bar{V}^4_c &=& i\frac{f_{\pi NN}g_{\sigma\pi\pi}}{m_\pi^2}\slas{q}\gamma_5
\tau^i\frac{q\cdot k}{q^2-m^2_\pi}\ . 
\end{eqnarray}

\subsection{$\eta(k)+ N (p)\to \sigma(k^\prime)+ N (p^\prime)$ }

\begin{eqnarray}
\bar{V}(5) = \bar{V}^5_a +\bar{V}^5_b\ ,
\end{eqnarray}
with
\begin{eqnarray}
\bar{V}^5_a&=&i g_{\sigma NN}\frac{f_{\eta NN}}{m_\eta}S_N(p+k)\slas{k}
\gamma_5\ , \\
\bar{V}^5_b&=&i g_{\sigma NN}\frac{f_{\eta NN}}{m_\eta}\slas{k}\gamma_5
S_N(p-k^\prime) \ .
\end{eqnarray}

\subsection{$\sigma(k)+ N(p) \to \sigma(k^\prime)+ N(p^\prime)$ }

\begin{eqnarray}
\bar{V}(6) = \bar{V}^6_a +\bar{V}^6_b\ ,    
\end{eqnarray}
with
\begin{eqnarray}
\bar{V}^6_a&=&g_{\sigma NN}^2 S_N(p+k)\ ,  \\
\bar{V}^6_b&=&g_{\sigma NN}^2 S_N(p-k^\prime)\ . 
\end{eqnarray}

\subsection{$\pi(k,i)+ N(p) \to \rho'(k^\prime,j)+ N(p^\prime)$}

\begin{eqnarray}
\bar{V}(7) = \bar{V}^7_a +\bar{V}^7_b+\bar{V}^7_c+\bar{V}^7_d+\bar{V}^7_e\ ,
\end{eqnarray}
with
\begin{eqnarray}
\bar{V}^7_a &=& i\frac{f_{\pi NN}}{m_\pi}g_{\rho NN}\Gamma_{\rho^\prime}
 S_N(p+k) \slas{k}\gamma_5 \tau^i \ , \\
\bar{V}^7_b &=& i\frac{f_{\pi NN}}{m_\pi}g_{\rho NN}
\slas{k}\gamma_5 \tau^i S_N(p-k^\prime)\Gamma_{\rho^\prime}
\ , \\
\bar{V}^7_c &=&\frac{f_{\pi NN}}{m_\pi} g_{\rho\pi\pi}
\epsilon_{ij{\it l}}\tau^{\it l}\frac{(q-k)\cdot\epsilon_{\rho^\prime}^*
\slas{q}\gamma_5}{q^2-m_\pi^2} \ , \\   
\bar{V}^7_d&=&-\frac{f_{\pi NN}}{m_\pi}g_{\rho NN}
\slas{\epsilon_{\rho^\prime}}^*\gamma_5 \epsilon_{ji{\it l}}\tau^{\it l}
\ , \\
\bar{V}^7_e&=& \frac{g_{\omega NN}g_{\omega \pi\rho}}{m_\omega}\delta_{ij}
\frac{\epsilon_{\alpha\beta\gamma\delta}\epsilon_{\rho^\prime}^{*\alpha}
 k^{\prime\beta}
k^\gamma}{q^2-m^2_\omega}
\left[
\gamma^\delta+\frac{\kappa_\omega}{4m_N}
(\gamma^\delta\slas{q}-\slas{q}\gamma^\delta)
\right]\ ,
\end{eqnarray}
where
\begin{eqnarray}
\Gamma_{\rho^\prime}& =&\frac{\tau^j}{2}
\left[
\slas{\epsilon_{\rho'}}^*+
\frac{\kappa_\rho}{4m_N}
(\slas{\epsilon_{\rho'}}^*\slas{k}^\prime
-\slas{k}^\prime\slas{\epsilon_{\rho'}}^*)
\right] \ .
\end{eqnarray}

\subsection{$\eta(k)+ N(p) \to \rho'(k^\prime,j)+ N(p^\prime)$}

\begin{eqnarray}
\bar{V}(8) = \bar{V}^8_a +\bar{V}^8_b\ ,
\end{eqnarray}
with
\begin{eqnarray}
\bar{V}^8_a &=& i\frac{f_{\eta NN}}{m_\eta}g_{\rho NN}\Gamma_{\rho^\prime}
 S_N(p+k) \slas{k}\gamma_5\ ,  \\
\bar{V}^8_b &=& i\frac{f_{\eta NN}}{m_\eta}g_{\rho NN}
\slas{k}\gamma_5  S_N(p-k^\prime)\Gamma_{\rho^\prime}\ .
\end{eqnarray}

\subsection{$\sigma(k)+ N(p) \to \rho'(k^\prime,j)+ N(p^\prime)$}

\begin{eqnarray}
\bar{V}(9) = \bar{V}^9_a +\bar{V}^9_b\ ,
\end{eqnarray}
with
\begin{eqnarray}
\bar{V}^9_a&=& g_{\rho NN} g_{\sigma NN}\Gamma_{\rho^\prime}
S_N(p+k)\ , \\
\bar{V}^9_b&=& g_{\rho NN} g_{\sigma NN} 
S_N(p-k^\prime) \Gamma_{\rho^\prime}\ .
\end{eqnarray}

\subsection{$\rho(k,i)+ N(p) \to \rho^\prime(k^\prime,j)+ N(p^\prime)$}

\begin{eqnarray}
\bar{V}(10) = \bar{V}^{10}_a +\bar{V}^{10}_b+ \bar{V}^{10}_c\ ,
\end{eqnarray}
with
\begin{eqnarray}
\bar{V}^{10}_a +\bar{V}^{10}_b &=&
g_{\rho NN}^2[\Gamma_{\rho^\prime} S_N(p+k)\Gamma_\rho 
+\Gamma_{\rho} S_N(p-k^\prime)\Gamma_{\rho^\prime}] \ ,
\end{eqnarray}
where
\begin{eqnarray}
\Gamma_{\rho}& =&\frac{\tau^i}{2}
\left[
\slas{\epsilon_{\rho}}-
\frac{\kappa_\rho}{4m_N}
(\slas{\epsilon_{\rho}}
\slas{k}-\slas{k}\slas{\epsilon_{\rho}})
\right] \ ,
\end{eqnarray}
and
\begin{eqnarray}
\bar{V}^{10}_c &=& i\frac{\kappa_\rho g^2_{\rho NN}}{8 m_N}
[\slas{\epsilon_\rho}\slas{\epsilon_{\rho^\prime}}^*-
\slas{\epsilon_{\rho^\prime}}^*\slas{\epsilon_{\rho}} ]
\epsilon_{i j {\it l}}\tau^{\it l}\ . 
\end{eqnarray}

\subsection{$\pi(k,i)+ N(p) \to \pi(k^\prime,j)+ \Delta(p^\prime)$}

\begin{eqnarray}
\bar{V}(11) = \bar{V}^{11}_a +\bar{V}^{11}_b+ \bar{V}^{11}_c
+\bar{V}^{11}_d+ \bar{V}^{11}_e\ ,
\end{eqnarray}
with
\begin{eqnarray}
\bar{V}^{11}_a&=&\frac{f_{\pi NN}f_{\pi N\Delta}}{m^2_\pi}
T^j\epsilon_\Delta^*\cdot k^\prime S_N(p+k)\slas{k}\gamma_5\tau^i \ , 
\\
\bar{V}^{11}_b&=&\frac{f_{\pi NN}f_{\pi N\Delta}}{m^2_\pi}
T^i\epsilon_\Delta^*\cdot k S_N(p-k^\prime)\slas{k}^\prime\gamma_5 
\tau^j \ , 
\\
\bar{V}^{11}_c&=&i \frac{f_{\rho N\Delta}f_{\rho\pi\pi}}{m_\rho}
\frac{\epsilon_{ji{\it l}}T^{\it l}}{q^2-m^2_\rho}
[\epsilon_\Delta^*\cdot q (\slas{k}+\slas{k}^\prime)\gamma_5
-\epsilon_\Delta^*\cdot(k+k^\prime)\slas{q}\gamma_5] \ , 
\\
\bar{V}^{11}_d&=& -\frac{f_{\pi\Delta\Delta}f_{\pi N\Delta}}{m^2_\pi}
[\epsilon_\Delta^*]_\mu \slas{k}^\prime\gamma_5 T^j_\Delta
 S^{\mu\nu}_{\Delta}(p' + k') T^i k_\nu \ , 
\label{eq:v11d}
\\
\bar{V}^{11}_e&=& -\frac{f_{\pi\Delta\Delta}f_{\pi N\Delta}}{m^2_\pi}
[\epsilon_\Delta^*]_\mu \slas{k}\gamma_5 T^i_\Delta
 S^{\mu\nu}_{\Delta}(p- k^\prime) T^j k_\nu^\prime\ . 
\end{eqnarray}

\subsection{$\eta(k)+ N(p) \to \pi(k^\prime,j)+ \Delta(p^\prime)$}

\begin{eqnarray}
\bar{V}(12) 
 =\frac{f_{\eta NN}f_{\eta N\Delta}}{m_\pi m_\eta}
T^j\epsilon_\Delta^*\cdot
k^\prime S_N(p+k) \slas{k}\gamma_5\ .
\end{eqnarray}

\subsection{$\sigma(k)+ N(p) \to \pi(k^\prime,j)+ \Delta(p^\prime)$}

\begin{eqnarray}
\bar{V}(13)
 =-i g_{\sigma NN}\frac{f_{\pi N\Delta}}{m_\pi}
T^j\epsilon_\Delta^*\cdot
k^\prime S_N(p+k) \ .
\end{eqnarray}

\subsection{$\rho(k,i)+ N(p) \to \pi(k^\prime,j)+ \Delta(p^\prime)$}

\begin{eqnarray}
\bar{V}(14) = \bar{V}^{14}_a +\bar{V}^{14}_b+ \bar{V}^{14}_c
+\bar{V}^{14}_d\ ,
\end{eqnarray}
with
\begin{eqnarray}
\bar{V}^{14}_a &=& -i\frac{f_{\pi N\Delta}g_{\rho NN}}{m_\pi}
T^j\epsilon_\Delta^*\cdot k^\prime S_N(p+k)\Gamma_\rho\ , 
\\
\bar{V}^{14}_b &=&i\frac{f_{\pi NN}f_{\rho N\Delta}}{m_\pi m_\rho}
T^i[
  \epsilon_\Delta^*\cdot k \slas{\epsilon}_\rho \gamma_5
- \epsilon_\Delta^*\cdot \epsilon_\rho \slas{k} \gamma_5]
 S_N(p-k^\prime)\slas{k}^\prime \gamma_5 \tau^j \ , 
\\
\bar{V}^{14}_c &=& -i \frac{f_{\pi \Delta\Delta}f_{\rho N\Delta}}{m_\pi m_\rho}
[\epsilon_\Delta^*]_\alpha\slas{k}^\prime \gamma_5 T^j_\Delta 
S^{\alpha\beta}_{\Delta}(p'+k')
[k_\beta\slas{\epsilon_\rho}\gamma_5 -
[\epsilon_\rho]_\beta \slas{k}\gamma_5]T^i \ , \\
\bar{V}^{14}_d &=&-i \frac{g_{\rho \Delta\Delta}f_{\pi N\Delta}}{m_\pi}
[\epsilon_\Delta^*]_\alpha 
\left[
\slas{\epsilon_\rho}
-\frac{\kappa_{\rho\Delta\Delta}}{4m_\Delta}
(\slas{\epsilon_\rho}\slas{k}-\slas{k}\slas{\epsilon_\rho})
\right]
T_\Delta^i S^{\alpha\beta}_\Delta(p-k^\prime) T^j k^{\prime}_\beta\ .
\end{eqnarray}

\subsection{$\pi(k,i)+ \Delta(p) \rightarrow \pi(k^\prime,j)+ \Delta^\prime (p^\prime)$}

\begin{eqnarray}
\bar{V}(15) = \bar{V}^{15}_a +\bar{V}^{15}_b+ \bar{V}^{15}_c +\bar{V}^{15}_d\ ,
\end{eqnarray}
with
\begin{eqnarray}
\bar{V}^{15}_a &=& \left(\frac{f_{\pi N\Delta}}{m_\pi}\right)^2
\epsilon^*_{\Delta^\prime}\cdot k^\prime T^j S_N(p+k) \epsilon_\Delta \cdot 
k (T^i)^\dagger \ , \\
\bar{V}^{15}_b &=& \left(\frac{f_{\pi \Delta\Delta}}{m_\pi}\right)^2
\slas{k}^\prime \gamma_5 T^j_\Delta [\epsilon^*_{\Delta^{\prime}}]_\mu
S^{\mu\nu}_\Delta(p+k) [\epsilon_{\Delta}]_\nu \slas{k}\gamma_5 T^i_\Delta
\ , \\
\bar{V}^{15}_c &=& \left(\frac{f_{\pi \Delta\Delta}}{m_\pi}\right)^2
\slas{k} \gamma_5 T^i_\Delta [\epsilon^*_{\Delta^{\prime}}]_\mu
S^{\mu\nu}_\Delta(p-k^\prime) [\epsilon_{\Delta}]_\nu 
\slas{k}^\prime \gamma_5 T^j_\Delta
\ , \\
\bar{V}^{15}_d&=&i g_{\rho \Delta\Delta}g_{\rho\pi\pi}
\frac{\epsilon_{ji{\it l}}T_\Delta^{\it l}}{q^2-m^2_\rho}
\left[
(\slas{k}+\slas{k}^\prime)+
\frac{\kappa_{\rho\Delta\Delta}}{4m_\Delta}
\bm{(}(\slas{k}+\slas{k}^\prime)\slas{q}-\slas{q}(\slas{k}+\slas{k}^\prime) \bm{)}
\right]
\epsilon^*_{\Delta^{\prime}}\cdot\epsilon_{\Delta}\ .
\end{eqnarray}

\subsection{$\pi(k,i)+ N(p) \to K(k^\prime)+ \Lambda(p^\prime)$}

\begin{eqnarray}
\bar{V}(16) = \bar{V}^{16}_a + \bar{V}^{16}_b + \bar{V}^{16}_c + \bar{V}^{16}_d + \bar{V}^{16}_e\ ,
\end{eqnarray}
with
\begin{eqnarray}
\bar{V}^{16}_a&=& 
{f_{K\Lambda N} f_{\pi NN}\over m_K m_\pi}
\slas{k}^\prime
\gamma_5 S_N(p+k) \slas{k}\gamma_5\tau^i\ , 
\\
\bar{V}^{16}_b&=& 
{f_{\pi\Lambda\Sigma}f_{K \Sigma N}\over m_K m_\pi}
\slas{k}
\gamma_5 S_\Sigma(p-k') \slas{k}^\prime\gamma_5\tau^i\ , 
\\
\bar{V}^{16}_c &=& \frac{f_{\pi \Lambda\Sigma^*}f_{K N\Sigma^*}}{m_K m_\pi} 
{k}_\alpha S^{\alpha\beta}_{\Sigma^*}(p-k^\prime){k}^\prime_{\beta} \tau^i \ , 
\\
\bar{V}^{16}_d&=& -g_{K^* N\Lambda}g_{K^*K\pi}
{-g_{\mu\rho} + q_\mu q_\rho/m^2_{K^*} \over q^2-m^2_{K^*}} 
\left(
\gamma^\mu - i\frac{\kappa_{{K^*}N\Lambda}}{m_N+m_\Lambda} \sigma^{\mu\nu}q_\nu
\right)
(k+k')^\rho \tau^i\ , 
\\
\bar{V}^{16}_e &=& - \frac{g_{\kappa\Lambda N}g_{\kappa K\pi}}{m_\pi}
\frac{k\cdot k^\prime}{q^2-m_\kappa^2} \tau^i \ .
\end{eqnarray}

\subsection{$\eta(k)+ N(p) \to K(k^\prime)+ \Lambda(p^\prime)$}

\begin{eqnarray}
\bar{V}(17) = \bar{V}^{17}_a + \bar{V}^{17}_b + \bar{V}^{17}_c + \bar{V}^{17}_d \ ,
\end{eqnarray}
with
\begin{eqnarray}
\bar{V}^{17}_a&=& 
{f_{K\Lambda N} f_{\eta NN}\over m_K m_\eta}
\slas{k}^\prime
\gamma_5 S_N(p+k) \slas{k}\gamma_5 \ , 
\\
\bar{V}^{17}_b&=& 
{f_{\eta\Lambda\Lambda}f_{K \Lambda N}\over m_K m_\eta}
\slas{k}
\gamma_5 S_\Lambda(p-k') \slas{k}^\prime\gamma_5 \ , 
\\
\bar{V}^{17}_c&=& -g_{K^* N\Lambda}g_{K^*K\eta}
{-g_{\mu\rho} + q_\mu q_\rho/m^2_{K^*} \over q^2-m^2_{K^*}} 
\left(
\gamma^\mu - i\frac{\kappa_{{K^*}N\Lambda}}{m_N+m_\Lambda} \sigma^{\mu\nu}q_\nu
\right)
(k+k')^\rho \ , 
\\
\bar{V}^{17}_d &=& - \frac{g_{\kappa\Lambda N}g_{\kappa K\eta}}{m_\pi}
\frac{k\cdot k^\prime}{q^2-m_\kappa^2} \ .
\end{eqnarray}

\subsection{$K(k)+ \Lambda(p) \to K(k^\prime)+ \Lambda(p^\prime)$}

\begin{eqnarray}
\bar{V}(18) = \bar{V}^{18}_a + \bar{V}^{18}_b + \bar{V}^{18}_c +
 \bar{V}^{18}_d\ ,
\end{eqnarray}
with
\begin{eqnarray}
\bar{V}^{18}_a&=& 
\left( \frac{f_{K\Lambda N}}{m_K} \right)^2
\slas{k}^\prime
\gamma_5 S_N(p+k) \slas{k}\gamma_5\ , \\
\bar{V}^{18}_b&=& 
\left( \frac{f_{K\Xi\Lambda}}{m_K} \right)^2
\slas{k}
\gamma_5 S_\Xi(p-k') \slas{k}^\prime\gamma_5\ , \\
\bar{V}^{18}_c&=&
- g_{\omega\Lambda\Lambda}g_{\omega KK}
{-g_{\mu\rho} + q_\mu q_\rho/m^2_{\omega} \over q^2-m^2_{\omega}} 
\left(
\gamma^\mu - i\frac{\kappa_{{\omega}\Lambda\Lambda}}{2 m_\Lambda} \sigma^{\mu\nu}q_\nu
\right)
(k+k')^\rho \ , \\
\bar{V}^{18}_d&=& 
-g_{\phi\Lambda\Lambda}g_{\phi KK}
{-g_{\mu\rho} + q_\mu q_\rho/m^2_{\phi} \over q^2-m^2_{\phi}} 
\left(
\gamma^\mu - i\frac{\kappa_{{\phi}\Lambda\Lambda}}{2 m_\Lambda} \sigma^{\mu\nu}q_\nu
\right)
(k+k')^\rho\ .
\end{eqnarray}

\subsection{$\pi(k,i)+ N(p) \to K(k^\prime)+ \Sigma(p^\prime,j)$}

\begin{eqnarray}
\bar{V}(19) = \bar{V}^{19}_a + \bar{V}^{19}_b + \bar{V}^{19}_c +
              \bar{V}^{19}_d + \bar{V}^{19}_e + \bar{V}^{19}_f\ ,
\end{eqnarray}
with
\begin{eqnarray}
\bar{V}^{19}_a&=& 
{f_{K\Sigma N} f_{\pi NN}\over m_K m_\pi}
\slas{k}^\prime
\gamma_5\tau^j S_N(p+k) \slas{k}\gamma_5\tau^i\ , 
\\
\bar{V}^{19}_b&=& 
{f_{\pi \Lambda\Sigma}f_{K\Lambda N}\over m_K m_\pi}
\slas{k}
\gamma_5 S_\Lambda(p-k') \slas{k}^\prime\gamma_5\delta^{ij}\ , 
\\
\bar{V}^{19}_c&=& 
{f_{\pi \Sigma\Sigma}f_{K\Sigma N}\over m_K m_\pi}
\slas{k}
\gamma_5 S_\Sigma(p-k') \slas{k}^\prime\gamma_5 
i \epsilon^{ijk}\tau_k\ , 
\\
\bar{V}^{19}_d &=& 
\frac{f_{\pi \Sigma\Sigma^*}f_{K N\Sigma^*}}{m_K m_\pi} 
{k}_\alpha S^{\alpha\beta}_{\Sigma^*}(p-k^\prime){k}^\prime_{\beta} 
i \epsilon^{ijk}\tau_k\ , 
\\
\bar{V}^{19}_e&=& -g_{K^* N\Sigma}g_{K^*K\pi}
{-g_{\mu\rho} + q_\mu q_\rho/m^2_{K^*} \over q^2-m^2_{K^*}} 
\left(
\gamma^\mu - i\frac{\kappa_{{K^*}N\Sigma}}{m_N+m_\Sigma} \sigma^{\mu\nu}q_\nu
\right)
(k+k')^\rho \tau^i\tau^j\ ,
\\
\bar{V}^{19}_f &=& - \frac{g_{\kappa\Sigma N}g_{\kappa K\pi}}{m_\pi}
\frac{k\cdot k^\prime}{q^2-m_\kappa^2} \tau^i \tau^j \ .
\end{eqnarray}

\subsection{$\eta(k)+ N(p) \to K(k^\prime)+ \Sigma(p^\prime,j)$}

\begin{eqnarray}
\bar{V}(20) = \bar{V}^{20}_a + \bar{V}^{20}_b + \bar{V}^{20}_c + \bar{V}^{20}_d \ ,
\end{eqnarray}
with
\begin{eqnarray}
\bar{V}^{20}_a&=& 
{f_{K\Sigma N} f_{\eta NN}\over m_K m_\eta}
\slas{k}^\prime
\gamma_5 S_N(p+k) \slas{k}\gamma_5\tau^j \ , 
\\
\bar{V}^{20}_b&=& 
{f_{\eta \Sigma\Sigma}f_{K\Sigma N}\over m_K m_\eta}
\slas{k}
\gamma_5 S_\Sigma(p-k') \slas{k}^\prime\gamma_5  \tau^j \ ,
\\
\bar{V}^{20}_c&=& 
-g_{K^* N\Sigma}g_{K^*K\eta}
{-g_{\mu\rho} + q_\mu q_\rho/m^2_{K^*} \over q^2-m^2_{K^*}} 
\left(
\gamma^\mu - i\frac{\kappa_{{K^*}N\Sigma}}{m_N+m_\Sigma} \sigma^{\mu\nu}q_\nu
\right)
(k+k')^\rho \tau^j\ ,
\\
\bar{V}^{20}_d &=& 
-\frac{g_{\kappa\Sigma N}g_{\kappa K\eta}}{m_\pi}
\frac{k\cdot k^\prime}{q^2-m_\kappa^2} \tau^j \ .
\end{eqnarray}

\subsection{$K(k)+ \Sigma(p,i) \to K(k^\prime)+ \Lambda(p^\prime)$}

\begin{eqnarray}
\bar{V}(21) = \bar{V}^{21}_a + \bar{V}^{21}_b + \bar{V}^{21}_c \ ,
\end{eqnarray}
with
\begin{eqnarray}
\bar{V}^{21}_a&=& 
\frac{f_{K\Lambda N} f_{K\Sigma N}}{m_K^2}
\slas{k}^\prime
\gamma_5 S_N(p+k) \slas{k}\gamma_5\tau^i\ , \\
\bar{V}^{21}_b&=& 
\frac{f_{K\Xi\Sigma}f_{K\Xi\Lambda}}{m_K^2}
\slas{k}
\gamma_5\tau^i S_\Xi(p-k') \slas{k}^\prime\gamma_5\ , \\
\bar{V}^{21}_c&=& 
-g_{\rho\Sigma\Lambda}g_{\rho KK}
{-g_{\mu\rho} + q_\mu q_\rho/m^2_{\rho} \over q^2-m^2_{\rho}}
\left(
\gamma^\mu - i\frac{\kappa_{{\rho}\Sigma\Lambda}}{m_\Sigma+m_\Lambda} \sigma^{\mu\nu}q_\nu
\right)
(k+k')^\rho \tau_i\ .
\end{eqnarray}

\subsection{$K(k)+ \Sigma(p,i) \to K(k^\prime)+ \Sigma(p^\prime,j)$}

\begin{eqnarray}
\bar{V}(22) = \bar{V}^{22}_a + \bar{V}^{22}_b + \bar{V}^{22}_c +
 \bar{V}^{22}_d + \bar{V}^{22}_e\ ,
\end{eqnarray}
with
\begin{eqnarray}
\bar{V}^{22}_a&=& 
\left(\frac{f_{K\Sigma N}}{m_K} \right)^2
\slas{k}^\prime
\gamma_5\tau^j S_N(p+k) \slas{k}\gamma_5\tau^i\ , \\
\bar{V}^{22}_b&=& 
\left( \frac{f_{K\Xi\Sigma}}{m_K} \right)^2
\slas{k}
\gamma_5\tau^i S_\Xi(p-k') \slas{k}^\prime\gamma_5\tau^j\ , \\
\bar{V}^{22}_c&=& 
-g_{\omega\Sigma\Sigma}g_{\omega KK}
{-g_{\mu\rho} + q_\mu q_\rho/m^2_{\omega} \over q^2-m^2_{\omega}} 
\left(
\gamma^\mu - i\frac{\kappa_{{\omega}\Sigma\Sigma}}{2 m_\Sigma} \sigma^{\mu\nu}q_\nu
\right)
(k+k')^\rho\delta^{ij} \ , \\
\bar{V}^{22}_d&=& 
-g_{\phi\Sigma\Sigma}g_{\phi KK}
{-g_{\mu\rho} + q_\mu q_\rho/m^2_{\phi} \over q^2-m^2_{\phi}} 
\left(
\gamma^\mu - i\frac{\kappa_{{\phi}\Sigma\Sigma}}{2 m_\Sigma} \sigma^{\mu\nu}q_\nu
\right)
(k+k')^\rho\delta^{ij}\ , \\
\bar{V}^{22}_e&=& 
-g_{\rho\Sigma\Sigma}g_{\rho KK}
{-g_{\mu\rho} + q_\mu q_\rho/m^2_{\rho} \over q^2-m^2_{\rho}} 
\left(
\gamma^\mu - i\frac{\kappa_{{\rho}\Sigma\Sigma}}{2 m_\Sigma} \sigma^{\mu\nu}q_\nu
\right)
(k+k')^\rho i \epsilon^{jik}\tau_k\ .
\label{eq:v22e}
\end{eqnarray}

The baryon propagators for the spin-1/2 octet baryon $B$
and the spin-3/2 decuplet baryon $D$ appearing in Eqs.~(\ref{eq:v1})-(\ref{eq:v22e}) are
given by
\begin{eqnarray}
S_B(p) &=& \frac{1}{\slas{p} - m_B} \,,\\
S^{\mu\nu}_D(p)&=&
\frac{1}{3(\slas{p}-m_D)}
\left[
2\left(-g^{\mu\nu} +\frac{p^\mu p^\nu}{m_D^2}\right)
+\frac{\gamma^\mu\gamma^\nu-\gamma^\nu\gamma^\mu}{2}
-\frac{p^\mu\gamma^\nu-p^\nu\gamma^\mu}{m_D}
\right] \,.
\label{eq:delta-p}
\end{eqnarray}
Equation~(\ref{eq:delta-p}) is the simplest choice of many possible
definitions of the spin-3/2 propagator. It is part of our
phenomenology for this rather complex coupled-channels calculations.

Although the expressions~(\ref{eq:v1})-(\ref{eq:v22e})
look like the usual Feynman amplitudes, the
unitary transformation method~\cite{sl,sko,jklmss09a,sljpg} 
defines definite procedures in evaluating the
time component of each propagator. For each propagator, the vertex
interactions associated with its ends define  
either a ``virtual'' process or a ``real'' process.
The real process is the process that can occur in free space such as 
$\Delta \rightarrow \pi N$. The virtual processes,
such as the  $\pi N \rightarrow N$,
$\pi\Delta \rightarrow \Delta$, and $\pi \Delta \rightarrow N$
transitions, are not allowed by the energy-momentum conservation.
The consequences of the unitary transformation is the following.
When both vertex interactions  
are ``virtual,'' the propagator is the average of
the propagators calculated with two different momenta specified by
the initial and final external momenta. For example, the propagator of
$\bar{V}^1_a$ of Eq.~(\ref{eq:v1a}), which corresponds to 
the $s$ channel $\pi (k) + N (p) \to N \to \pi (k') + N(p')$, should be evaluated by
\begin{eqnarray}
S_N(p+k) &\to&\frac{1}{2}[S_N(p+k) + S_N(p'+k')] 
\nonumber \\
&=& \frac{1}{2}
\left[ 
\frac{(E_N(p)+E_\pi(k))\gamma^0 - \vec{\gamma}\cdot (\vec{p}+\vec{k})+ m_N}
{(E_N(p) +E_\pi(k))^2- (\vec{p}+\vec{k})^2 - m_N^2} 
\right.
\nonumber \\
& & 
\quad
\left.
+\frac{(E_N(p')+E_\pi(k'))\gamma^0 - \vec{\gamma}\cdot (\vec{p'}+\vec{k'}) + m_N}
{(E_N(p')+E_\pi(k'))^2- (\vec{p'}+\vec{k'})^2 - m_N^2} 
\right] \,.
\end{eqnarray}
One sees clearly that the denominators of the above expression
are independent of
the collision energy $E$ of scattering equation and 
finite in the all real momentum region. 
This is the essence of the unitary transformation method 
in deriving the interactions from Lagrangian.
When only one of the vertex interactions is ``real,'' the propagator
is evaluated by using the momenta associated with the ``virtual'' vertex.
For example, the propagator of $\bar V^{11}_d$ of Eq.~(\ref{eq:v11d}), 
which corresponds to the $s$ channel 
$\pi(k) + N(p) \to \Delta \rightarrow \pi (k') + \Delta (p')$, is
$S^{\mu\nu}_\Delta (p'+k')$, 
but neither $S^{\mu\nu}_\Delta (p+k)$ nor 
$[S^{\mu\nu}_\Delta (p'+k')+S^{\mu\nu}_\Delta (p+k)]/2$. 
We  note that there is no propagator in Eqs.~(\ref{eq:v1})-(\ref{eq:v22e}),
which is attached by two real processes such as $\pi N \to \Delta \to \pi N$.
Such real processes are included in the resonant term $t^R_{M'B',MB}$ of Eq.~(\ref{eq:tmbmb}).

\section{Matrix elements of $\gamma N \rightarrow MB$ transitions}
\label{app:pot-em}

To include the final meson-baryon interactions in the
photo-production, it is only necessary to
perform the partial-wave decomposition of the final $MB$ state.
We thus introduce the following helicity-$LSJ$ mixed representation
\begin{eqnarray}
{\it v}^{JT}_{L' S' M'B', \lambda_\gamma \lambda_N}(k',q) &=& 
\sum_{ \lambda'_M\lambda'_B } \sqrt{\frac{(2L'+1)}{2J+1}}
\inp{j'_M j'_B \lambda'_M  (-\lambda'_B)}{S' S'_z} \inp{ L'S'0S'_z }{ J S'_z } 
\nonumber \\
& & 
\qquad \qquad \qquad \qquad
\times 
\bra{ J,k' \lambda'_M (-\lambda'_B) } 
v_{M'B',\gamma N}
\ket{ J, q\lambda_\gamma (-\lambda_N) } \,,
\label{eq:e-1}
\end{eqnarray}
where 
$\bra{ J,k' \lambda'_M (-\lambda'_B) } v_{M'B',\gamma N} \ket{ J,q\lambda_\gamma (-\lambda_N) }$ 
can be evaluated using the same expression as Eq.~(\ref{eq:c-2}) but replacing $v_{M'B',MB}$
with $v_{M'B',\gamma N}$. 
To evaluate these quantities with our  normalizations of states, we define
for a photon four-momentum $q=(\omega,\vec{q})$
\begin{eqnarray}
\bra{ (k'j),p' } v_{M'B',\gamma N} \ket{ q,p } 
&=&
\frac{1}{\sqrt{2q_0}} 
\bra{(k'j),p'} \sum_{n}
J^\mu(n) \epsilon_\mu \ket{ q,p } 
\nonumber \\
&=& 
\frac{1}{(2\pi)^3}\sum_{n}
\sqrt{\frac{m_{B'}}{E_{B'}(k')}}\sqrt{\frac{1}{2 E_{M'}(k')}}
\bar{u}_{B'}(\vec{p'}) 
e I(n) u_N(\vec{p}) \sqrt{\frac{m_N}{E_N(q)}}
\frac{1}{\sqrt{2q_0}}\,,
\nonumber\\
\end{eqnarray}
where $\epsilon_\mu $ is the photon polarization vector, 
and $n$  denotes a given considered process
\begin{eqnarray}
I(n) = \epsilon \cdot \bar{j}(n)\,.
\end{eqnarray}
Here $\bar{j}(n)$ can be constructed by using the Feynman rules. 
The resulting expressions for each of 
$\gamma N \rightarrow \pi N, \eta N, \sigma N, \rho N, \pi\Delta, K\Lambda, K\Sigma$ 
are listed below.

\subsection{$\gamma(q)+N(p) \rightarrow \pi(k',j)+ N(p')$ }

\begin{eqnarray}
I(1)=I^1_a+I^1_b+I^1_c+I^1_d+I^1_e+I^1_f+I^1_g + I^1_h\ ,
\end{eqnarray}
with 
\begin{eqnarray}
I^1_a &=&
+i\frac{f_{\pi NN}}{m_\pi}
\slas{k'}\gamma_5\tau^j\frac{1}{\slas{p'}+\slas{k'} -m_N }\Gamma_N \ , 
\\
I^1_b &=&
+i\frac{f_{\pi NN}}{m_\pi}
\Gamma_N \frac{1}{\slas{p}-\slas{k'} -m_N }\slas{k'}\gamma_5\tau^j \ , 
\\
I^1_{c}&=& 
- \frac{f_{\pi N \Delta}}{m_\pi}
{\Gamma^{\mathrm{em},\Delta}_\nu}^\dagger S^{\nu\mu}_\Delta(p-k') k'_\mu T^j \ , 
\\
I^1_d &=& 
+\frac{f_{\pi NN}}{m_\pi}\epsilon_{ij3}\tau^i \slas{\epsilon_\gamma}\gamma_5 \ , 
\\
I^1_e&=&
-\frac{f_{\pi NN}}{m_\pi}
\frac{\slas{\tilde{k}} \gamma_5}{\tilde{k}^2 -m^2_\pi}\epsilon_{ij3}\tau^i(\tilde{k}+k')
\cdot\epsilon_\gamma \ , 
\\
I^1_f&=&
-\frac{g_{\rho NN}g_{\rho\pi\gamma}}{m_\pi}
\frac{\tau^j}{2} 
\left[ 
\gamma^\delta+\frac{\kappa_\rho}{4m_N}(\gamma^\delta\slas{\tilde{k}}- \slas{\tilde{k}} \gamma^\delta)
\right] 
\epsilon_{\alpha\beta\eta\delta} {\tilde{k}}^\eta q^\alpha \epsilon^\beta_\gamma 
\frac{1}{\tilde{k}^2-m^2_\rho} \ , 
\\
I^1_g&=&
-\frac{g_{\omega NN}g_{\omega\pi\gamma}}{m_\pi}
\left[
\gamma^\delta+\frac{\kappa_\omega}{4m_N}(\gamma^\delta\slas{\tilde{k}}- \slas{\tilde{k}} \gamma^\delta)
\right] 
\epsilon_{\alpha\beta\eta\delta} {\tilde{k}}^\eta q^\alpha \epsilon^\beta_\gamma \delta_{j3}
\frac{1}{\tilde{k}^2-m^2_\omega} \ . 
\end{eqnarray}
In the above equations, we introduced
$\tilde{k}=p-p'$,
$\Gamma_N = \hat{e}_N\slas{\epsilon_\gamma}-(\hat\kappa_{NN}/4m_N) 
[\slas{\epsilon_\gamma}\slas{q}-\slas{q}\slas{\epsilon_\gamma}]$,
and
$\Gamma^{\mathrm{em},\Delta}_\mu = \Gamma^{\mathrm{em},\Delta}_{\mu\nu}\epsilon^\nu_\gamma$.

\subsection{$\gamma(q)+ N(p) \rightarrow \eta(k')+ N(p')$ }

\begin{eqnarray}
I(2)=I^2_a+I^2_b+I^2_c\ ,
\end{eqnarray}
with
\begin{eqnarray}
I^2_a &=&
+i\frac{f_{\eta NN}}{m_\eta}
\slas{k'}\gamma_5\frac{1}{\slas{p'}+\slas{k'} -m_N }\Gamma_N \ , 
\\
I^2_b &=&
+i\frac{f_{\eta NN}}{m_\eta}
\Gamma_N \frac{1}{\slas{p}-\slas{k'} -m_N }\slas{k'}\gamma_5 \ , 
\\
I^2_c&=&
-\frac{g_{\rho NN}g_{\rho\eta\gamma}}{m_\rho}\frac{\tau^3}{2}
\left[
\gamma^\nu+\frac{\kappa_\rho}{4m_N}(\gamma^\nu\slas{\tilde{k}}- \slas{\tilde{k}} \gamma^\nu)
\right] 
\epsilon_{\mu\nu\alpha\beta} {\tilde{k}}^\mu q^\alpha \epsilon^\beta_\gamma
\frac{1}{\tilde{k}^2-m^2_\rho}\ . 
\end{eqnarray}

\subsection{$\gamma(q)+ N(p) \rightarrow \sigma(k')+ N(p')$ }

\begin{eqnarray}
I(3)=I^3_a+I^3_b\ ,
\end{eqnarray}
with
\begin{eqnarray}
I^3_a &=& 
-g_{\sigma NN}\frac{1}{\slas{p'}+\slas{k'} -m_N }\Gamma_N \ , 
\\
I^3_b &=&
-g_{\sigma NN}\Gamma_N \frac{1}{\slas{p}-\slas{k'} -m_N } \ .
\end{eqnarray}

\subsection{$\gamma(q)+N(p) \rightarrow \rho'(k',j)+ N(p')$ }

\begin{eqnarray}
I(4)=I^4_a+I^4_b+I^4_c+I^4_d+I^4_e+I^4_f+I^4_g\ ,
\end{eqnarray}
with
\begin{eqnarray}
I^4_a &=& 
-g_{\rho NN}\Gamma_{\rho^\prime}\frac{1}{\slas{p'}+\slas{k'} -m_N }\Gamma_N \ , 
\\
I^4_b &=&
-g_{\rho NN}\Gamma_N \frac{1}{\slas{p}-\slas{k'}-m_N } \Gamma_{\rho^\prime} \ , 
\\
I^4_{c}&=& 
+\frac{f_{\rho N\Delta}}{m_\rho}
( k'_\mu \slas{\epsilon_{\rho^\prime}}^* -\slas{k'}\epsilon_{\rho^\prime \mu}^* ) \gamma_5
{T^\dagger}^j S^{\mu\nu}_\Delta(p'+k') \Gamma^{\mathrm{em},\Delta}_\nu  \ , 
\\
I^4_{d}&=& 
-\frac{f_{\rho N\Delta}}{m_\rho}
[\Gamma^{\mathrm{em},\Delta}_\mu]^\dagger S^{\mu\nu}_\Delta(p-k')  T^j
(k'_\nu \slas{\epsilon_\rho}^*- \slas{k'} \epsilon_{\rho^\prime \nu}^*)\gamma_5  \ , \\
& & \nonumber \\
I^4_e &=& 
+i \frac{g_{\rho NN}\kappa_\rho}{8m_N}\epsilon_{ij3}\tau_i
(\slas{\epsilon_{\rho^\prime}}^*\slas{\epsilon_\gamma}
-\slas{\epsilon_\gamma}\slas{\epsilon_{\rho^\prime}}^*) \ , 
\\
I^4_f&=& 
-i\frac{g_{\rho NN}}{2}
\left[
\gamma_\mu+\frac{\kappa_\rho}{2m_N}(\gamma_\mu\slas{\tilde{k}}-\slas{\tilde{k}} \gamma_\mu)
\right] \nonumber \\
& &\times 
\left[
  \epsilon^{\mu *}_{\rho^\prime}(\tilde{k}+k')\cdot \epsilon_\gamma 
-(\tilde{k}\cdot \epsilon_{\rho^\prime}^*)\epsilon^\mu_\gamma 
-(\epsilon_\gamma \cdot\epsilon_{\rho^\prime}^*)k^{'\mu}
\right]
\frac{\epsilon_{ij3}\tau^i}{\tilde{k}^2-m^2_\rho} \ , 
\\
I^4_g&=&
+i\frac{f_{\pi NN} g_{\rho\pi\gamma}}{m_\pi^2}
     \tau^j\slas{\tilde{k}}\gamma_5\epsilon_{\alpha\beta\eta\delta}k^{'\eta}
     \epsilon_{\rho^\prime}^{\delta *} q^\alpha\epsilon^\beta_\gamma
     \frac{1}{\tilde{k}^2-m_\pi^2} \ . 
\end{eqnarray}

\subsection{$\gamma(q)+N(p) \rightarrow \pi(k',j)+ \Delta(p')$ }

\begin{eqnarray}
I(5)=I^5_a+I^5_b+I^5_c+I^5_d+I^5_e+I^5_f+I^5_g\ ,
\end{eqnarray}
with
\begin{eqnarray}
I^5_a&=& 
+i\frac{f_{\pi N\Delta}}{m_\pi}\epsilon_\Delta^*\cdot k'T^j S_N(p'+k')\Gamma_N \ , 
\\
I^5_b&=& 
+i\frac{f_{\pi N\Delta}}{m_\pi}
\Gamma^{\mathrm{em},\Delta}_\nu \epsilon^{\nu *}_\Delta S_N(p-k')\slas{k'}\gamma_5\tau^j \ , 
\\
I^5_c &=& 
-\frac{f_{\pi \Delta\Delta}}{m_\pi}
\epsilon^{*}_{\Delta \mu} \slas{k'}\gamma_5 T_\Delta^j S^{\mu\nu}_\Delta(p'+k')  
\Gamma^{\mathrm{em},\Delta}_\nu \ , 
\\
I^5_d&=&
+i\frac{f_{\pi N\Delta}}{m_\pi}
\epsilon^{*}_{\Delta\eta} \left(\frac{1}{2} + T_\Delta^3\right) 
[ - g^{\eta\mu} \slas{\epsilon}_\gamma+ (\epsilon_\gamma)^\eta \gamma^\mu]
S^\Delta_{\mu\nu}(p-k')k^{'\nu} T^j \ , 
\\
I^5_e&=&
+\frac{f_{\pi N\Delta}}{m_\pi}\epsilon_{ij3}T^i \epsilon_\gamma\cdot \epsilon_\Delta^*\ , 
\\
I^5_g&=&
-\frac{f_{\pi N\Delta}}{m_\pi}\epsilon_{ij3}T^i [V^5_g + Z^5_g]
\ , \\
I^5_g&=&
-\frac{f_{\rho N\Delta}}{m_\rho} \frac{g_{\rho\pi\gamma}}{m_\pi}T^j
\frac{1}{\tilde{k}^2-m^2_\rho}
[\tilde{k}\cdot\epsilon^*_\Delta\gamma^\mu
- \slas{\tilde{k}} \epsilon_\Delta^{*\mu}]\gamma_5\epsilon_{\alpha\beta\eta\mu}
q^\alpha\epsilon^\beta_\gamma\tilde{k}^\eta \ ,
\end{eqnarray}
where the pion pole term $I^5_g$ consists of energy independent
interaction $V^5_g$ and energy dependent interaction $Z^5_g$ given as
\begin{eqnarray}
V^5_g & = & 
\frac{1}{2E_\pi(k'-q)}
\frac{\epsilon_\Delta^*\cdot k_1 (k_1 + k')\cdot\epsilon_\gamma}{E_N(q) - E_\Delta(k') - E_\pi(k'-q)}
 +\epsilon_\Delta^{0 *}\epsilon_\gamma^0 \ , \\
Z^5_g & = &  
\frac{1}{2E_\pi(k'-q)}
\frac{\epsilon_\Delta^*\cdot k_2 (k_2 + k')\cdot\epsilon_\gamma}{E - E_N(q) - E_\pi(k') - E_\pi(k'-q)+i\epsilon}\ ,
\end{eqnarray}
with $k_1=\bm{(}E_\pi(k'-q),\vec{k}'-\vec{q}\bm{)}$ 
and $k_2=\bm{(}-E_\pi(k'-q),\vec{k}'-\vec{q}\bm{)}$.
The on-shell matrix element of $V^5_g+Z^5_g$  is given as
\begin{eqnarray}
V^5+Z^5 & = & \epsilon_\Delta^*\cdot \tilde{k}
 (\tilde{k}+k')\cdot\epsilon_\gamma\frac{1}
 {\tilde{k}^2-m^2_\pi}\ .
\end{eqnarray}

\subsection{$\gamma(q) + N(p) \rightarrow  K(k')+\Lambda(p')$}

\begin{eqnarray}
I(6) 
 & = & 
 I^6_a(1/2) + I^6_b + I^6_c  + I^6_d + I^6_e + g_{K^*K\gamma}^c I^6_f
\qquad {\rm for}\qquad \gamma p \rightarrow \Lambda K^+ \ ,
\\
 & = & 
 I^6_a(-1/2)+ I^6_b - I^6_c + g_{K^*K\gamma}^0 I^6_f
\qquad\qquad\quad\ \:  {\rm for}\qquad \gamma n \rightarrow \Lambda K^0\ ,
\end{eqnarray}
with
\begin{eqnarray}
I^6_a(t_N) &=& 
+i\frac{f_{K N \Lambda}}{m_K} \slas{k'}\gamma_5 S_N(p'+k')\Gamma_N(t_N) \ , 
\\
I^6_b &=& 
+i\frac{f_{KN\Lambda}}{m_K} \Gamma_\Lambda S_\Lambda(p-k') \slas{k'}\gamma_5 \ , 
\\
I^6_c &=&  
+i\frac{f_{KN\Sigma}}{m_K} \Gamma_{\Lambda\Sigma} S_\Sigma(p-k') \slas{k'}\gamma_5 \ , 
\\
I^6_d &=& -i\frac{f_{KN\Lambda}}{m_K}\slas{\epsilon_\gamma}\gamma_5 \ , 
\\
I^6_e&=& 
+i\frac{f_{KN\Lambda}}{m_K}
\frac{\slas{\tilde{k}}\gamma_5} {\tilde{k}^2-m^2_K}(\tilde{k}+k')\cdot\epsilon_\gamma \ , 
\\
I^6_f&=& 
-\frac{g_{K^*N\Lambda}}{m_K}
\left[
\gamma^\delta +\frac{\kappa_{K^*N\Lambda}}{2(m_N+m_\Lambda)}
 (\gamma^\delta\slas{\tilde{k}}-\slas{\tilde{k}} \gamma^\delta)
\right] 
\epsilon_{\alpha\beta\eta\delta}
{\tilde{k}}^\eta q^\alpha \epsilon^\beta_\gamma 
\frac{1}{\tilde{k}^2-m^2_{K^*}}\ ,
\end{eqnarray}
where $t_N=+(-)1/2$ for the proton (neutron).

\subsection{$\gamma(q) + N(p) \rightarrow  K(k') + \Sigma(p')$}

\begin{align}
I(7) &= I^7_a(1/2) + I^7_b + I^7_c  + I^7_d + I^7_e + g_{K^*K\gamma}^c I^7_f + I^7_g(0)
     &{\rm for}\quad \gamma p \to \Sigma^0 K^+\ , 
\\
     &= -\sqrt{2}[I^7_a(1/2)+  I^7_c + g_{K^*K\gamma}^0 I^7_f + I^7_g(+1) ]
     &{\rm for}\quad \gamma p \to \Sigma^+ K^0\ ,
\\
     &= \sqrt{2}[I^7_a(-1/2)  + I^7_c + I^7_d + I^7_e + g_{K^*K\gamma}^c I^7_f +I^7_g(-1)]
     &{\rm for}\quad \gamma n \to \Sigma^- K^+\ ,
\\
     &= -I^7_a(-1/2) + I^7_b - I^7_c - g_{K^*K\gamma}^0 I^7_f - I^7_g(0)
     &{\rm for}\quad \gamma n \rightarrow \Sigma^0 K^0\ ,
\end{align}
with
\begin{eqnarray}
I^7_a(t_N) &=& 
+i\frac{f_{K N \Sigma}}{m_K} \slas{k'}\gamma_5 S_N(p'+k')\Gamma_N(t_N) \ , 
\\
I^7_b &=& 
+i\frac{f_{KN\Lambda}}{m_K} \Gamma_{\Sigma\Lambda} S_\Lambda(p-k') \sla{k'}\gamma_5 \ , 
\\
I^7_c &=&  
+i\frac{f_{KN\Sigma}}{m_K} \Gamma_{\Sigma} S_\Sigma(p-k')\sla{k'}\gamma_5 \ , 
\\
I^7_d &=& 
-i\frac{f_{K\Lambda\Sigma}}{m_K}\sla{\epsilon_\gamma}\gamma_5 \ , 
\\
I^7_e&=& 
+i\frac{f_{K\Lambda\Sigma}}{m_K}
\frac{\sla{\tilde{k}}\gamma_5} {\tilde{k}^2-m^2_K}(\tilde{k}+k')\cdot\epsilon_\gamma\ , 
\\
I^7_f&=& 
-\frac{g_{K^*N\Sigma}}{m_K}
\left[
\gamma^\delta+\frac{\kappa_{K^*N\Lambda}}{2(m_N+m_\Lambda)}
(\gamma^\delta\sla{\tilde{k}}-\sla{\tilde{k}} \gamma^\delta)
\right] 
\epsilon_{\alpha\beta\eta\delta} {\tilde{k}}^\eta q^\alpha \epsilon^\beta_\gamma 
\frac{1}{\tilde{k}^2-m^2_{K^*}}\ .
\\
I^7_g(t_\Sigma)&=& 
- \frac{f_{K N \Sigma^\ast}}{m_K}
[\Gamma^{\mathrm{em},\Sigma^*\Sigma}_\nu(t_\Sigma)]^\dagger S^{\nu\mu}_{\Sigma^*}(p-k') k'_\mu \ , 
\end{eqnarray}
where $\Gamma^{\mathrm{em},\Sigma^*\Sigma}_\nu(t_\Sigma)$ is the matrix element 
defined in Eq.~(\ref{eq:vgDB}), which contains $G_{M,E,C}^{(\Sigma^*)^0\Sigma^0}$ for $t_\Sigma=0$
and $G_{M,E,C}^{(\Sigma^*)^\pm\Sigma^\pm}$ for $t_\Sigma=\pm 1$.

The isospin projections for the matrix elements are given as
\begin{eqnarray}
j_{1/2p} & = & \frac{1}{\sqrt{3}}\bra{K^+ \Sigma^0}j\ket{p}
            - \frac{\sqrt{2}}{\sqrt{3}}\bra{K^0 \Sigma^+}j\ket{p} \,, 
\\
j_{3/2p} & = & \frac{\sqrt{2}}{\sqrt{3}}\bra{K^+ \Sigma^0}j\ket{p}
            + \frac{1}{\sqrt{3}}\bra{K^0 \Sigma^+}j\ket{p} \,, 
\\
j_{1/2n} & = & \frac{\sqrt{2}}{\sqrt{3}}\bra{K^+ \Sigma^-}j\ket{n}
            - \frac{1}{\sqrt{3}}\bra{K^0 \Sigma^0}j\ket{n} \,, 
\\
j_{3/2n} & = & \frac{1}{\sqrt{3}}\bra{K^+ \Sigma^-}j\ket{n}
            + \frac{\sqrt{2}}{\sqrt{3}}\bra{K^0 \Sigma^0}j\ket{n} \, .
\end{eqnarray}

\section{Model parameters}
\label{app:model-para}

In this appendix, we list the model parameters determined by 
our global fits to the data of pion- and photon-induced
$\pi N$, $\eta N$, $K\Lambda$, and $K\Sigma$ reactions
in Tables~\ref{tab:model-para-mass}-\ref{tab:res-em}.

Here it will be worthwhile to mention that
the bare $N^\ast$ masses (Table~\ref{tab:res-mass}) resulting from the current analysis
are larger than those in our early analysis~\cite{jlms07} in general.
This increase of the value of the bare masses is mainly attributable to the $KY$ channels
newly included in this analysis.
The coupling of a resonance to a meson-baryon
channel produces an attractive (repulsive) mass shift
for the resonance which locates
in the complex energy plane below (above) the threshold energy
of the meson-baryon channel. 
Therefore, as a general tendency, the value of the bare mass becomes larger 
when we include a new channel whose threshold energy is higher than 
the resulting resonance masses.
\begin{table}[h]
\caption{\label{tab:model-para-mass}
Masses appearing in the meson-exchange potentials.
All the masses are kept as constant during the fit
except for the $\sigma$ and $\kappa$ masses, $m_\sigma$ and $m_\kappa$.
}
\begin{ruledtabular}
\begin{tabular}{lr}
Mass & (MeV) \\
\hline
$m_N$  & 938.5 \\
$m_\Lambda$ & 1115.7\\
$m_\Sigma$ & 1193.2\\
$m_{\Xi}$&    1318.1  \\
$m_\Delta$ & 1236.0\\
$m_{\Sigma^*}$&    1385.0  \\
$m_\pi$ & 138.5 \\
$m_\eta$ & 547.5 \\
$m_K$ & 495.7 \\
$m_\rho$ & 769.0 \\
$m_{K^*}$ & 893.9 \\
$m_\omega$ & 782.6 \\
$m_\phi$ & 1019.5 \\
$m_{f_0}$   &     974.1   \\
$m_{a_1}$   &    1260.0   \\
$m_\sigma$  &     326.2   \\
$m_{\kappa}$&     803.6 
\end{tabular}
\end{ruledtabular}
\end{table}

\begin{table}
\caption{\label{tab:nres-cc}
Fitted values of coupling constants associated with the meson-exchange potentials.
Here only the coupling constants varied and adjusted in the fit are listed.
The values of fixed coupling constants can be found in Appendix~\ref{app:lag}.
}
\begin{ruledtabular}
\begin{tabular}{lrlr}
Couplings & & Couplings&\\
\hline
$f_{\eta NN}$                                      &    0.050      &$f_{KN\Sigma}$                                     &   $-$0.140    \\
$g_{\rho NN}$                                      &    4.724      &$g_{\kappa \Lambda N}$                             &    2.674      \\
$\kappa_{\rho NN}$                                 &    1.177      &$g_{\kappa \Sigma N}$                              &   11.823      \\
$g_{\omega NN}$                                    &    5.483      &$f_{\pi\Lambda\Sigma^*}$                           &   23.060      \\
$\kappa_{\omega NN}$                               &    0.944      &$f_{K N\Sigma^*}$                                  &    0.039      \\
$g_{\sigma NN}$                                    &   13.453      &$f_{\pi\Sigma\Sigma^*}$                            &   87.891      \\
$f_{\pi N\Delta}$                                  &    1.256      &$g_{\kappa K\pi}$                                  &    0.094      \\
$g_{\rho N\Delta}$                                 &    8.260      &$g_{\kappa K\eta}$                                 &   $-$0.147    \\
$f_{\pi\Delta\Delta}$                              &    0.415      &&\\
$g_{\rho\Delta\Delta}$                             &    7.576      &$f_{KN\Sigma^*} \times G_M^{(\Sigma^*)^0\Sigma^0}$ & $-$0.286      \\
$\kappa_{\rho\Delta\Delta}$                        &    4.799      &$f_{KN\Sigma^*} \times G_M^{(\Sigma^*)^+\Sigma^+}$ & $-$0.156      \\
$g_{a_1 NN}$                                       &    8.247      &$f_{KN\Sigma^*} \times G_M^{(\Sigma^*)^-\Sigma^-}$ & $-$2.648      \\
$g_{f_0 NN}\times g_{f_0\pi\pi}$                   &  182.490      &$c_{\gamma \pi NN}$                                &    0.896      \\
$c_{\pi\rho NN}$                                   &    6.910      &$c_{\gamma KN\Lambda}$                             &   $-$0.003    \\   
$c_{\rho\rho NN}$                                  &   $-$1.052    &$c_{\gamma KN\Sigma}$                              &    0.001      \\   
$C_{V1}$                                           &    0.786      &$g_{\gamma\rho\pi}$                                &    0.128      \\
$C_{V2}$                                           &    1.531      &$g_{\gamma\omega\pi}$                              &    0.211      \\
$C_{S}\times m_\sigma^2$                           &    1.683      &$g_{\gamma\rho\eta}$                               &    1.150      \\
$c_{S31}$                                          &   -0.152      &$g_{\gamma\omega\eta}$                             &    0.237      \\
$g_{\rho\pi\pi}$                                   &    6.938      &\\
$g_{\sigma\pi\pi}$                                 &    1.173      &\\
$\tilde g_{\sigma\pi\pi}$                                &   $-$3.015    &\\
$g_{\omega\pi\rho}$                                &    4.486      &
\end{tabular}
\end{ruledtabular}
\end{table}
\begin{table}
\caption{\label{tab:nres-cu}
Fitted values of cutoff parameters associated with the meson-exchange potentials.
}
\begin{ruledtabular}
\begin{tabular}{lrlr}
Cutoffs & (MeV) & Cutoffs & (MeV)\\
\hline
$\Lambda_{\pi NN}$               &     656       &$\Lambda_{\pi\Lambda\Sigma}$     &     674       \\
$\Lambda_{\eta NN}$              &     494       &$\Lambda_{\pi\Sigma\Sigma}$      &    1716       \\
$\Lambda_{\rho NN}$              &     920       &$\Lambda_{KN\Sigma}$             &    1142       \\
$\Lambda_{\omega NN}$            &     768       &$\Lambda_{KN\Lambda}$            &     500       \\
$\Lambda_{\sigma NN}$            &    1209       &$\Lambda_{K\Xi\Lambda}$          &     538       \\
$\Lambda_{f_0 NN}$               &     680       &$\Lambda_{K\Xi\Sigma}$           &     619       \\
$\Lambda_{a_1 NN}$               &     658       &$\Lambda_{\eta\Lambda\Lambda}$   &     880       \\
$\Lambda_{\pi N\Delta}$          &     709       &$\Lambda_{\eta\Sigma\Sigma}$     &    1676       \\
$\Lambda_{\rho N\Delta}$         &    1611       &$\Lambda_{\rho\Lambda\Sigma}$    &    1699       \\
$\Lambda_{\pi\Delta\Delta}$      &     703       &$\Lambda_{\rho\Sigma\Sigma}$     &    1066       \\
$\Lambda_{\rho\Delta\Delta}$     &     755       &$\Lambda_{K^*N\Sigma}$           &     536       \\
$\Lambda_{V}$                    &    1296       &$\Lambda_{K^*N\Lambda}$          &    1492       \\
$\Lambda_{S}$                    &    1147       &$\Lambda_{\omega\Lambda\Lambda}$ &     672       \\
$\Lambda_{S31}$                  &     646       &$\Lambda_{\omega\Sigma\Sigma}$   &     601       \\
$\Lambda_{\rho\pi\pi}$           &     868       &$\Lambda_{\phi\Lambda\Lambda}$   &     587       \\
$\Lambda_{\sigma\pi\pi}$         &    1242       &$\Lambda_{\phi\Sigma\Sigma}$     &    1089       \\
$\Lambda'_{\sigma\pi\pi}$        &    1986       &$\Lambda_{\kappa N\Lambda}$      &    1381       \\
$\Lambda_{f_0 \pi\pi}$           &    1268       &$\Lambda_{\kappa N\Sigma}$       &     800       \\
$\Lambda_{\omega\pi\rho}$        &     620       &$\Lambda_{\pi\Lambda\Sigma^*}$   &    1582       \\
$\Lambda^{\text{em}}_{\pi NN}$   &     527       &$\Lambda_{KN\Sigma^*}$           &     722       \\
$\Lambda_{\gamma\pi NN}$         &     883       &$\Lambda_{\pi\Sigma\Sigma^*}$    &     814       \\
&&$\Lambda_{\rho KK}$              &    1600       \\
&&$\Lambda_{K^*K\pi}$              &    1455       \\
&&$\Lambda_{\omega KK}$            &     500       \\
&&$\Lambda_{\phi KK}$              &     500       \\
&&$\Lambda_{K^*K\eta}$             &     500       \\
&&$\Lambda_{\kappa K\pi}$          &    1131       \\
&&$\Lambda_{\kappa K\eta}$         &    1490       \\
\end{tabular}
\end{ruledtabular}
\end{table}
\begin{table}
\caption{\label{tab:res-mass}
Fitted values of bare mass $M_{N^\ast}^0$ of the $N^\ast$ states.
}
\begin{ruledtabular}
\begin{tabular}{lr}
$L_{2I2J}$  & $M_{N^\ast}^0$ (MeV) \\ 
\hline
$S_{11}$ (1)& 2400      \\
$S_{11}$ (2)& 2878      \\
$P_{11}$ (1)& 2210      \\
$P_{11}$ (2)& 2440      \\
$P_{13}$ (1)& 2087      \\
$P_{13}$ (2)& 2901      \\
$D_{13}$ (1)& 2480      \\
$D_{13}$ (2)& 3926      \\
$D_{15}$ (1)& 2058      \\
$D_{15}$ (2)& 3258      \\
$F_{15}$    & 2292      \\
$F_{17}$    & 2629      \\
$G_{17}$    & 2785      \\
$G_{19}$    & 2733      \\
$H_{19}$    & 2967      \\
$S_{31}$    & 2100      \\
$P_{31}$ (1)& 2374      \\
$P_{31}$ (2)& 3287      \\
$P_{33}$ (1)& 1600      \\
$P_{33}$ (2)& 2397      \\
$D_{33}$ (1)& 2403      \\
$D_{33}$ (2)& 2564      \\
$D_{35}$ (1)& 2231      \\
$D_{35}$ (2)& 3700      \\
$F_{35}$ (1)& 3131      \\
$F_{35}$ (2)& 3710      \\
$F_{37}$ (1)& 2076      \\
$F_{37}$ (2)& 2698      \\
$G_{37}$    & 3218      \\
$G_{39}$    & 3541      \\
$H_{39}$    & 3400       
\end{tabular}
\end{ruledtabular}
\end{table}

\begin{table}
\caption{\label{tab:res-mb}
Fitted values of cutoffs and coupling constants of the bare $N^\ast \to MB$ vertices 
($MB = \pi N, \eta N, \pi \Delta, \sigma N, \rho N, K\Lambda, K\Sigma$). 
The corresponding $(LS)$ quantum numbers of each $MB$ state are shown in Table~\ref{tab:pw}.
The cutoff $\Lambda_{N^*}$ is listed in the unit of MeV.
}
\begin{ruledtabular}
\begin{tabular}{lrrrrrrrrrrr}
$L_{2I2J}$ &$\Lambda_{N^*}$ & \multicolumn{10}{c}{$C_{MB(LS),N^*}$}\\
\cline{3-12}
  & &$\pi N$&$\eta N$&$(\pi\Delta)_1$&$(\pi\Delta)_2$&$\sigma N$&$(\rho N)_1$&$(\rho N)_2$&$(\rho N)_3$& $K\Lambda$& $K\Sigma$\\
\hline
$S_{11}$ (1)&2000&  12.412&   6.374&$-$0.021&     -  &   0.621&    0.457&$-$0.117&      - &   2.324&   0.352\\      
$S_{11}$ (2)& 750&$-$3.562&   7.234&   1.124&     -  &$-$8.903&    5.207&$-$5.024&      - &$-$0.371&   4.968\\      
$P_{11}$ (1)&1179&   2.809&   1.286&   0.743&     -  &   0.293&    3.209&$-$0.600&      - &   1.124&   0.441\\      
$P_{11}$ (2)& 517&  12.312&$-$0.479&   9.266&     -  &   9.984&   12.498&   3.210&      - &   4.545&   2.500\\      
$P_{13}$ (1)& 518&   7.139&$-$0.248&   8.003&$-$0.929&$-$2.089&    1.341&   0.833&   1.487&   1.237&   0.820\\      
$P_{13}$ (2)&1170&   1.426&   0.089&$-$4.735&   0.147&$-$0.707& $-$0.931&$-$0.300&$-$0.069&   0.543&   0.406\\      
$D_{13}$ (1)&1519&   0.178&   0.125&   4.563&$-$0.040&   0.658&    0.126&$-$9.865&$-$0.230&   0.157&$-$0.063\\      
$D_{13}$ (2)&1614&   0.254&   0.234&   2.063&   0.164&   0.333& $-$0.392&   1.554&$-$0.417&   0.154&$-$0.134\\      
$D_{15}$ (1)& 613&   1.006&$-$0.475&$-$2.414&   0.001&$-$0.772& $-$0.808&   1.749&   0.215&   0.011&$-$0.065\\      
$D_{15}$ (2)&1286&   0.309&$-$0.004&$-$0.523&$-$0.001&$-$0.084&    0.283&$-$0.182&   0.011&   0.000&   0.086\\      
$F_{15}$    &1003&   0.145&$-$0.010&   0.836&$-$0.178&   0.567& $-$0.161&   1.463&$-$0.005&   0.007&$-$0.017\\      
$F_{17}$    &1028&   0.004&   0.007&$-$0.141&   0.004&$-$0.005&    0.064&   0.025&$-$0.002&   0.046&   0.030\\      
$G_{17}$    &1191&   0.008&   0.001&$-$0.385&$-$0.009&   0.055& $-$0.004&$-$0.436&   0.001&   0.001&   0.003\\      
$G_{19}$    & 874&   0.013&   0.000&$-$0.081&   0.000&$-$0.000& $-$0.017&   0.019&$-$0.000&$-$0.009&   0.012\\      
$H_{19}$    &1110&   0.002&   0.000&$-$0.043&$-$0.002&   0.010& $-$0.001&$-$0.127&   0.000&$-$0.000&   0.001\\      
$S_{31}$    & 657&   0.000&     -  &$-$3.829&     -  &     -  &$-$20.000&   1.355&      - &      - &$-$1.927\\      
$P_{31}$ (1)& 725&   1.205&     -  &   7.353&     -  &     -  &    4.252&   0.476&      - &      - &   2.386\\      
$P_{31}$ (2)&2000&   1.651&     -  &   1.691&     -  &     -  &    0.607&   0.653&      - &      - &   1.259\\      
$P_{33}$ (1)& 878&   1.038&     -  &$-$2.662&   0.183&     -  &    0.544&   6.000&   0.033&      - &$-$0.005\\      
$P_{33}$ (2)& 739&  10.069&     -  &   5.907&   0.308&     -  &    1.621&   5.346&   0.246&      - &   0.063\\      
$D_{33}$ (1)&1086&   0.512&     -  &  13.498&$-$0.060&     -  & $-$0.272&$-$8.482&   0.963&      - &   0.193\\      
$D_{33}$ (2)& 629&   0.144&     -  &$-$1.599&$-$0.260&     -  &    0.933&   3.901&$-$2.794&      - &$-$0.523\\      
$D_{35}$ (1)& 641&   0.573&     -  &$-$0.527&   0.029&     -  &    0.917&$-$0.312&   0.213&      - &   0.490\\      
$D_{35}$ (2)&1098&   0.602&     -  &$-$0.268&$-$0.004&     -  & $-$0.606&   0.543&$-$0.008&      - &$-$0.201\\      
$F_{35}$ (1)&1026&   0.032&     -  &   3.181&$-$0.284&     -  &    0.114&   3.558&   0.088&      - &   0.083\\      
$F_{35}$ (2)&1631&   0.048&     -  &   2.872&   0.001&     -  & $-$0.001&$-$0.233&$-$0.005&      - &   0.006\\      
$F_{37}$ (1)& 778&   0.216&     -  &   0.308&   0.004&     -  &    0.019&$-$0.072&   0.002&      - &   0.026\\      
$F_{37}$ (2)& 903&   0.145&     -  &   0.208&$-$0.001&     -  &    0.131&   0.110&$-$0.000&      - &$-$0.052\\      
$G_{37}$    &1203&   0.006&     -  &$-$0.384&$-$0.016&     -  &    0.002&   0.480&   0.002&      - &   0.002\\      
$G_{39}$    & 874&   0.022&     -  &$-$0.108&   0.000&     -  &    0.004&$-$0.003&   0.000&      - &$-$0.003\\      
$H_{39}$    & 973&   0.001&     -  &$-$0.086&$-$0.010&     -  &    0.001&   0.167&   0.001&      - &   0.004
\end{tabular}
\end{ruledtabular}
\end{table}
\begin{table}
\caption{\label{tab:res-em}
Fitted values of model parameters associated with the bare $\gamma N \to N^*$ helicity 
amplitudes defined in Eqs.~(\ref{eq:nstar-gn})-(\ref{eq:E-para}).
In the last column, fixed values of $M_{N^*}$ that are used for
computing $q_R$ in Eq.~(\ref{eq:nstar-gn}) are presented.
The helicity amplitudes $\tilde A_{1/2}$ and $\tilde A_{3/2}$ shown 
in the third and the fourth columns are $A_{1/2}$ and $A_{3/2}$ 
calculated at $q=m_\pi$, where $\tilde M^{N^*}_{l\pm}$ and $\tilde E^{N^*}_{l\pm}$
are obtained with the use of the relations in Eqs.~(\ref{eq:a32-me+})-(\ref{eq:a12-me-}).
As for the 1st bare $P_{33}$ state [$P_{33}(1)$], we list
the value of $x_{A_{1/2}}$ and $x_{A_{3/2}}$ in the columns
of $\tilde A^{N^\ast}_{1/2}$ and $\tilde A^{N^\ast}_{3/2}$, respectively.
}
\begin{ruledtabular}
\begin{tabular}{lrrrr}
$L_{2I2J}$  & 
$\Lambda_{N^\ast}^{\text{e.m.}}$ (MeV) &
$\tilde A^{N^\ast}_{1/2}$ ($10^{-3}$ GeV$^{-1/2}$) &
$\tilde A^{N^\ast}_{3/2}$ ($10^{-3}$ GeV$^{-1/2}$) &
$M_{N^\ast}$ (MeV) \\
\hline
$S_{11}$ (1)&  634&   78.79&      -  &1535\\      
$S_{11}$ (2)& 1595& $-$2.92&      -  &1650\\      
$P_{11}$ (1)& 1035& $-$1.83&      -  &1440\\      
$P_{11}$ (2)& 1558&   56.60&      -  &1710\\      
$P_{13}$ (1)& 1208&    3.07&     1.67&1720\\      
$P_{13}$ (2)&  510&  258.77& $-$22.46&1900\\      
$D_{13}$ (1)&  538&   27.24& $-$59.49&1520\\      
$D_{13}$ (2)&  986& $-$7.53&     4.00&1700\\      
$D_{15}$ (1)&  655&    0.62&     7.40&1675\\      
$D_{15}$ (2)& 1099&    5.39&     5.03&2200\\      
$F_{15}$    &  569&    7.22& $-$10.18&1680\\      
$F_{17}$    &  870&    0.35&     0.12&1990\\      
$G_{17}$    &  510&    4.71&  $-$5.89&2190\\      
$G_{19}$    &  505& $-$3.05&  $-$5.09&2250\\      
$H_{19}$    & 1589&    0.03&     0.03&2220\\      
$S_{31}$    &  500&$-$291.50&    -   &1620\\      
$P_{31}$ (1)& 1599&    6.49&     -   &1750\\      
$P_{31}$ (2)& 1600&    4.04&     -   &1910\\      
$P_{33}$ (1)&    -&    1.10&     1.36&1238\\      
$P_{33}$ (2)& 1670&$-$32.08&$-$59.12 &1600\\      
$D_{33}$ (1)& 1615&$-$11.53&$-$18.14 &1700\\      
$D_{33}$ (2)& 1374& $-$6.31&$-$44.47 &1940\\      
$D_{35}$ (1)&  826&    8.41&    8.40 &1935\\      
$D_{35}$ (2)&  594& $-$1.97&$-$30.00 &1935\\      
$F_{35}$ (1)& 1601&    1.08&    1.84 &1905\\      
$F_{35}$ (2)& 1306& $-$1.01& $-$1.81 &2000\\      
$F_{37}$ (1)& 1562& $-$0.49& $-$0.55 &1950\\      
$F_{37}$ (2)&  703& $-$5.36& $-$8.43 &2390\\      
$G_{37}$    &  500&   12.07&   10.74 &2200\\      
$G_{39}$    &  803&    0.75&    1.39 &2400\\      
$H_{39}$    &  518& $-$2.34& $-$1.67 &2300 
\end{tabular}
\end{ruledtabular}
\end{table}


\clearpage




\begin{thebibliography}{99}
\bibitem{msl07}
A.~Matsuyama, T.~Sato, and T.-S.~H.~Lee,
Phys. Rep. {\bf 439}, 193 (2007).

\bibitem{jlms07}
B.~Juli\'a-D\'iaz, T.-S.~H.~Lee, A.~Matsuyama, and T.~Sato,
Phys. Rev. C {\bf 76}, 065201 (2007).

\bibitem{jlmss08}
B.~Juli\'a-D\'iaz, T.-S. H. Lee,A. Matsuyama, T.~Sato, and L.~C.~Smith,
Phys. Rev.  C {\bf 77}, 045205 (2008).

\bibitem{jklmss09}
B.~Juli\'a-D\'iaz, H. Kamano, T.-S.~H.~Lee, A.~Matsuyama, T.~Sato, and N. Suzuki,
Phys. Rev. C {\bf 80}, 025207 (2009).

\bibitem{djlss08}
J.~Durand, B.~Juli\'a-D\'iaz, T.-S.~H.~Lee, B.~Saghai, and T.~Sato,
Phys. Rev. C {\bf 78}, 025204 (2008).

\bibitem{kjlms09-1}
H.~Kamano, B.~Juli\'a-D\'iaz, T.-S.~H.~Lee, A.~Matsuyama, and T.~Sato,
Phys. Rev. C {\bf 79}, 025206 (2009).

\bibitem{kjlms09-2}
H.~Kamano, B.~Juli\'a-D\'iaz, T.-S.~H.~Lee, A.~Matsuyama, and T.~Sato,
Phys. Rev. C {\bf 80}, 065203 (2009).

\bibitem{ssl09}
N.~Suzuki, T.~Sato, and T.-S.~H.~Lee,
Phys. Rev. C {\bf 79}, 025205 (2009).

\bibitem{ssl10}
N. Suzuki, T. Sato, and T.-S.H. Lee,
Phys. Rev. C {\bf 82}, 045206 (2010).

\bibitem{sjklms10}
N. Suzuki, B. Juli\'a-D\'iaz, H. Kamano, T.-S. H. Lee, A. Matsuyama, and T. Sato,
Phys. Rev. Lett. {\bf 104}, 042302 (2010).

\bibitem{knls10}
H.~Kamano, S.~X.~Nakamura, T.-S.~H.~Lee, and T.~Sato,
Phys. Rev. C {\bf 81}, 065207 (2010).

\bibitem{pj-91}
B.~C.~Pearce and B.~K.~Jennings, 
Nucl. Phys. {\bf A528}, 655 (1991).

\bibitem{ntuanl}
C.~C.~Lee, S.~N.~Yang, and T.-S.~H.~Lee, J. Phys. {\bf G17} L131 (1991);
C.~T.~Hung, S.~N.~Yang, and T.-S.~H.~Lee, Phys. Rev. C {\bf 64}, 034309 (2001).

\bibitem{gross}
F.~Gross and Y.~Surya, 
Phys. Rev. C {\bf 47}, 703 (1993).

\bibitem{sl}
T.~Sato and T.-S.~H.~Lee, 
Phys. Rev. C {\bf 54}, 2660 (1996); {\bf 63}, 055201 (2001);
B.~Juli\'a-D\'iaz, T.-S.~H.~Lee, T.~Sato, and L.~C.~Smith,
{\it ibid.} {\bf 75}, 015205 (2007).

\bibitem{fuda}
Y.~Elmessiri and M.~G.~Fuda, Phys. Rev. C {\bf 60}, 044001 (1999);
M.~G.~Fuda and H.~Alharbi, Phys. Rev. C {\bf 68}, 064002 (2003).

\bibitem{pasc}
V.~Pascalutsa and J.~A.~Tjon, 
Phys. Rev. C {\bf 61}, 054003 (2000);
G.~L.~Caia, L.~E.~Wright, and V.~Pascalutsa, 
Phys. Rev. C {\bf 72}, 035203 (2005).

\bibitem{pitt-ky}
W.-T. Chiang, F.~Tabakin, T.-S.~H.~Lee, B.~Saghai, 
Phys. Lett. {\bf B517}, 101 (2001); 
B.~Juli\'a-D\'iaz, B.~Saghai, T.-S.~H.~Lee, F.~Tabakin,
Phys. Rev. C {\bf 73}, 055204 (2006).

\bibitem{juelich}
A.~M.~Gasparyan, J.~Haidenbauer, C.~Hanhart, and J.~Speth,
Phys. Rev. C {\bf 68}, 045207 (2003);
M.~D\"oring, C.~Hanhart, F.~Huang, S.~Krewald, and U.-G.~Mei{\ss}ner,
Nucl. Phys. {\bf A829}, 170 (2009);
M.~D\"oring, C.~Hanhart, F.~Huang, S.~Krewald, U.-G.~Mei{\ss}ner, and D.~R\"onchen,
Nucl.\ Phys.\ A {\bf 851}, 58 (2011);
F.~Huang, M.~D\"oring, H.~Haberzettl, J.~Haidenbauer, C.~Hanhart, S.~Krewald, U.-G.~Mei{\ss}ner, and K.~Nakayama,
Phys.\ Rev.\ C {\bf 85}, 054003 (2012).

\bibitem{juelich13-1}
D.~R\"onchen, M.~D\"oring, F.~Huang, H.~Haberzettl, J.~Haidenbauer, C.~Hanhart, S.~Krewald, U.-G.~Mei{\ss}ner, and K. Nakayama,
Eur.\ Phys.\ J.\ A {\bf 49}, 44 (2013).

\bibitem{said}
R.~A.~Arndt, J.~M.~Ford, and L.~D.~Roper,
Phys.\ Rev.\ D {\bf 32}, 1085 (1985);
R.~A.~Arndt, W.~J.~Briscoe, I.~I.~Strakovsky, and R.~L.~Workman,
Phys.\ Rev.\ C {\bf 74}, 045205 (2006); references therein.

\bibitem{bonn}
A.~V.~Anisovich, A.~V.~Sarantsev, O.~Bartholomy, E.~Klempt, V.~A.~Nikonov, and U.~Thoma, 
Eur. Phys. J. A {\bf 25}, 427 (2005);
A.~V.~Sarantsev {\it et al.},
Phys.\ Lett.\ {\bf B659}, 94 (2008);
A.~V.~Anisovich, E.~Klempt, V.~Kuznetsov, V.~A.~Nikonov, M.~V.~Polyakov, A.~V.~Sarantsev, and U.~Thoma,
Phys.\ Lett.\ B {\bf 719}, 89 (2013).
A.~V.~Anisovich, R.~Beck, E.~Klempt, V.~A.~Nikonov, A.~V.~Sarantsev, and U.~Thoma;
Eur.\ Phys.\ J.\ A {\bf 48}, 88 (2012);
A.~V.~Anisovich, E.~Klempt, V.~A.~Nikonov, A.~V.~Sarantsev, H.~Schmieden, and U.~Thoma,
Phys.\ Lett.\ B {\bf 711}, 162 (2012);
A.~V.~Anisovich, E.~Klempt, V.~A.~Nikonov, A.~V.~Sarantsev, and U.~Thoma,
Phys.\ Lett.\ B {\bf 711}, 167 (2012);
A.~V.~Anisovich, E.~Klempt, V.~A.~Nikonov, A.~V.~Sarantsev, and U.~Thoma,
Eur.\ Phys.\ J.\ A {\bf 47}, 153 (2011);
A.~V.~Anisovich, E.~Klempt, V.~A.~Nikonov, A.~V.~Sarantsev, and U.~Thoma,
Eur.\ Phys.\ J.\ A {\bf 47}, 27 (2011);
A.~V.~Anisovich, E.~Klempt, V.~A.~Nikonov, M.~A.~Matveev, A.~V.~Sarantsev, and U.~Thoma,
Eur.\ Phys.\ J.\ A {\bf 44}, 203 (2010).

\bibitem{bg2012}
A.~V.~Anisovich, R.~Beck, E.~Klempt, V.~A.~Nikonov, A.~V.~Sarantsev, and U. Thoma, 
Eur. Phys. J. A {\bf 48}, 15 (2012); references therein.

\bibitem{ksu}
D.~M.~Manley, R.~A.~Arndt, Y.~Goradia, and V.~L.~Teplitz,
Phys. Rev. D {\bf 30}, 904 (1984);
D.~M.~Manley, E.~M.~Saleski,
{\it ibid.} {\bf 45}, 4002 (1992).

\bibitem{maid}
D. Drechsel, O. Hanstein, S.S. Kamalov, and L. Tiator, Nucl. Phys. {\bf A645},
145 (1999);
S.S. Kamalov, S. N. Yang, D. Drechsel, O. Hanstein, and L. Tiator,
Phys, Rev, C {\bf 64}, 032201(R) (2001).

\bibitem{jlab-yeve}
I. G. Aznauryan, Phys. Rev. C {\bf 68}, 065204 (2003).

\bibitem{mokeev}
V. I. Mokeev, V. D. Burkert, T. -S. H. Lee, L. Elouadrhiri, G. V. Fedotov, 
and B. S. Ishkhanov
Phys. Rev. C{\bf  80}, 045212 (2009)

\bibitem{giessen}
T.~Feuster and U.~Mosel, 
Phys. Rev. C {\bf 58}, 457 (1998); {\bf 59}, 460-491 (1999);
V.~Shklyar, H.~Lenske, U.~Mosel, and G.~Penner,
{\it ibid.} {\bf 71}, 055206 (2005).

\bibitem{kvi}
A. Usov and O. Scholten, Phys. Rev. C {\bf 72}, 025205 (2005).

\bibitem{cmb}
R.~E.~Cutkosky, C.~P.~Forsyth, R.~E.~Hendrick, and R.~L.~Kelly,
Phys. Rev. D {\bf 20}, 2839 (1979);
R.~E.~Cutkosky and S.~Wang,
Phys. Rev. D {\bf 42}, 235 (1990).

\bibitem{zegrab}
M.~Batinic, I.~Slaus, A.~Svarc, and B.~M.~K.~Nefkens,
Phys.\ Rev.\ C {\bf 51}, 2310 (1995);
M.~Batinic, I.~Dadic, I.~Slaus, A.~Svarc, B.~M.~K.~Nefkens,
and T.-S.~H.~Lee,
Phys. Scr. {\bf 58}, 15 (1998);
S.~Ceci, A.~Svarc, B.~Zauner,
Phys.\ Rev.\ Lett.\ {\bf 97}, 062002 (2006).

\bibitem{pitt-anl}
T.~P.~Vrana, S.~A.~Dytman, T.-S.~H.~Lee,
Phys. Rep. {\bf 328}, 181 (2000).

\bibitem{rccr11}
C.~D.~Roberts, I.~C.~Clo\"et, L.~Chang, and H.~L.~L.~Roberts, 
AIP Conf. Proc. {\bf 1432}, 309 (2012).

\bibitem{feshbach}
H. Feshbach, {\it Theoretical Nuclear Physics, Nuclear Reactions}
(Wiley, New York, 1992)

\bibitem{sko}
M.~Kobayashi, T.~Sato, and H.~Ohtsubo, 
Prog. Theor. Phys. {\bf 98}, 927 (1997).

\bibitem{jklmss09a}
B.~Juli\'a-D\'iaz, H.~Kamano, T.-S.~H.~Lee, A.~Matsuyama, T.~Sato, and N.~Suzuki,
Chin. J. Phys. {\bf 47}, 142 (2009).

\bibitem{sljpg}
T.~Sato and T.-S.~H.~Lee,
J. Phys. G {\bf 36}, 073001 (2009).

\bibitem{bohm}
A.~Bohm,
{\it Quantum mechanics: foundations and applications} (Springer-Verlag, New York, 1993).

\bibitem{dalitz} 
R.~H.~Dalitz and R.~G.~Moorhouse,
Proc. Roy. Soc. Lond. {\bf A318}, 279 (1970);
A.~J.~F. Siegert, Phys. Rev. {\bf 56}, 750 (1939).

\bibitem{saiddb}
CNS Data Analysis Center, George Washington University University,
http://gwdac.phys.gwu.edu

\bibitem{hohler}
G. Hohler, F. Kaiser, R. Koch, and E. Pietarinen,
{\it Handbook of Pion Nucleon Scattering}, Physics Data Vol.12 (Karlsruhe, 1979);
G. H{\"o}hler,
{\it Pion-nucleon scattering} (Springer-Verlag, Berlin 1983), Vol. I/92;
G. Hoehler and A. Schulte,
$\pi N$ Newsletter {\bf 7} (1992) 94;
G. Hoehler,
$\pi N$ Newsletter {\bf 9} (1993) 1.

\bibitem{cmb-pin}
R.~E.~Cutkosky, R.~E.~Hendrick, J.~W.~Alcock, Y.~A.~Chao, R.~G.~Lipes, J.~C.~Sandusky, and R.~L.~Kelly, 
Phys. Rev. D {\bf 20}, 2804 (1979); 
R.~L.~Kelly and R.~E.~Cutkosky, {\it ibid.} {\bf 20}, 2782 (1979).

\bibitem{cbelsa2012}
A.~Thiel \textit{et al.} (CBELSA/TAPS Collaboration), 
Phys. Rev. Lett. {\bf 109}, 102001 (2012).

\bibitem{shkl11}
A.~M.~Sandorfi, S.~Hoblit, H.~Kamano, and T.-S.~H.~Lee,
J. Phys. G {\bf 38}, 053001 (2011).

\bibitem{rosetta}
A.~M.~Sandorfi, B.~Dey, A.~Sarantsev,~L.~Tiator, and R.~Workman,
AIP Conf. Proc. {\bf 1432}, 219 (2012);
see arXiv:1108.5411v2 for the updated version.

\bibitem{ehat}
J.~Ahrens {\it et al.} (GDH and A2 Collaboration), 
Phys. Rev. Lett. {\bf 88} 232002 (2002);
Eur. Phys. J. A {\bf 21}, 323 (2004);
Phys. Rev. C {\bf 74} 045204 (2006).

\bibitem{jparc-p45}
K.~Hicks and H.~Sako {\it et al.},
Proposal for measurements of $\pi N\to\pi\pi N$ and $\pi N \to K Y$ at J-PARC
(J-PARC P45), http://j-parc.jp/researcher/Hadron/en/pac\_1207/pdf/P45\_2012-3.pdf .

\bibitem{andy-proc}
A.~Sandorfi,
J. Phys.: Conf. Ser. {\bf 424}, 012001 (2013).

\bibitem{pdg}
J. Beringer {\it et al.} (Particle Data Group), 
Phys. Rev. D{\bf 86}, 010001 (2012).

\bibitem{inna}
I.~G.~Aznauryan, V.~D.~Burkert, and T.-S.~H.~Lee,
arXiv:0810.0997.

\bibitem{maid07}
D.~Drechsel, S.~S.~Kamalov, and L.~Tiator, 
Eur. Phys. J. A {\bf 34}, 69 (2007).

\bibitem{gwu-said}
M.~Dugger {\it et al.} (CLAS Collaboration),
Phys. Rev. C {\bf 76}, 025211 (2007).

\bibitem{dse13}
J.~Segovia, C.~Chen, C.~D.~Roberts, and S.~Wan,
Phys. Rev. C {\bf 88}, 032201(R) (2013).

\bibitem{tabakin}
W.-T.~Chiang and F.~Tabakin, 
Phys. Rev. C {\bf 55} (1997) 2054.

\bibitem{cmu-b}
B.~Dey, M.~E.~McCracken, D.~G.~Ireland, and C.~A.~Meyer,
Phys. Rev. C {\bf 83}, 055208 (2011).

\bibitem{maid-said}
R.~L.~Workman, M.~W.~Paris, W.~J.~Briscoe, L.~Tiator, S.~Schumann, M.~Ostrick, and S.~S.~Kamalov,
Eur. Phys. J. A {\bf 47} 143 (2011).

\bibitem{knls11}
H.~Kamano, S.~X.~Nakamura, T.-S.~H.~Lee, and T.~Sato, 
Phys. Rev. D{\bf 84}, 114019 (2011).

\bibitem{knls12}
S.~X.~Nakamura, H.~Kamano, T.-S.~H.~Lee, and T.~Sato,
Phys. Rev. D{\bf 86}, 114012 (2012).

\bibitem{lee-review}
V.~D.~Burkert and T.-S.~H.~Lee, 
Int. J. of Mod. Phys. E {\bf 13}, 1035 (2004).

\bibitem{emnn}
I.~G.~Aznauryan {\it et al.}, 
Int. J. Mod. Phys. E {\bf 22} 1330015 (2013); arXiv:0907.1901.

\bibitem{neuint}
S.~X.~Nakamura, Y.~Hayato, M.~Hirai, H.~Kamano, S.~Kumano, M.~Sakuda, K.~Saito, and T.~Sato,
arXiv:1303.6032.

\bibitem{sul03}
T.~Sato, D.~Uno, and T.-S.~H. Lee,
Phys. Rev. C {\bf 67}, 065201 (2003);

\bibitem{msl05}
K.~Matsui, T.~Sato, and T.-S.~H.~Lee,
Phys. Rev. C {\bf 72}, 025204 (2005).

\bibitem{knls-neutrino}
H.~Kamano, S.~X.~Nakamura, T.-S.~H.~Lee, and T.~Sato,
Phys. Rev. D{\bf 86}, 097503 (2012).

\bibitem{kl12}
H.~Kamano and T.-S.~H.~Lee, 
AIP Conf. Proc. {\bf 1432}, 74 (2012).

\bibitem{betz-lee}
M.~Betz and T.-S.~H.~Lee,
Phys. Rev. C {\bf 23}, 375 (1981).

\bibitem{bj}
J.~D.~Bjorken and S.~D.~Drell, {\it Relativistic Quantum Field Theory}
(McGraw-Hill, New York, 1964).

\bibitem{onl}
Y.~Oh, K.~Nakayama, and T.-S.~H.~Lee,
Phys. Rep. {\bf 423}, 49 (2006).

\bibitem{jw}
M.~Jacob and G.~C.~Wick, 
Ann. Phys. {\bf 7}, 404 (1959); {\bf 281}, 774 (2000).


\end{thebibliography}
\end{document}